\documentstyle[epsfig,12pt,preprint,tighten,aps]{revtex}
\begin{document}

\draft

\title{\rightline{{\tt April 1999}}
\rightline{{\tt UM-P-99/06}}
\rightline{{\tt RCHEP-99/06}}
\ \\
Implications of mirror neutrinos for Early Universe Cosmology}
\author{R Foot and R. R. Volkas}
\address{School of Physics\\
Research Centre for High Energy Physics\\
The University of Melbourne\\
Parkville 3052 Australia\\
(foot@physics.unimelb.edu.au, r.volkas@physics.unimelb.edu.au)}

\maketitle

\begin{abstract}
The Exact Parity Model (EPM) is, in part, a theory 
of neutrino mass and mixing that can solve
the atmospheric, solar and LSND anomalies. 
The central feature of the neutrino sector is three pairs
of maximally mixed ordinary and mirror neutrinos. 
It has been shown that ordinary-mirror
neutrino oscillations can generate large neutrino asymmetries 
in the epoch of the early universe
immediately prior to Big Bang Nucleosynthesis (BBN). 
The large neutrino asymmetries generically 
suppress the production of mirror neutrinos, and 
a sufficiently large $\nu_e$ asymmetry can
directly affect light element synthesis through nuclear 
reaction rates. In this paper we present
a detailed calculation of neutrino asymmetry evolution driven 
by the six-flavour EPM neutrino
sector, focusing on implications for BBN.

\end{abstract}

\newpage
\section{Introduction}

It has been known for a long time, but not widely appreciated, that parity can be a symmetry
of nature if the particle content is doubled\cite{ly,blin,flv,flv2}. In this
circumstance, for each ordinary particle there is a mirror particle of exactly the same mass
as the corresponding ordinary particle. The set of mirror particles interact with each other
in exactly the same way that ordinary particles interact with themselves.  The mirror
particles are not copiously produced in any laboratory experiments because they either do
not couple, or couple extremely weakly, to the ordinary particles.  In the modern language
of gauge theory, the mirror particles are all gauge singlets under the standard model $G =
\text{SU}(3)_c \otimes \text{SU}(2)_L \otimes \text{U}(1)_Y$ gauge interactions. Instead,
the mirror particles interact with a set of mirror gauge particles. This is mathematically
described by a doubled gauge symmetry of the theory, that is $G$ is extended to $G \otimes
G$. (The ordinary particles are of course singlets under the mirror gauge symmetry.) Parity
is conserved because the mirror particles experience $V+A$ mirror weak interactions instead
of the usual $V-A$ weak interactions.

The ordinary and mirror sectors can interact with each other in a number of ways. All of
these interactions apart from gravity can be controlled by an {\it a priori} arbitrary
parameter. Apart from the irremovable gravitational interaction, there are three other ways
in which ordinary and mirror particles can interact with each other. Two of these are:
photon - mirror photon (and $Z$ - mirror $Z$) kinetic mixing \cite{flv,gl} and Higgs -
mirror Higgs mass mixing \cite{flv}. If one demands that the reasonably successful Big Bang
Nucleosynthesis (BBN) predictions not be greatly disturbed then it seems unlikely that these
interactions can have observable laboratory implications. See Refs.\cite{cg,lew} for
details.\footnote{In particular extensions of the exact parity idea, neutral gauge boson
kinetic mixing and/or Higgs mixing may not be controlled by an independent arbitrary
parameter. For instance, in exact parity extensions of grand unified models such as SU(5)
the kinetic mixing parameter is calculable as a function of other parameters in the 
theory\cite{gl}.}

Neutrinos provide a third possible interaction between the ordinary and mirror sectors. {\it
If neutrinos have mass then mass mixing between ordinary and mirror neutrinos is possible.
This leads to very important experimental tests of the exact parity idea.} 
We call the
$G \otimes G$ extension of the Standard Model the Exact Parity Model (EPM) \cite{flv,flv2}.
It is, in part, an explicit theory of neutrino mass and mixing. It is a candidate for the
Standard Model extension called for by the solar, atmospheric and accelerator neutrino
experiments that strongly suggest the existence of neutrino oscillations. Ongoing and future
terrestrial experiments, such as SuperKamiokande, the Sudbury Neutrino Observatory (SNO),
Borexino, the long and short baseline neutrino oscillation searches and other experiments,
will in the next few years provide important new clues in the search for a theory of
neutrino mass and mixing, and will further test the proposed EPM resolution of all of the
anomalies.
 
Of course, if mirror matter exists then there should be dramatic implications for
astrophysics and cosmology as well as particle physics. Some studies \cite{blin} suggest
that mirror matter is an interesting candidate for dark matter. In fact, there is some
evidence that mirror stars may have already been discovered in the MACHO
experiments\cite{mac}. Another exciting possibility is that Gamma Ray Bursts may be due to
collapsing or merging mirror stars\cite{blin2}. Of direct relevance to the present paper,
however, is the observation that early universe cosmology, through Big Bang Nucleosynthesis,
structure formation and in the near future through detailed cosmic microwave background
measurements, should provide important new information about the cosmological role of
neutrino physics. This information may thus also provide a test of the EPM. Indeed, the
purpose of this paper is to perform a detailed study of the early universe cosmology of the
EPM, with particular emphasis on BBN. Before embarking on this analysis, we will briefly
review why the EPM supplies an interesting theory of neutrino mass and mixing, and therefore
why some effort to study its early universe cosmology is justified.

It was pointed out several years ago\cite{flv2} that the EPM provides an interesting theory
of neutrino mass for one simple reason: the exact parity symmetry between the ordinary and
mirror sectors forces an ordinary neutrino $\nu_{\alpha}$ ($\alpha = e, \mu, \tau$)  to be
maximally mixed\footnote{Of course this result only holds if the parity symmetry is {\it not
broken} by the vacuum. In Ref.\cite{flv} it was shown that this occurs for a large range of
parameters with just the minimal Higgs sector of one Higgs doublet and one mirror Higgs
doublet. It was explained in Ref.\cite{bmoh} that if additional Higgs scalars exist, then
the parity symmetry can be spontaneously broken with the mirror electroweak symmetry
breaking scale left as a free parameter. It was argued in Ref.\cite{bmoh} that such a
scenario
could be motivated by the neutrino anomalies. Of course the implications for neutrino
experiments and early universe cosmology of the model in Ref.\cite{bmoh} are quite different
from the minimal case considered in the present paper, where the parity 
symmetry is {\it not} broken
by the vacuum.} with its mirror partner $\nu'_{\alpha}$. It is certainly {\it very
interesting} that the atmospheric neutrino observations of SuperKamiokande\cite{skatmos} and
other experiments\cite{otherexp} point to the muon-neutrino being maximally mixed with
another flavour $\nu_x$. It is known that $\nu_x$ cannot be the $\nu_e$\cite{fvy}, which
leaves $\nu_x = \nu_{\tau}$ and $\nu_x = \nu_s$ as the viable possibilities\cite{fvy2,ll},
where the subscript $s$ denotes a sterile neutrino. The EPM provides a natural candidate for
$\nu_x$, namely the mirror muon-neutrino $\nu'_{\mu}$. As far as terrestrial experiments are
concerned, the $\nu'_{\mu}$ is a sterile flavour.

Since the confirmation of atmospheric $\nu_{\mu}$ disappearence by SuperKamiokande, a
significant amount of theoretical effort has gone into trying to explain the large mixing
angle observed. This work has focussed almost entirely on the $\nu_x = \nu_{\tau}$
possibility. It is interesting to note that in the immediate past, small interfamily mixing,
as observed for the quark sector, was considered to be natural also for the lepton sector.
With the advent, in particular, of the beautiful SuperKamiokande results, this theoretical
prejudice is now being criticised. Our proposal is completely different from any current
effort to realise large $\nu_{\mu} - \nu_{\tau}$ mixing. {\it We simply argue that the
connection between exact parity symmetry and ordinary-mirror neutrino maximal mixing is an
especially elegant and simple explanation of the large mixing angle observed in the
atmospheric neutrino experiments.} Since this points to $\nu_x$ being $\nu_s$ rather than
$\nu_{\tau}$, a neutral current atmospheric neutrino measurement is vital\cite{kk}.

Intriguingly, there is independent experimental evidence for large angle neutrino
oscillations from another set of measurements: maximal mixing between the electron-neutrino
and some other flavour is well motivated by the solar neutrino problem\cite{flv2,fvun,oth}.  
In the EPM, the ``other flavour'' is of course the mirror electron-neutrino $\nu'_e$.  Such
a scenario leads to an energy independent $50\%$ solar $\nu_e$ day-time flux
reduction\footnote{In a very interesting recent paper, Guth, Randall and Serna \cite{guth}
have pointed out that an energy-dependent day-night effect in general exists for solar
neutrinos even if the vacuum mixing is maximal, thus correcting a misconception shared by
the present authors and some of the rest of the community. It is {\it not correct} to
conclude that maximal oscillations out of the ``just-so'' regime always
leads to a completely energy
independent suppression, because the night-time rate is in general energy-dependent due to
matter effects in the Earth.} for a squared mass difference greater than about $3 \times
10^{-10}$ eV$^2$, and to a ``just-so'' picture for a squared mass difference in the
approximate range,\footnote{There is also an interesting `window' around $\delta m^2 \sim
5\times 10^{-10}\ \text{eV}^2$\cite{flor} which leads to an approximate energy integrated
flux reduction of $50\%$ and can also explain the distortion of high energy $E
\stackrel{>}{\sim} 13 \ \text{MeV}$ boron neutrinos suggested by recent SuperKamiokande
data\cite{sksolar}.} $\text{few} \times 10^{-11} \to 3\times 10^{-10}$ eV$^2$. The most
recent solar neutrino data, compared with the most recent solar model
calculations\cite{bah}, show that four out of the five solar neutrino experiments observe
close to a $50\%$ flux deficit. (The Chlorine experiment sees a greater than $50\%$
deficit.) The detailed implications of the solar neutrino situation, though surely
indicative of $\nu_e$ oscillations, is not at present as clear as the atmospheric neutrino
situation. For various reasons, more experiments are needed:  the relatively low Chlorine
result needs to be checked by another experiment, and the cause/existence of the apparent
distortion of the ``boron'' neutrino energy spectrum requires further investigation. Of
particular relevance for the EPM, we will presumably soon find out from SNO whether the
solar neutrino flux contains a significant sterile component.

Notice that no mention was made of the Liquid Scintillator Neutrino Detector (LSND)
observations\cite{lsnd} in advocating the existence of what are essentially light sterile
neutrinos. It has become commonplace to motivate light sterile neutrinos from the inability
of three-flavour oscillations to simultaneously resolve the atmospheric, solar and LSND
anomalies.  We have used this argument ourselves. We would, however, like to emphasise that
our obsession with the EPM arose from the maximal mixing feature, long before the advent of
LSND.

Let us turn, then, to the cosmological implications of mirror neutrinos.  The distinction
between mirror neutrinos and strictly sterile neutrinos, which is totally unimportant for
terrestrial, atmospheric and solar neutrinos, is of some significance in the early universe.
This issue will be discussed in depth in later sections. For the purposes of these
introductory remarks, however, the distinction need not be made. Many of the qualitative
features of sterile neutrino early universe cosmology pertain also to mirror neutrinos.

In recent years, the physics of active-sterile neutrino oscillations during and before the
BBN epoch has been re-examined\cite{ftv,fv1,fv2,bfv,f,bvw}.  Prior to this re-analysis,
it had been concluded that light sterile neutrinos were cosmologically disfavoured for much
of parameter space\cite{thomo}.  Focusing on the $\nu_{\mu} \to \nu_s$ solution to the
atmospheric neutrino problem by way of concrete example, it had been concluded that the
oscillation parameters required would lead to the $\nu_s$ being thermally equilibrated prior
to BBN, thus increasing the expansion rate of the universe and worsening agreement between
theory and primordial light element abundance measurements.  However, it was subsequently
realised\cite{ftv} that the explosive production of large neutrino-antineutrino asymmetries
or chemical potentials by the active-sterile oscillations themselves had not been properly
taken into account in the early studies. Large neutrino asymmetries generically suppress
active-sterile oscillations by making the effective mixing angle in matter very
small\cite{fvprl}. Detailed numerical work has shown that, for a large region of parameter
space, the generation through the oscillations themselves of large neutrino asymmetries
suppresses the production of sterile neutrinos sufficiently for the expansion rate of the
universe during BBN to be essentially unaffected\cite{fv1,f}.

Furthermore, unless the mixing between the electron-neutrino and the other neutrinos is
really tiny, one expects an asymmetry to develop for $\nu_e$'s\cite{fv2}.  This has a direct
effect on the rates of the weak interactions processes $\nu_e n \to e^{-} p$ and
$\overline{\nu}_e p \to e^{+} n$ which help to determine the neutron to proton ratio during
BBN. A detailed calculation within a particular neutrino mass and mixing scenario is
required to work out the magnitude of this effect. It has been shown for two different ``3
active plus 1 sterile neutrino models'' that the generation of a $\nu_e$ asymmetry can be
important\cite{fv2,bfv}. In a Sec.V, we will for the first time explore this
effect for mirror rather than strictly sterile neutrinos.

Aspects of the early universe cosmology of mirror neutrinos were discussed in
Ref.\cite{fv3}. The present paper improves and extends this analysis.  Reference \cite{fv3}
focussed solely on what we can call ``high temperature neutrino asymmetry evolution''.  (We
will explain precisely what we mean by this designation later on.) It showed that the
$\nu_{\mu} \to \nu'_{\mu}$ and $\nu_e \to \nu'_e$ solutions to the atmospheric and solar
neutrino problems, respectively, were compatible with BBN for a large range of parameters.
In Ref.\cite{fv3}, the calculations were carried out in the ``static approximation''. In the
present work we improve on these calculations by using the full quantum kinetic equations,
rather than the above approximation. In addition, we also analyse the ``low temperature
neutrino asymmetry evolution'' that occurs immediately prior and during BBN. The size and
evolution of the $\nu_e$ asymmetry will be the main issue here.

Finally, let us remark that the neutrino phenomenology of the EPM is very similar to some
models employing pseudo-Dirac neutrinos\cite{pd}. Many of the implications for early
universe cosmology will be qualitatively similar to the EPM. There are of course
quantitative differences because the mirror weak interactions play an important role in the
early universe through their impact on the matter potential and also because they affect the
momentum distribution of the mirror neutrinos. We focus on the mirror neutrino scenario in
this paper because it is arguably much more elegant from a model building point of view.
(For example, the see-saw mechanism can be invoked to understand the smallness of both the
neutrino and mirror neutrino masses \cite{flv2}.) It is also theoretically very well
motivated because it restores parity as an unbroken symmetry of nature.

The outline of this paper is as follows: In Sec.II we define and motivate the neutrino mass
and mixing parameters which we will use in our subsequent analysis. We also very briefly
review the neutrino asymmetry amplification phenomenon. In Sec.III the quantum kinetic
equations for ordinary - mirror neutrino oscillations are defined and discussed. In Sec.IV
we compute the region of parameter space where the $\nu_\tau \leftrightarrow \nu'_\mu$
oscillations generate $L_{\nu_\tau}$ and $L_{\nu'_\mu}$ asymmetries in such a way that the
maximal $\nu_\mu \leftrightarrow \nu'_\mu$ oscillations cannot significantly populate the
mirror $\nu'_\mu$ states.  The main issue here is whether or not the $\nu_\mu
\leftrightarrow \nu'_\mu$ oscillations can produce compensating $L_{\nu_\mu}$ and
$L_{\nu'_\mu}$ asymmetries such that the matter term for $\nu_\mu \leftrightarrow \nu'_\mu$
oscillations becomes unimportant. In Sec.V the low temperature evolution of the neutrino
asymmetries is studied in detail. The main issue here is the effect of the oscillations on
BBN.  In Secs.VI and VII we comment on the implications of the EPM for the hot plus cold
dark matter scenario and the anisotropy of the cosmic microwave background. Section VIII is
a conclusion.

\section{Overview and orientation}

The analysis of neutrino oscillations in the 
early universe is complicated. In order to avoid
the pedagogical danger of becoming mired in the full technical detail, 
we present first a short overview. 

There are six light neutrino flavours in the Exact Parity 
Model: the three ordinary neutrinos
$\nu_{e,\mu,\tau}$ and their mirror 
partners $\nu'_{e,\mu,\tau}$, respectively. In the absence of interfamily 
mixing, the most general neutrino mass matrix consistent
with parity symmetry  
for each generation is contained in \cite{flv2}
\begin{equation}
{\cal L}_{\text{mass}} =
\left[ \overline{\nu}_L, \ \overline{(\nu'_R)^c}\right]\left(
\begin{array}{cc}
m_1&m_2\\
m_2&m_1^*
\end{array}\right)
\left[ \begin{array}{c}
(\nu_L)^c\\
\nu'_R
\end{array}\right] + \text{H.c.}
\end{equation}
We have assumed Majorana masses for definiteness and simplicity, and one should note
that the parity symmetry interchanges
$\nu_L$ with $\gamma_0 \nu'_R$. The quantity $m_2$ must be real, while $m_1$ may be complex.
However, the phase of $m_1$ can, without loss of generality, be absorbed by the neutrino
and mirror neutrino fields. In the phase redefined basis,
the mass matrix is diagonalised by the orthogonal transformation,
\begin{equation}
\left[\begin{array}{c}
\nu_{+}\\
\nu_{-}
\end{array}\right]_R = {1 \over \sqrt{2}}
\left(\begin{array}{cc}
1 & -1\\
1 & \ \ 1 
\end{array}\right)
\left[ \begin{array}{c}
(\nu_L)^c\\
\nu'_R
\end{array}\right].
\end{equation}
We see that the mass eigenstates ($\nu_{\pm}$)
are maximal combinations of the weak eigenstates (and vice-versa). 
Obviously it follows that if, as in the quark sector, the mixing between the generations
is nonzero but small, then each pair of weak eigenstates,
\begin{equation}
(\nu_e, \nu'_e),\ (\nu_\mu, \nu'_\mu), \
(\nu_\tau, \nu'_\tau)
\end{equation}
is approximately given by an orthogonal pair of maximal mixtures of the appropriate pair
of mass eigenstates. We use the notation 
\begin{equation}
\nu_{e+}, \ \nu_{e-},
\ \nu_{\mu+}, \ \nu_{\mu-},
\ \nu_{\tau+}, \ \nu_{\tau-},
\end{equation}
for the mass eigenstates. The subscript in the above
equation is used to indicate the pair of states which
relate to the corresponding weak eigenstates.
In the limit of no mixing between the generations,
\begin{eqnarray}
& \nu_\tau = {1 \over \sqrt{2}}\left(
|\nu_{\tau+}\rangle + 
|\nu_{\tau-}\rangle \right),\ 
\nu'_\tau = {1 \over \sqrt{2}}\left(
|\nu_{\tau+}\rangle - 
|\nu_{\tau-}\rangle \right),&\  
\nonumber \\
& \nu_\mu = {1 \over \sqrt{2}}\left(
|\nu_{\mu+}\rangle + 
|\nu_{\mu-}\rangle \right),\ 
\nu'_\mu = {1 \over \sqrt{2}}\left(
|\nu_{\mu+}\rangle - 
|\nu_{\mu-}\rangle \right),&\  
\nonumber \\
& \nu_e = {1 \over \sqrt{2}}\left(
|\nu_{e+}\rangle + 
|\nu_{e-}\rangle \right),\ 
\nu'_e = {1 \over \sqrt{2}}\left(
|\nu_{e+}\rangle - 
|\nu_{e-}\rangle \right).&\ 
\end{eqnarray}
Of course the exact expressions for $\nu_{\alpha}$ and $\nu'_{\alpha}$
($\alpha = e, \mu, \tau$) will in general be
a linear combination of all possible mass eigenstates when mixing between generations
exists. This
means that all possible oscillations modes amongst the six neutrino flavours are in
general expected to occur.
The assumption of small mixing between the generations, together with the necessarily
maximal
mixing between the ordinary and mirror neutrinos of a given generation, implies that
intergenerational modes such as $\nu_\tau \leftrightarrow \nu'_\mu$ or $\nu_\mu 
\leftrightarrow \nu_e$ will have much smaller amplitudes than the $\nu_{\alpha}
\leftrightarrow \nu'_{\alpha}$ modes (in vacuum). The analysis to follow will only consider
the region of parameter space where vacuum mixing between generations is small.

In order to proceed, we also have to make a guess about the pattern of 
mass eigenvalues. We will suppose that the 
neutrino sector is qualitatively identical to the quark and charged-lepton sectors,
with the masses displaying the 
standard hierarchy. We will further assume, most of the time, 
that the mass splitting between the parity partners within a
given family is smaller than the interfamily mass splitting. 
Putting this together, we have the mass pattern
\begin{equation}
m_{\nu_{\tau +}} \simeq m_{\nu_{\tau -}} \gg m_{\nu_{\mu +}} \simeq 
m_{\nu_{\mu -}} \gg m_{\nu_{e +}} \simeq m_{\nu_{e -}}.
\end{equation}
The LSND result suggests that the 
$e-\mu$ mass splittings are of the order of an eV or so, although we 
will also consider smaller mass splittings. If the $e-\mu$
mass difference is of the order of an eV, then to maintain the 
assumed mass hierarchy the $\nu_{\tau}$ and $\nu'_{\tau}$ masses 
should be larger than or about a few eV. A mass in
the few eV range would of course make $\nu_{\tau}$ a hot dark matter 
particle. Cosmological closure puts an upper bound of about 
40 eV on $m_{\nu_{\tau}}$. Analogy with
the quark sector suggests that neutrinos in adjacent families, 
$e-\mu$ and $\mu-\tau$, should mix more
strongly than $e-\tau$. Furthermore, one might guess that 
$\alpha-\beta/\alpha'-\beta'$ mixing should be
stronger than $\alpha'-\beta/\alpha-\beta'$ mixing if one believes 
that the more ``closely related'' are the neutrinos the more strongly 
they should mix. (Also observe that the parity symmetry
forces the $\alpha - \beta$ and
$\alpha' - \beta'$ mixing angles to be equal, similarly the $\alpha' - \beta$ and $\alpha -
\beta'$ mixing angles.)  Putting these guesses together with the 
$\nu_e \to \nu'_e$ solution to the solar
neutrino problem and the $\nu_{\mu} \to \nu'_{\mu}$ solution to the 
atmospheric neutrino problem, we arrive at the 
parameter space region\footnote{If $3\times 10^{-5} \stackrel{<}{\sim} 
|m^2_{ee'}|/\text{eV}^2 
\stackrel{<}{\sim} 10^{-3}$ then the electron neutrino oscillations
can have potentially observable effects 
for atmospheric neutrinos. see Ref.\cite{bunn} for details.}
\begin{eqnarray}
& m_{\nu_{e +}},\ m_{\nu_{e -}}  \ll  1\ \text{eV},\quad 
10^{-11}\ \text{eV}^2 \stackrel{<}{\sim} 
|\delta m^2_{ee'}|
\equiv |m^2_{\nu_{e +}} - m^2_{\nu_{e -}}| 
\stackrel{<}{\sim} 10^{-3}\ \text{eV}^2,&\ \nonumber\\
& m_{\nu_{\mu +}},\ m_{\nu_{\mu -}} \stackrel{<}{\sim} \
 \text{few} \ \text{eV},\quad 
10^{-3}\ \text{eV}^2 \stackrel{<}{\sim} 
|\delta m^2_{\mu \mu'}| \equiv
|m^2_{\nu_{\mu +}} - m^2_{\nu_{\mu -}}| 
\stackrel{<}{\sim} 10^{-2}\ \text{eV}^2,&\  \nonumber\\
& \text{few\ eV} \stackrel{<}{\sim} m_{\nu_{\tau +}},\ 
m_{\nu_{\tau -}} \stackrel{<}{\sim} 40\
\text{eV},\quad 
|\delta m^2_{\tau \tau'}| \equiv
|m^2_{\nu_{\tau +}} - m^2_{\nu_{\tau -}}| \ll 1\ 
\text{eV}^2, &\ 
\label{massranges}
\end{eqnarray}
with a mixing angle pattern as described above.

We wish to calculate the effect on early universe cosmology of 
neutrino oscillations within the EPM. A full six-flavour analysis 
is a daunting task, even with the parameter space restrictions
discussed above. Fortunately, the physics of the problem allows some 
simplifications to be made without sacrificing too 
much in the way of rigor. In particular, we can build on what we already
know about the early universe cosmology of active-sterile neutrino 
oscillations.

It is useful to start by identifying four qualitatively different epochs:
\begin{enumerate}
\item the Quantum Zeno Epoch, where neutrino oscillations are completely   
damped; 
\item the High-Temperature Epoch, where large neutrino 
asymmetries are initially generated;
\item the Low-Temperature Epoch, where decoherence can be 
neglected; and,
\item the Big Bang Nucleosynthesis Epoch, where neutrino oscillations 
impact on light element synthesis.
\end{enumerate} 
We now very briefly, and qualitatively, discuss these epochs in 
turn. The mathematics needed to fully explain this cosmological 
history is available in previous publications and in later
sections of this paper. 

\subsection{The Quantum Zeno Epoch}

Neutrino oscillations in the early universe are always to some 
extent damped through collisions with the background medium. 
As we look back toward the Big Bang, the collision rate increases as
$T^5$ (below the electroweak phase transition). At sufficiently 
high temperatures, collisions occur so frequently that the quantally 
coherent oscillatory behaviour cannot develop. The
neutrino ensemble is frozen with respect to its flavour 
content (Quantum Zeno Effect). In addition, the finite 
temperature contributions to the effective matter potentials for many of
the oscillation modes are high enough to render the associated 
matter mixing angles extremely small. So even with collisions 
artificially switched off, many of the oscillation modes would
have tiny amplitudes.

\subsection{The High-Temperature Epoch}

As the temperature decreases, collisional damping is reduced, 
and partially incoherent evolution of the neutrino ensemble begins.
For simplicity, we will in this and the next subsection very briefly review the evolution of
the $\alpha$-like lepton number in the somewhat artificial case where
only the $\nu_{\alpha} \leftrightarrow \nu_s$ mode is operative. It was shown in
Ref.\cite{ftv} that under the influence of this mode\footnote{For the purposes
of this introductory discussion, the distinction between mirror and sterile
neutrinos will often be neglected.} the $\alpha$-like lepton number $L_{\nu_{\alpha}}$
evolves as per
\begin{equation}
{dL_{\nu_\alpha} \over dt} \simeq C\left(L_{\nu_\alpha} +
{\eta \over 2}\right).
\label{sterileL}
\end{equation}
The $\alpha$-like lepton number is defined by
\begin{equation}
L_{\nu_{\alpha}} \equiv {n_{\nu_{\alpha}} - n_{\overline{\nu}_{\alpha}} \over n_{\gamma}}
\end{equation}
and is synonymously called the ``$\alpha$-like neutrino asymmetry''. The quantity $n_i$ is
the number density for species $i$.
Equation (\ref{sterileL}) holds provided that (i) the squared mass difference $\delta
m^2_{\alpha s}$ between the neutrinos obeys $|\delta m^2_{\alpha s}| \stackrel{>}{\sim}
10^{-4} \ \text{eV}^2$, and (ii) $L_{\nu_\alpha}$ is small.  
The quantity $\eta$ is set by the relic nucleon number densities
and is expected to be small: $\eta/2 \sim 10^{-10}$. The term $C$ is a
function of time $t$ (or equivalently temperature $T$).
At high temperature it turns out that $C$ is negative, so that
$(L_{\nu_\alpha} + \eta/2) \simeq 0$ is an approximate
fixed point. However if $\delta m^2_{\alpha s} < 0$ [our $\delta m^2$ convention is defined
in Eq.(\ref{zwig}) below], then $C$ changes
sign at a particular temperature $T = T_c$, estimated to be\cite{ftv} 
\begin{equation}
T_c \sim 
16\left({-\delta m^2_{\alpha s} \cos2\theta_{\alpha s} \over 
\text{eV}^2}\right)^{1 \over 6}\ \text{MeV}.
\end{equation}
At this temperature, rapid exponential growth of neutrino asymmetry
occurs, unless $\sin^2 2\theta_{\alpha s}$ is very tiny [see Eq.(\ref{larges}) below].
The generation of neutrino asymmetry occurs because the
$\nu_\alpha \to \nu_s$ oscillation probability is 
different from the $\overline \nu_\alpha \to \overline \nu_s$ oscillation
probability due to the matter effects in a $CP$ asymmetric background.
As the asymmetry is created, the background becomes more $CP$
asymmetric because the neutrino asymmetries contribute to the
$CP$ asymmetry of the background. This leads to a period of runaway
exponential growth of the neutrino asymmetry
for a large range of parameters\cite{fv1,fv3}\footnote{In the region of parameter space where
$|\delta m^2_{\alpha s}| \ll 10^{-4} \ \text{eV}^2$, the 
evolution of the neutrino asymmetry is dominated by oscillations
between collisions and the lepton number tends to be
oscillatory\cite{ee,kir,shi}.} summarised by:
\begin{eqnarray}
&\delta m^2_{\alpha s} < 0 \ \text{with} \ |\delta m^2_{\alpha s}| 
\stackrel{>}{\sim} 10^{-4}\ \text{eV}^2,& \nonumber \\ 
&10^{-10} \ \stackrel{<}{\sim} \sin^2 2\theta_{\alpha s} 
\stackrel{<}{\sim} \ \text{few} \ \times 10^{-5}\left[{\text{eV}^2 \over
|\delta m^2_{\alpha s}|}\right]^{1 \over 2}\ \text{for ordinary - sterile
oscillations},& \nonumber \\ 
&10^{-10} \stackrel{<}{\sim} \sin^2 2\theta_{\alpha s} 
\stackrel{<}{\sim} \ \text{few} \  \times 10^{-4}\left[{\text{eV}^2 \over
|\delta m^2_{\alpha s}|}\right]^{1 \over 2}
\ \text{for ordinary - mirror oscillations}.&\  
\label{larges}
\end{eqnarray}
(The upper bound in the above equation comes from a constraint on the effective number of
neutrino flavours, $N_{\nu,\text{eff}}$, during BBN. We have used $N_{\nu,\text{eff}} - 3 
\stackrel{<}{\sim} 0.6$ in this equation for illustrative purposes.)
{\it We want to emphasise and to state very clearly the following fact: Provided the
oscillation parameters are in the large range given in
Eq.(\ref{larges}), the ordinary - sterile (or mirror) neutrino oscillations will generate,
at the temperature $T_c$, a significant neutrino asymmetry
(or chemical potential) from the tiny seed $CP$ asymmetry of 
the background plasma. There is no choice about this, a point sometimes misunderstood in the
literature: the large neutrino asymmetry will inevitably be generated.} Once
generated, neutrino asymmetries
in turn contribute to the effective matter potentials and generically
suppress oscillations by inducing small effective mixing angles.
For typical oscillation parameter values within our scenario, 
the explosive neutrino asymmetry growth begins while
collisions still dominate the evolution (though they now do not 
completely damp the oscillations).
Note that the evolution of lepton number for $T < T_c$ 
is approximately independent of the initial neutrino asymmetries
provided that they are not too big (that is, less than about $10^{-5}$).
This is because of the approximate fixed point structure which
sees $L_{\nu_\alpha} \to - \eta/2$ for $T > T_c$.

\subsection{The Low-Temperature Epoch}

While neutrino asymmetries develop and evolve, the collision rate 
continues to decrease in a $T^5$ fashion. Eventually the flavour 
evolution of the neutrino ensemble becomes dominated by
coherent processes rather than decoherence-inducing collisions. This 
observation is of practical importance, because the evolution equations 
then reduce to MSW form. If the dynamics satisfies the
adiabatic condition,
then the evolution becomes particularly simple. Actually, it turns out that adiabaticity
indeed holds for the parameter space of Eq.(\ref{larges}).
The low temperature evolution of the asymmetry is then approximately independent
of the vacuum mixing angle, in the small vacuum mixing angle region. Staying with our
example of a $\nu_{\alpha} - \nu_s$ system in isolation, it has been computed that
the `final' value of the asymmetry arises at the temperature\cite{fv2}
\begin{equation}
T^f_\nu \simeq 0.5 \left( {|\delta m^2_{\alpha s}| \over \text{eV}^2} \right)^{1 \over 4}\
\text{MeV}.
\label{nau}
\end{equation}
The magnitude of the final value was calculated to be\cite{fv2},
\begin{eqnarray}
L_{\nu_{\alpha}}^f &\simeq & 0.29 h\ \text{for}
\ |\delta m^2_{\alpha s}|/\text{eV}^2 \stackrel{>}{\sim} 1000, \nonumber \\
L_{\nu_{\alpha}}^f &\simeq & 0.23 h\ \text{for}
\ 3 \stackrel{<}{\sim} |\delta m^2_{\alpha s}|/\text{eV}^2 \stackrel{<}{\sim} 1000,
\nonumber \\
L_{\nu_{\alpha}}^f &\simeq & 0.35 h\ \text{for}
\ 10^{-4}\stackrel{<}{\sim} |\delta m^2_{\alpha s}|/\text{eV}^2 \stackrel{<}{\sim} 3,
\label{go}
\end{eqnarray}
where $h \equiv (T_\nu/T_\gamma)^3$.
Similar results also hold for ordinary - mirror neutrino oscillations.

\subsection{The Big Bang Nucleosynthesis Epoch}

At temperatures of a few MeV, weak interaction rates start to become 
smaller than the expansion rate of the universe. This causes the 
ordinary neutrinos to fall out of kinetic and chemical
equilibrium with the background plasma. It also signals the onset 
of the BBN epoch because of the end of nuclear statistical equilibrium.
For the typical parameter space of interest in the EPM, we will
show that a significant electron neutrino asymmetry is
generated by and during during
this epoch. This will have important implications for BBN, 
and one of the major goals of this paper to compute this effect.

\section{Quantum kinetic equations for ordinary-mirror neutrino oscillations}

Before we begin in earnest, we need to say something about the 
thermodynamics of the mirror particles.
Because the mirror particles interact amongst themselves just
like the ordinary particles,
the mirror particles can be described by a temperature $T'$ (and
chemical potentials, which we assume are initially negligible).
In fact, the ordinary and mirror particles form two weakly
coupled thermodynamic systems.
As in our previous paper\cite{fv3}, we will suppose that there is an 
{\it asymmetry} between the temperature of the mirror plasma and the 
temperature of the ordinary plasma so that $T' \ll T$.
Of course if $T' = T$ then a neutrino asymmetry would not
be expected to develop. The energy density
of the mirror sector would then double the expansion
rate of the universe. In this case the reasonably successful BBN
predictions would be lost. However one should remember that exact 
microscopic symmetry does not imply exact macroscopic symmetry.
In reality if the ordinary and mirror particles are only in very weak
thermal contact there is no compelling reason for $T' =T$.
Note that the assumption that $T' \ll T$ does {\it not} imply
that the amount of mirror baryonic matter in the universe today
is less than the ordinary baryonic matter. The origin of baryon 
number (and mirror baryon number) is not understood at the moment, 
so no definite conclusions can be drawn regarding the amount
of mirror baryonic matter (and hence mirror stars and so on) in the universe today.
Actually there are strong astrophysical arguments for
the existence of a large amount of dark matter in the universe, and
this suggests that the mirror baryon number is comparable or even
greater than the ordinary baryon number.
However, as far as the early\footnote{By `early' 
we mean the time during and earlier than the BBN epoch} universe is concerned, 
the precise value of the mirror baryon number should be
unimportant since the energy density will be dominated by the 
relativistic degrees of freedom (neutrinos, electrons/positrons 
and photons).  Thus, when we use the term `mirror matter' below, 
we will be referring to the `light' mirror particles, that is
the mirror electrons/positrons, mirror photons and
mirror neutrinos, since these are the mirror particles which
affect the expansion rate of the {\it early} universe.

We now discuss the Quantum Kinetic Equations (QKEs) for a 
two-flavour subsystem consisting of an ordinary neutrino 
$\nu_{\alpha}$ and a mirror neutrino $\nu'_{\beta}$. We
will not, in this work, provide an exhaustive discussion of the 
derivation of the QKEs or their meaning,
since this territory is well covered in previous 
papers\cite{bvw,stod,stod2,bm}. 
We will, however, provide a complete discussion of the special 
features mirror neutrinos bring to the QKEs (by contrast
to strictly sterile neutrinos). Note that two-flavour subsystems
will be used as building blocks for the full six-flavour 
system in a later section.

We will focus on evolution
during the temperature regime $m_e \stackrel{<}{\sim} T \stackrel{<}{\sim} 
m_{\mu}$. The plasma 
therefore consists of (i) the relativistic ordinary particles 
$\nu_e$, $\overline{\nu}_e$, $\nu_{\mu}$, $\overline{\nu}_{\mu}$, 
$\nu_{\tau}$, $\overline{\nu}_{\tau}$ $e^{-}$, $e^{+}$ and $\gamma$; 
(ii) the nonrelativistic ordinary protons and neutrons (and the nonrelativistic mirror
protons and neutrons discussed above); and 
(iii) whatever amount of mirror matter gets created through
ordinary-mirror neutrino oscillations. The character of the mirror 
matter in the plasma depends on how much of it is 
created through oscillations. If a sufficiently tiny amount is
created, then the mirror electromagnetic and mirror weak 
interactions amongst the mirror neutrinos will take place at a rate 
that is smaller than the expansion rate of the universe.
In this case, the mirror neutrino distributions will not be of 
Fermi-Dirac form, and mirror electrons, positrons and photons will 
not be created. When the amount of mirror matter
exceeds a certain level, mirror electromagnetic and mirror weak 
interactions amongst the mirror neutrinos become larger than the 
expansion rate. In this case, the mirror
neutrinos produced through oscillations quickly assume a distribution of   
Fermi-Dirac form, and equilibrium distributions of mirror electrons, mirror 
positrons and mirror photons get excited in the plasma. The
full plasma thus consists of two weakly coupled thermodynamic systems: 
the aforementioned ordinary particles at temperature $T$, and the 
corresponding mirror particles at a smaller
temperature $T'$. For the case where mirror species contribute 
negligibly to the expansion rate of the universe, we have earlier 
shown\cite{fv3} that the inequality
\begin{equation}
T' \stackrel{>}{\sim} 2 \left({T \over \text{MeV}}\right)^{\frac{2}{5}}\   
\text{MeV,}
\label{Tprime}
\end{equation}
must be obeyed to ensure that the mirror self-interactions are 
sufficiently fast to thermally equilibrate the mirror species.

Our notation/convention for ordinary-mirror 
neutrino two state mixing is as follows. The weak
eigenstates $\nu_{\alpha}$ and $\nu'_{\beta}$
are linear combinations
of two mass eigenstates $\nu_a$ and $\nu_b$,
\begin{equation}
\nu_{\alpha} = \cos\theta_{\alpha\beta'} \nu_a + 
\sin\theta_{\alpha\beta'} \nu_b,\
\nu'_{\beta} = - \sin\theta_{\alpha\beta'} \nu_a + 
\cos\theta_{\alpha\beta'} \nu_b,
\label{zwig}
\end{equation}
where $\theta_{\alpha\beta'}$ is the vacuum mixing angle.
We define $\theta_{\alpha\beta'}$ so that 
$\cos2 \theta_{\alpha\beta'} > 0$ and
we adopt the convention that $\delta m^2_{\alpha\beta'} 
\equiv m^2_b - m^2_a$.

Recall that the $\alpha$-type neutrino asymmetry is defined by 
\begin{equation}
L_{\nu_\alpha} \equiv 
{n_{\nu_\alpha} - n_{\overline \nu_\alpha} \over n_\gamma}.
\label{def}
\end{equation}
We also need to define an $\alpha-$type mirror neutrino
asymmetry,
\begin{equation}
L_{\nu'_\alpha} \equiv 
{n_{\nu'_\alpha} - n_{\overline \nu'_\alpha} \over n_\gamma},
\end{equation}
In the above equation, $n_{\gamma}$ is the
number density of {\it ordinary} photons.

Note that when we refer to ``neutrinos'', sometimes we
will mean neutrinos and/or antineutrinos and/or mirror neutrinos and/or mirror
anti-neutrinos. We hope the correct meaning will be clear from context.

The evolution of the ensemble of $\nu_{\alpha}$ and $\nu'_{\beta}$ 
neutrinos is described by a density matrix $\rho_{\alpha \beta'}$ which 
obeys the QKEs. A similar density matrix
$\overline{\rho}_{\alpha \beta'}$ describes the antineutrinos. 
These density matrices \cite{stod,bm} are conveniently parameterised by
\begin{equation}
\rho_{\alpha\beta'} (p) = {1 \over 2} [P_0(p)I + {\bf P}(p)\cdot
{\bf \sigma}],\quad
\overline{\rho}_{\alpha\beta'}(p) = 
{1 \over 2} [\overline{P}_0(p)I + {\bf \overline{P}}(p)\cdot
{\bf \sigma}],
\label{kdf}
\end{equation}
where $I$ is the $2 \times 2$ identity matrix, the ``polarisation vector''
${\bf P}(p) = P_x(p){\bf \hat x} + P_y (p) {\bf \hat y}
+ P_z(p){\bf \hat z}$ and ${\bf \sigma} = \sigma_x {\bf \hat x} + 
\sigma_y {\bf \hat y} + \sigma_z {\bf \hat z}$, with 
$\sigma_i$ being the Pauli matrices.\footnote{Note that 
our previous papers used a different
definition of ${\bf P}$ through the equation $\rho = 
{1 \over 2} P_0(p) [I + {\bf P}(p)\cdot {\bf \sigma}]$
rather than Eq.(\ref{kdf}). The difference is just a matter
of convention.}
 
The quantity $p$ is the magnitude of the neutrino
3-momentum or energy.
It will be understood that the density matrices
and the quantities $P_i(p)$ also depend on time $t$ or, equivalently, 
temperature $T$. (For the situation of negligible mirror energy density, the
time-temperature relation 
for $m_e \stackrel{<}{\sim} T \stackrel{<}{\sim} m_\mu$ is
$dt/dT \simeq -M_P/5.5T^3$, where $M_P \simeq 1.22 \times 10^{22} \
\text{MeV}$ is the Planck mass).  

We will normalise the density matrices so that
the momentum distributions of $\nu_{\alpha}(p)$ 
and $\nu'_{\beta}(p)$ are given by
\begin{equation}
N_{\nu_{\alpha}}(p) = {1 \over 2}[P_0(p) + P_z(p)]N^{\text{eq}}(p,T,0),
\quad N_{\nu'_{\beta}}(p) = {1 \over 2}[P_0(p) - P_z(p)]N^{\text{eq}}(p,T,0),
\label{c}
\end{equation}
where
\begin{equation}
N^{\text{eq}}(p,T,\mu) \equiv {1 \over 2\pi^2}{p^2 
\over 1 + \exp\left({p-\mu \over T}\right) },
\end{equation}
is the Fermi-Dirac distribution with chemical potential $\mu$ 
and temperature $T$. Note that $P_0$ is related to the 
total number of $\nu_{\alpha}$'s and $\nu'_{\beta}$'s of momentum $p$,
\begin{equation}
P_0(p) = {N_{\nu_{\alpha}}(p) + N_{\nu'_{\beta}}(p) \over N^{\text{eq}}(p,T,0)},
\label{sum}
\end{equation}
while $P_z(p)$ is related to difference,
\begin{equation}
P_z(p) = {N_{\nu_{\alpha}}(p) - N_{\nu'_{\beta}}(p) \over 
N^{\text{eq}}(p,T,0)}.
\label{diff}
\end{equation}
Similar expressions pertain to antineutrinos. The ``transverse'' 
components $P_{x,y}(p)$ and $\overline{P}_{x,y}(p)$ measure the degree 
of quantal coherence in the ensemble. Note that in subsequent 
expressions we will suppress the independent variables for 
notational clarity unless there is a chance of confusion.

The time evolution of $P_0$ and ${\bf P}$ is governed 
by three effects: coherent
$\nu_{\alpha} \leftrightarrow \nu'_{\beta}$
oscillations, decoherence inducing collisions, and 
repopulation of $\nu_{\alpha}$ and
$\nu'_{\beta}$
states from the background plasma. These effects are 
incorporated in the Quantum Kinetic
Equations \cite{bm,bvw},
\begin{eqnarray}
{\partial {\bf P} \over \partial t} & = &
\left. {\partial {\bf P} \over \partial t}\right|_{\nu_{\alpha} \leftrightarrow
\nu'_{\beta}} +
\left. {\partial {\bf P} \over \partial t}\right|_{\text{coll}} +
\left. {\partial {\bf P} \over \partial t}\right|_{\text{repop}},\nonumber\\
{\partial P_0 \over \partial t} & = &
\left. {\partial P_0 \over \partial t}\right|_{\nu_{\alpha} \leftrightarrow \nu'_{\beta}}
+
\left. {\partial P_0 \over \partial t}\right|_{\text{coll}} +
\left. {\partial P_0 \over \partial t}\right|_{\text{repop}},  
\end{eqnarray}
where
\begin{eqnarray}
\left. {\partial {\bf P} \over \partial t}\right|_{\nu_{\alpha} \leftrightarrow
\nu'_{\beta}} 
& = & {\bf V_{\alpha\beta'}} \times {\bf P},\nonumber\\
\left. {\partial {\bf P} \over \partial t}\right|_{\text{coll}} 
& = & - D {\bf P_T}\quad \text{where}\quad {\bf P_T} \equiv P_x 
{\bf \hat{x}} + P_y \bf{\hat{y}},\nonumber\\
\left. {\partial {\bf P} \over \partial t}\right|_{\text{repop}}
& = & (R_{\nu_{\alpha}} - R_{\nu'_{\beta}}) \bf {\hat{z}},
\label{b1}
\end{eqnarray}
and
\begin{equation}
\left. {\partial P_0 \over \partial t}\right|_{\nu_{\alpha} 
\leftrightarrow \nu'_{\beta}}  =  0,\
\left. {\partial P_0 \over \partial t}\right|_{\text{coll}}  
=  0,\ \left. 
{\partial P_0 \over \partial t}\right|_{\text{repop}}  
= R_{\nu_{\alpha}} + R_{\nu'_{\beta}}.
\label{b2}
\end{equation}
We will explicitly define the new terms appearing above shortly. 
But before doing so, we remark that the general form of the 
above equations is reasonably easy to understand. The
${\bf V} \times {\bf P}$ term leads to the precession of the 
polarisation vector without change in its length. 
The $-D{\bf P_T}$ decoherence term causes $P_x$ and $P_y$ to 
decrease in length ($D > 0$), which quantifies the rate of 
loss of quantal coherence. The function
$R_i(p)$ is related to the repopulation rate for a particle 
of species $i$ with momentum $p$. The
functions $P_{x,y}$ are unaffected by repopulation because 
they measure quantal coherence only. On the other hand $P_z$, 
being proportional to the difference in the
momentum distributions of the two neutrino flavours 
as per Eq.(\ref{diff}), receives a contribution proportional to the
difference in the repopulation rates. The function $P_0$ obviously 
remains unchanged under $\nu_{\alpha} \leftrightarrow
\nu'_{\beta}$ oscillations, and it plays no role in quantifying 
loss of coherence. Since it is related to the sum of momentum 
distributions as per Eq.(\ref{sum}), its time derivative
from repopulation is related to the sum of the repopulation rates.

Similar equations are satisfied for the antineutrino functions 
$\overline{P}_0$ and $\overline{{\bf P}}$, with the substitutions
\begin{equation}
{\bf V_{\alpha\beta'}} 
\to {\bf \overline{V}_{\alpha\beta'}},\quad D \to \overline{D},\quad
R_i \to R_{\overline{i}}.
\end{equation}
We now explicitly define the terms appearing in these equations.

The function ${\bf V_{\alpha\beta'}}$, which is related to the 
effective matter potential, drives the
coherent aspect of the evolution of the density matrix.
Importantly, ${\bf V_{\alpha\beta'}}$ depends on the 
neutrino and mirror neutrino asymmetries. It is given by \cite{stod,bm}
\begin{equation} 
{\bf V_{\alpha\beta'}} = \beta {\bf \hat {x}} + \lambda 
{\bf \hat {z}},
\end{equation}
where $\beta$ and $\lambda$ are
\begin{equation}
\beta(p) = {\delta m_{\alpha\beta'}^2 \over 2p}\sin 2\theta_{\alpha\beta'},
\quad \lambda(p) = -{\delta m^2_{\alpha\beta'} \over 2p}
[\cos2\theta_{\alpha\beta'} - b(p) \pm a(p)],
\label{sf}
\end{equation}
in which the $+(-)$ sign corresponds to neutrino (antineutrino)
oscillations. The dimensionless variables $a(p)$ and $b(p)$ 
contain the matter effects \cite{msw}, being the
matter potential divided by $\delta m_{\alpha\beta'}^2/2p$.
For $\nu_{\alpha} \leftrightarrow \nu'_{\beta}$ oscillations 
$a(p)$ and $b(p)$ are given by \cite{rn} 
\begin{equation}
a(p) \equiv - {4\zeta(3)\sqrt{2}G_FT^3L^{(\alpha\beta')}p
\over \pi^2\delta m^2_{\alpha\beta'}},
\quad b(p) \equiv - {4\zeta (3) \sqrt{2} G_F T^4 A_{\alpha} p^2 \over
\pi^2 \delta m^2_{\alpha\beta'} M_W^2},
\label{sal}
\end{equation}
where $\zeta (3) \simeq 1.202$ is the Riemann zeta function
of 3, $G_F$ is the Fermi constant, $M_W$ is the 
$W-$boson mass, $A_e \simeq 17$ and $A_{\mu, \tau} \simeq 4.9$ 
(for $m_e \stackrel{<}{\sim} T \stackrel{<}{\sim} m_\mu$). The expression 
for $b(p)$ is valid provided that the plasma has a negligible 
component of mirror energy density.  The quantity $L^{(\alpha\beta')}$ 
is given by
\begin{equation}
L^{(\alpha\beta')} = L^{(\alpha)} - L'^{(\beta)},
\end{equation}
where
\begin{equation}
L^{(\alpha)} = L_{\nu_\alpha} + L_{\nu_e} + L_{\nu_\mu}
+ L_{\nu_\tau} + \eta,
\quad L'^{(\beta)} = L_{\nu'_\beta} + L_{\nu'_e} + L_{\nu'_\mu}
+ L_{\nu'_\tau} + \eta'.
\end{equation}
Recall that the term $\eta$ is due to the 
asymmetry of the electrons and nucleons and is expected
to be very small, $\eta \sim 5\times 10^{-10}$. The mirror analogue, 
$\eta'$, will also be taken
to be very small. For antineutrinos, the corresponding function 
${\bf \overline{V}_{\alpha\beta'}}$ is obtained by the
substitution $L^{(\alpha\beta')} \to - L^{(\alpha\beta')}$. 
The MSW resonance conditions are given by 
\begin{equation}
\lambda(p_{\text{res}}) = 0,
\end{equation}
where $p_{\text{res}}$ is the resonance momentum.

The term $D(p)$ is the decoherence or damping function. When the 
number density of mirror species is much less than the number 
density of ordinary species, it is given by\cite{stod2}
\begin{equation}
D(p) \simeq {\Gamma(p) \over 2},
\end{equation} 
where $\Gamma(p)$ is
the total collision rate of a $\nu_{\alpha}$
neutrino of momentum $p$ with the background 
plasma.\footnote{If the number density of mirror species is significant,   
$D(p)$ must also include
the collision rate of $\nu'_{\beta}$'s with the background 
mirror particles.} From Refs.\cite{bvw,ekt} it is given by 
\begin{equation}
\Gamma(p) = 
y_{\alpha} G_F^2 T^5 \left({p \over \langle p \rangle}\right), 
\end{equation}
where $\langle p \rangle \simeq 3.15 T$ is
the average momentum of the ordinary neutrinos, $y_{e} \simeq 4.0$ and 
$y_{\mu,\tau} \simeq 2.9$ (for the $m_e 
\stackrel{<}{\sim} T \stackrel{<}{\sim} m_\mu$ epoch we are considering). 
The total collision rate for a $\nu'_{\beta}$ mirror
neutrino of momentum $p$ is roughly,
\begin{equation}
\Gamma'(p) \simeq \left\{ \begin{array}{ll}
           \left({T' \over T}\right)^4 
\Gamma(p)\quad  & \mbox{if $T'$\
                           \text{obeys}\ \text{Eq}.(\ref{Tprime})} \\
                         \quad\quad 0\quad  & \mbox{\text{otherwise}}
                      \end{array} \right. .
\end{equation}
In the presence of neutrino asymmetries, the collision rates 
for neutrinos and antineutrinos differ. The collision rates quoted 
above hold when the asymmetries are small,
with the antineutrino rate being approximately equal to the 
neutrino rate in that limit.  Note that in the parameter space 
regime we are considering, neutrino asymmetries do not
become large until temperatures are sufficiently low that collisions 
can be approximately neglected. Therefore, the dependence of the 
collision rates on the neutrino asymmetries is
never of practical importance.
 
The repopulation functions $R_{\nu_{\alpha}}$ and 
$R_{\nu'_{\beta}}$ are given by
\begin{eqnarray}
R_{\nu_{\alpha}} & \simeq &
\Gamma \left[K_{\nu_{\alpha}} - {1\over 2}(P_0 + P_z)\right],\nonumber\\
R_{\nu'_{\beta}} & \simeq &
\Gamma' \left[K_{\nu'_{\beta}} - {1\over 2}(P_0 - P_z)\right],
\label{RK}
\end{eqnarray}
where
\begin{eqnarray}
K_{\nu_{\alpha}}(p) & \equiv & {N^{\text{eq}}(p,T,\mu_{\nu_{\alpha}}) \over
N^{\text{eq}}(p,T,0)},\nonumber\\ 
K_{\nu'_{\beta}}(p) & \equiv & {N^{\text{eq}}(p,T',\mu_{\nu'_{\beta}}) \over
N^{\text{eq}}(p,T,0)},
\label{Ks}
\end{eqnarray}
with $\mu_{i}$ being the chemical potential for species $i$. 
For antineutrinos, $\mu_{\nu_\alpha}$ is replaced by 
$\mu_{\overline{\nu}_{\alpha}}$ 
in the above equation. The approximate equality sign in 
Eq.(\ref{RK}) indicates that the
right-hand side is not an exact result. It holds when all species
are in thermal equilibrium apart from $\nu_{\alpha}$ and 
$\nu'_{\beta}$, which are instead approximately in
equilibrium. [See Ref.\cite{bm} for the exact form of Eq.(\ref{b2}).] 
The two terms $R_{\nu_{\alpha}}$ and $R_{\nu'_{\beta}}$
are due to the repopulation of $\nu_{\alpha}$ states by ordinary
weak interactions, and the repopulation of $\nu'_{\beta}$ states by 
mirror weak interactions, respectively.

In order to integrate Eqs.(\ref{b1}) and (\ref{b2}), we need to 
relate the chemical potentials appearing in Eqs.(\ref{RK}) and 
(\ref{Ks}) to the asymmetries appearing in Eq.(\ref{b1}).  
In general, for a distribution in thermal equilibrium,
\begin{equation}
L_{\nu_{\alpha}} = {1 \over 4\zeta (3)}\int^{\infty}_0
{x^2 dx \over 1 + e^{x - \tilde{\mu}_{\alpha}}} - 
{1 \over 4\zeta (3)}\int^{\infty}_0
{x^2 dx \over 1 + e^{x - \tilde{\mu}_{\overline{\alpha}}}},  
\end{equation}
where $\tilde{\mu}_{\alpha} \equiv \mu_{\nu_\alpha}/T$
and $\tilde{\mu}_{\overline{\alpha}} \equiv 
\mu_{\overline{\nu}_{\alpha}}/T$. Expanding out the above equation,
\begin{equation}
L_{\nu_{\alpha}} \simeq {1 \over 24\zeta (3)}\left[
\pi^2 (\tilde{\mu}_{\alpha} - 
\tilde{\mu}_{\overline{\alpha}})
+ 6(\tilde{\mu}_{\alpha}^2 - 
\tilde{\mu}_{\overline{\alpha}}^2)\ln 2
+ (\tilde{\mu}_{\alpha}^3 - 
\tilde{\mu}_{\overline{\alpha}}^3) \right].
\label{j1}
\end{equation}
This is an exact equation for $\tilde{\mu}_{\alpha} =
- \tilde{\mu}_{\overline{\alpha}}$, otherwise it holds to a good
approximation provided that $\tilde{\mu}_{\alpha,\overline{\alpha}}
\stackrel{<}{\sim} 1$. For $T \stackrel{>}{\sim} T^{\alpha}_{\text{dec}}$,
where $T^e_{\text{dec}} \approx 2.5$ MeV and $T^{\mu,\tau}_{\text{dec}} \approx 3.5
$ MeV are the chemical decoupling temperatures,
$\mu_{\nu_{\alpha}} \simeq - \mu_{\overline{\nu}_{\alpha}}$
because inelastic processes such as $\nu_{\alpha}\overline{\nu}_{\alpha}
\leftrightarrow e^+ e^-$ and $e^{+} e^{-} \leftrightarrow \gamma \gamma$    
are rapid enough to make
$\tilde{\mu}_{\alpha} +\tilde{\mu}_{\overline{\alpha}} 
\ \simeq \ \tilde{\mu}_{e^+} + \tilde{\mu}_{e^-} 
\simeq 0$.
However, for $1\ \text{MeV} \stackrel{<}{\sim} T \stackrel{<}{\sim}
T^{\alpha}_{\text{dec}}$, weak interactions are rapid enough to 
approximately thermalise the neutrino momentum distributions,
but not rapid enough to keep the neutrinos in chemical 
equilibrium.\footnote{The chemical and thermal decoupling temperatures are so
different because the inelastic collision rates are much
less than the elastic collision rates. See, for example, Ref.\cite{ekt} for
a list of the collision rates.}
In this case, the value of $\tilde{\mu}_{\alpha}$ is
approximately frozen at $T \simeq T^{\alpha}_{\text{dec}}$ (taking
for definiteness $L_{\nu_\alpha} > 0$), while the anti-neutrino 
chemical potential $\tilde{\mu}_{\overline{\alpha}}$
continues evolving until $T \simeq 1$ MeV. For $T \stackrel{<}{\sim} 1$ MeV, the exact form
for the right-hand side of Eq.(\ref{b2}) should be used.

The neutrino asymmetries that appear in $\lambda$, $R$ and their antineutrino
analogues are in principle
calculated from the density matrices.
Recall that the neutrino asymmetry is defined in Eq.(\ref{def}).
The number density of $\nu_\alpha$ is
\begin{equation}
n_{\nu_\alpha} = \int^{\infty}_{0} N_{\nu_\alpha}dp =
\int^{\infty}_0 {1 \over 2} (P_0 + P_z) N^{\text{eq}}(p,T,0) dp,
\end{equation}
so that 
\begin{equation}
L_{\nu_{\alpha}} = {1 \over 2n_{\gamma}}\int^{\infty}_{0} \left[ (P_0 + P_z) -
(\overline{P}_0 +
\overline{P}_z) \right] N^{\text{eq}}(p,T,0)dp.
\end{equation}

Although it is in a strict technical sense redundant, it is useful to 
derive an equation for the rate of change of lepton number. It is given by
\begin{equation}
{dL_{\nu_\alpha} \over dt} =
{d \over dt}\left( {n_{\nu_\alpha} - n_{\overline{\nu}_\alpha}\over
n_\gamma}\right).
\end{equation}
Thus, using Eq.(\ref{c}),
\begin{equation}
{dL_{\nu_\alpha} \over dt} =
{1 \over 2n_\gamma}\int^{\infty}_{0} 
\left[{\partial P_0 \over \partial t} + 
 {\partial P_z \over \partial t} 
- {\partial \overline{P}_0 \over \partial t} - 
{\partial\overline{P}_z \over \partial t} \right]N^{\text{eq}}(p, T, 0)dp.
\end{equation}
This equation can be further simplified using the QKEs and the 
fact that the repopulation does not directly affect the
lepton number to obtain,
\begin{equation}
{dL_{\nu_\alpha} \over dt} =
{1 \over 2n_\gamma}\int^{\infty}_{0} 
\beta \left( P_y - \overline{P}_y \right)N^{\text{eq}}(p, T, 0)dp.
\label{wed}
\end{equation}
In our numerical work, this equation is the one 
actually used to calculate the lepton number that appears in 
the QKEs. Note for future reference that a limiting case 
of these equations will take centre stage when we come to study the Low
Temperature Epoch.

The last piece of information needed is the 
evolution equation for the mirror sector temperature $T'$. 
This is obtained by using a conservation of energy argument 
that was first presented in Ref.\cite{fv3}. 
It goes as follows: Consider $\nu_{\alpha} \to \nu'_{\beta}$
oscillations with the mirror interactions felt by $\nu'_{\beta}$ 
artificially switched off.  The energy density of the 
$\nu'_{\beta}$ and $\overline{\nu}'_{\beta}$ states is then 
given by
\begin{eqnarray}
\rho_{\nu'_{\beta}} & = & \int_0^{\infty} \left( N_{\nu'_{\beta}} +
N_{\overline{\nu}'_{\beta}} \right) p dp \nonumber\\
& = & {1 \over 2} \int_0^{\infty} \left(P_0 - P_z + \overline{P}_0 - \overline{P}_z
\right) p N^{\text{eq}}(p, T, 0) dp.
\label{mirrorrho}
\end{eqnarray}
Now switch on the mirror self-interactions. 
They will quickly distribute this energy density
amongst all of the relevant mirror species: the three 
mirror neutrinos and antineutrinos,
the mirror electrons and positrons, and the mirror photon. 
However, the energy density that
is being fed into the mirror sector by 
ordinary-mirror oscillations is still given by the
righthand-side of Eq.(\ref{mirrorrho}). The rate at 
which energy density is being
transferred from the ordinary to the mirror 
sector is therefore equal to the time rate of
change of the righthand-side of Eq.(\ref{mirrorrho}) 
due to {\it oscillations only}. 
Therefore we conclude that
\begin{equation}
\left. {d\rho' \over dt} \right|_{\nu_{\alpha} \leftrightarrow 
\nu'_{\beta}} = {1 \over 2} \int_0^{\infty} 
\left. {\partial \over \partial t}\right|_{\nu_{\alpha}
\leftrightarrow \nu'_{\beta}} \left( P_0 - P_z + \overline{P}_0 -
\overline{P}_z \right) p N^{\text{eq}}(p, T, 0) dp
\label{rhoprime}
\end{equation}
where $\rho' \equiv 3\rho_{\nu'_{\beta}} + \rho_{e'} + \rho_{\gamma'}$ 
is the total energy density in mirror species. 
The complete evolution equation for $T'$ is obtained by combining
Eq.(\ref{rhoprime}) with the cosmological red-shifting of $T'$. 
To this end, consider the quantity $\gamma_{\rho}$ where
\begin{equation}
\gamma_{\rho} \equiv {\rho' \over \rho} = \left({T' \over T}\right)^4,
\label{gammarho}
\end{equation}
with $\rho = {43 \over 4}{\pi^2 \over 30} T^4$ 
being the total energy density due to ordinary species. 
This ratio of energy densities does not red-shift. Its total rate of
change can therefore be calculated
from Eqs.(\ref{rhoprime}), (\ref{b1}) and (\ref{b2}) to yield
\begin{equation}
{d\gamma_{\rho} \over dt} \simeq 
- {1 \over 2 \rho}
\int^{\infty}_{0} \beta \left( P_y + \overline{P}_y \right) 
p N^{\text{eq}}(p, T, 0) dp,
\label{Tprimeeq}
\end{equation}
which is the required evolution equation for $T'$. 
Note that we implicitly assumed in the
above derivation that the redistribution of 
mirror energy density from $\nu'_{\beta}$'s to
the other mirror species did not affect the 
expansion rate of the universe. This is a good
approximation provided that the mirror energy density is small, 
that is $\gamma_{\rho} \ll 1$.
This is generally expected to be the case since $\delta N_{\nu, \text{eff}} 
\stackrel{<}{\sim} 0.6 \Rightarrow \gamma_\rho \stackrel{<}{\sim} 0.1$.

We end this section with an example of the evolution 
of neutrino asymmetry generated by
two flavour ordinary - mirror neutrino oscillations.
In Fig.1 the evolution of the $\tau$-like asymmetry is plotted. For our
present illustrative
purpose, we have considered the evolution of $L_{\nu_{\tau}}$ under the influence of 
the $\nu_\tau - \nu'_\mu$ oscillation mode only. The parameter point 
$\delta m^2_{\tau\mu'} 
= - 50\ \text{eV}^2$ and $\sin^2 2\theta_{\tau\mu'} =
10^{-8}$ has been chosen. The initial $L_{\nu_\tau}$ is set to zero. 
Notice that $L_{\nu_\tau}$ evolves from
zero to a value which approximately cancels the baryon 
asymmetry by $T \simeq 70\
\text{MeV}$. The asymmetry then remains constant until 
the critical temperature $T_c \simeq 38\ \text{MeV}$ when explosive 
growth begins. Shortly thereafter, the explosive
growth phase gives way to power law $T^{-4}$ growth. 
During the power law phase, the High 
Temperature Epoch evolves into the Low Temperature Epoch. 
Of course the behaviour shown in
Fig.1 is quite general and in fact quite
similar to ordinary - sterile neutrino oscillations.
The latter have already been studied in some detail
in previous papers\cite{ftv,fv1,fv2,f,bvw}.
Finally note that we have plotted $|L_{\nu_\tau}|$.
This is because $L_{\nu_\tau}$ changes sign at the
critical temperature (the reason for this behaviour
has been discussed in Ref.\cite{fv1}). 
For values of $\sin^2 2\theta$ large enough, our
numerical results indicate that the sign of $L_{\nu_\tau}$ initially oscillates
and thus the final sign of the asymmetry may be random.
This may lead to different regions of space having different
neutrino asymmetries (as suggested earlier in Ref.\cite{ftv}).
Also it should be mentioned that the effect of statistical 
fluctuations on lepton number
asymmetry is an important open problem, and consequently it is 
also possible that the sign of the asymmetry may turn 
out to be random even for small values of $\sin^2 2\theta$.
For the purposes of the present paper, we acknowledge the
indeterminate nature of the asymmetry by considering
the two possible signs in all our numerical work.

In the next section we will discuss
the High Temperature Epoch in the EPM. Our main goal there 
will be to demonstrate the consistency of
the $\nu_{\mu} \to \nu'_{\mu}$ solution to the atmospheric 
neutrino problem with BBN for a range of parameters.

\section{High Temperature Epoch: consistency of $\nu_\mu \to \nu_\mu'$ 
solution of the atmospheric neutrino anomaly with BBN}

Previous work has shown that the QKEs for ordinary-mirror 
(or ordinary-sterile) oscillations will imply the explosive 
creation of neutrino asymmetries provided some fairly mild
restrictions on the oscillation parameter space are imposed. 
In particular, the $\delta m^2_{\alpha\beta'}$ involved must be 
negative, and the vacuum mixing angle $\theta_{\alpha\beta'}$ 
must be in the approximate range
\begin{equation}
10^{-10} \stackrel{<}{\sim} \sin^2 2\theta_{\alpha\beta'} \stackrel{<}{\sim} 
\ \text{few} \times 10^{-4} \left(
{\text{eV}^2 \over |\delta m^2_{\alpha \beta'}|}\right)^{{1 \over 2}}.
\label{thetarange}
\end{equation}
The lower bound comes from the requirement that the oscillation 
mode be sufficiently strong\footnote{But note that the vacuum 
oscillation amplitude can still be tiny!}, while the upper bound was 
derived in Ref.\cite{fv3} from the requirement that $\nu_{\alpha} \to
\nu'_{\beta}$ oscillations, considered in isolation, not spoil BBN 
[Eq.(\ref{thetarange}) takes for definiteness that $\delta N_{\nu,\text{eff}} 
\stackrel{<}{\sim} 0.6$]. 
Once created, the large neutrino asymmetry or asymmetries will suppress 
other ordinary-sterile oscillation modes for a range of parameters.

For the generic parameter space region considered here 
(see Sec.II), the oscillation modes
\begin{equation}
\nu_{\tau} \to \nu'_{\mu},\quad \nu_{\tau} \to 
\nu'_e,\quad \nu_{\mu} \to \nu'_e,
\end{equation}
could all satisfy the above criteria. We will call these 
the ``lepton number creating
modes''.  The other oscillation modes, including the
$\nu_{\mu} \to \nu'_{\mu}$ mode that hypothetically 
solves the atmospheric neutrino
problem, tend to destroy a linear combination of asymmetries.

Lepton number amplification begins at a critical temperature 
$T_c$, given roughly by
\begin{equation}
T_c \approx 16 \left( - {\delta m^2 \cos2\theta \over \text{eV}^2} 
\right)^{{1 \over 6}}\
\text{MeV},
\end{equation}
where the oscillation parameters pertain to the two-flavour lepton 
number creating mode responsible. The mode with the largest 
$|\delta m^2|$ will therefore be expected to create
lepton number first, provided its vacuum mixing angle is in the 
range of Eq.(\ref{thetarange}). Within the scenario of Sec.II, the 
$\nu_{\tau} \to \nu'_{\mu}$ and $\nu_{\tau} \to \nu'_e$ modes will
have the largest $\delta m^2$ values. Which, if either, of them 
dominates lepton number creation then depends on their specific 
oscillation parameters. For the sake of a plausible
example, we will suppose that $\nu_{\tau} \to \nu'_{\mu}$ dominates, 
even though $\nu_{\tau} \to \nu'_e$ has a slightly larger $\delta m^2$. 
This is because we expect $\theta_{\tau e'} \ll \theta_{\tau \mu'}$ 
as per Sec.II. Basically, we will work in the parameter space
region where $\theta_{\tau e'}$ is negligibly tiny.

So, we are led to consider the four flavour subsystem,
\begin{equation}
\begin{array}{ccc}
\nu_{\mu} & \leftrightarrow & \nu'_{\mu} \\
\updownarrow & {\nearrow \!\!\!\!\!\! \nwarrow} \!\!\!\!\!\! {\searrow \!\!\!\!\!\! 
\swarrow}
& \updownarrow \\
\nu_{\tau} & \leftrightarrow & \nu'_{\tau}
\end{array}.
\end{equation}
Further, we decompose this four flavour system into the two 
flavour subsystems indicated by the arrows above. Some discussion of 
the justification for this sort of decomposition can be
found in Ref.\cite{bvw}. 
Heuristically, it is expected that this simplifying assumption 
is justified because the MSW resonance momenta of each
of the oscillation modes are generally different.
The ordinary-ordinary and mirror-mirror modes are 
governed by the same mixing angle $\theta_{\mu\tau}$, the 
modes $\nu_{\tau} \to \nu'_{\mu}$ and $\nu_{\mu} \to
\nu'_{\tau}$ are both governed by another mixing angle 
$\theta_{\tau\mu'}$, while the $\nu_{\tau,\mu} \to \nu'_{\tau,\mu}$ 
modes are maximally mixed. The squared mass difference
$\delta m^2_{\mu\mu'}$ is set by the atmospheric neutrino data to 
be in the range quoted in Eq.(\ref{massranges}). The other mass 
parameters, $\delta m^2_{\tau\mu'}$ and $\delta
m^2_{\tau\tau'}$, are free, subject to the restrictions 
discussed in Sec.II and summarised in Eq.(\ref{massranges}).

For simplicity and the sake of the example, we will set 
$|\delta m^2_{\tau\tau'}|$ to be so small that the associated oscillation 
mode can be neglected.\footnote{This will be the case provided that
$|\delta m^2_{\tau \tau'}| \stackrel{<}{\sim} |\delta m^2_{\mu \mu'}|$.
If $|\delta m^2_{\tau \tau'}|$ is much larger than 
$|\delta m^2_{\mu \mu'}|$ then the resulting `allowed region' 
will be significantly reduced.} The precise value of the mixing
angle $\theta_{\mu\tau}$ is unimportant, provided it is small. 
The $\nu_{\mu} \leftrightarrow \nu_{\tau}$ mode has almost no effect 
until lepton number is large (here `large' means
greater than about $10^{-2}$), because of the
approximately equal number densities of the two species involved.
In the High Temperature Epoch being
considered in this Section, lepton number will always be 
small. For a given $\delta m^2_{\mu\mu'}$ we are therefore 
effectively left with two free parameters:
$\theta_{\tau\mu'}$ and $\delta m^2_{\tau\mu'}$. Our task 
is to find the region of this parameter space for which the 
$\nu_{\mu} \to \nu'_{\mu}$ solution to the atmospheric
neutrino problem is consistent with BBN. For the EPM\footnote{
For the case of strictly sterile neutrinos, this calculation
was done in the static approximation in Ref.\cite{fv1}
and by numerically integrating the quantum kinetic
equations in Ref.\cite{f}.}
this calculation was first performed in
Ref.\cite{fv3} within the static approximation. We improve on this approach 
here through the use of the QKEs.

We now write down the equations we must solve. We begin by 
introducing three two-flavour density matrices,
\begin{equation}
\rho_{\tau\mu'} \equiv {1 \over 2}(P_0 + {\bf \sigma} \cdot {\bf P}),\quad
\rho_{\mu\mu'} \equiv {1 \over 2}(Q_0 + {\bf \sigma} \cdot {\bf Q}),\quad
\rho_{\mu\tau'} \equiv {1 \over 2}(S_0 + {\bf \sigma} \cdot {\bf S}),
\end{equation}
for the three significant oscillation modes, 
\begin{equation}
\nu_{\tau} \leftrightarrow \nu'_{\mu},\quad \nu_{\mu} \leftrightarrow \nu'_{\mu},\quad
\nu_{\mu} \leftrightarrow \nu'_{\tau},
\end{equation}
respectively. Since $\nu'_{\mu}$ is common to the first pair of 
modes, and $\nu_{\mu}$ is common to the second pair, we have 
the constraints
\begin{eqnarray}
{N_{\nu'_{\mu}} \over N^{\text{eq}}(p,T,0)} & = & 
{1 \over 2}(P_0 - P_z) = {1 \over 2}(Q_0
- Q_z),\nonumber\\
{N_{\nu_{\mu}} \over N^{\text{eq}}(p,T,0)} & = & 
{1 \over 2}(Q_0 + Q_z) = {1 \over 2}(S_0 + S_z).
\label{constraints}
\end{eqnarray}
Extending the two flavour case discussed in the previous section,
the time derivatives of the functions $P_0$, $Q_0$, $S_0$, 
${\bf P}$, ${\bf Q}$ and ${\bf S}$ are observed to receive 
contributions from each of the three oscillation modes, 
from decohering collisions, and from repopulation. Denoting a 
generic function by $F$, we have that
\begin{equation}
{\partial F \over \partial t} = \left. 
{\partial F \over \partial t}\right|_{\nu_{\tau} \to
\nu'_{\mu}} +
\left. {\partial F \over \partial t}\right|_{\nu_{\mu} \to
\nu'_{\mu}} + 
\left. {\partial F \over \partial t}\right|_{\nu_{\mu} \to
\nu'_{\tau}} +
\left. {\partial F \over \partial t}\right|_{\text{coll}} +
\left. {\partial F \over \partial t}\right|_{\text{repop}}.
\end{equation}

 From the two-flavour formalism described in Sec.III we have that
\begin{eqnarray}
\left. {\partial {\bf P} \over \partial t}\right|_{\nu_{\tau} \to \nu'_{\mu}}
+ \left. {\partial {\bf P} \over \partial t}\right|_{\text{coll}} 
& = & {\bf V_{\tau\mu'}} \times {\bf P} - D {\bf P_T},\nonumber\\
\left. {\partial {\bf Q} \over \partial t}\right|_{\nu_{\mu} \to \nu'_{\mu}}
+ \left. {\partial {\bf Q} \over \partial t}\right|_{\text{coll}} 
& = & {\bf V_{\mu\mu'}} \times {\bf Q} - D {\bf Q_T},\nonumber\\
\left. {\partial {\bf S} \over \partial t}\right|_{\nu_{\mu} \to \nu'_{\tau}}
+ \left. {\partial {\bf S} \over \partial t}\right|_{\text{coll}} 
& = & {\bf V_{\mu\tau'}} \times {\bf S} - D {\bf S_T}.
\label{home}
\end{eqnarray}
It is also clear that
\begin{eqnarray}
\left. {\partial P_0 \over \partial t}\right|_{\nu_{\tau} \to \nu'_{\mu}} 
& = & \left. {\partial P_0 \over \partial t}\right|_{\text{coll}}
 =  0,\nonumber\\
\left. {\partial Q_0 \over \partial t}\right|_{\nu_{\mu} \to \nu'_{\mu}}    
& = & \left. {\partial Q_0 \over \partial t}\right|_{\text{coll}} = 
0,\nonumber\\
\left. {\partial S_0 \over \partial t}\right|_{\nu_{\mu} \to \nu'_{\tau}} 
& = & \left. {\partial S_0 \over \partial t}\right|_{\text{coll}} = 0.
\label{ano}
\end{eqnarray}
We also obviously know that
\begin{eqnarray}
\left. {\partial {\bf P} \over \partial t}\right|_{\nu_{\mu} \to 
\nu'_{\tau}} & = & 
\left. {\partial P_0 \over \partial t}\right|_{\nu_{\mu} \to 
\nu'_{\tau}}  = 0,\nonumber\\
\left. {\partial {\bf S} \over \partial t}\right|_{\nu_{\tau} \to 
\nu'_{\mu}} & = &
\left. {\partial S_0 \over \partial t}\right|_{\nu_{\tau} \to \nu'_{\mu}}  =
0.
\label{zero}
\end{eqnarray}
Consider now the contribution of $\nu_{\mu} \to \nu'_{\mu}$ 
oscillations to the evolution of ${\bf P}$ and $P_0$. First of 
all, the transverse components $P_{x,y}$ receive no contribution, 
\begin{equation}
\left. {\partial P_{x,y} \over \partial t}\right|_{\nu_{\mu} \to 
\nu'_{\mu}} = 0,
\end{equation}
because they are affected only by decohering collisions. 
The evolution of
$P_z$ and $P_0$ can be obtained from Eq.(\ref{constraints}) by noting that
\begin{eqnarray}
\left. {\partial \over \partial t} (P_0 - P_z) \right|_{\nu_{\mu} 
\to \nu'_{\mu}} & = & 
\left. {\partial \over \partial t} (Q_0 - Q_z) \right|_{\nu_{\mu} \to
\nu'_{\mu}}
 = - \left. {\partial Q_z \over \partial t}\right|_{\nu_{\mu} \to
\nu'_{\mu}},\nonumber\\
\left. {\partial \over \partial t} (P_0 + P_z) \right|_{\nu_{\mu} \to 
\nu'_{\mu}} & = &  0,
\end{eqnarray}
so that
\begin{eqnarray}
\left. {\partial P_0 \over \partial t} \right|_{\nu_{\mu} \to 
\nu'_{\mu}} & = & -{1 \over 2}
\left. {\partial Q_z\over \partial t}\right|_{\nu_{\mu} \to 
\nu'_{\mu}},\nonumber\\
\left. {\partial P_z \over \partial t} \right|_{\nu_{\mu} \to 
\nu'_{\mu}} & = & 
+ {1 \over 2}
\left. {\partial Q_z \over \partial t}\right|_{\nu_{\mu} \to \nu'_{\mu}},
\end{eqnarray}
having used Eq.(\ref{ano}). The expression for 
$\left. (\partial Q_z/\partial t)\right|_{\nu_{\mu} \to \nu'_{\mu}}$ 
is obtained from Eq.(\ref{home}). Finally, we have that
\begin{eqnarray}
\left. {\partial P_{x,y} \over \partial t} \right|_{\text{repop}} 
& = & 0,\nonumber\\
\left. {\partial \over \partial t} {1 \over 2}(P_0 + P_z) 
\right|_{\text{repop}} & = &
\Gamma\left[K_{\nu_{\tau}} - {1 \over 2}(P_0 + P_z) \right],\nonumber\\
\left. {\partial \over \partial t} {1 \over 2}(P_0 - P_z) \right|_{\text{repop}} & = &
\Gamma'\left[K_{\nu'_{\mu}} - {1 \over 2}(P_0 - P_z) \right].
\end{eqnarray}
This completes the specification of the evolution equations 
for ${\bf P}$ and $P_0$.

The evolution of ${\bf S}$ and $S_0$ due to $\nu_{\mu} 
\to \nu'_{\mu}$ oscillations and
repopulation is handled in a very similar manner to yield,
\begin{eqnarray}
\left. {\partial S_{x,y} \over \partial t}\right|_{\nu_{\mu} \to 
\nu'_{\mu}} & = & 0,
\nonumber\\
\left. {\partial S_0 \over \partial t} \right|_{\nu_{\mu} \to 
\nu'_{\mu}} & = & +{1 \over 2}
\left. {\partial Q_z\over \partial t}\right|_{\nu_{\mu} \to 
\nu'_{\mu}},\nonumber\\
\left. {\partial S_z \over \partial t} \right|_{\nu_{\mu} \to 
\nu'_{\mu}} & = & + {1 \over 2}
\left. {\partial Q_z \over \partial t}\right|_{\nu_{\mu} \to 
\nu'_{\mu}},\nonumber\\
\left. {\partial S_{x,y} \over \partial t} \right|_{\text{repop}} 
& = & 0,\nonumber\\
\left. {\partial \over \partial t} {1 \over 2}(S_0 + S_z) 
\right|_{\text{repop}} & = &
\Gamma\left[K_{\nu_{\mu}} - {1 \over 2}(S_0 + S_z) \right],\nonumber\\
\left. {\partial \over \partial t} {1 \over 2}(S_0 - S_z) 
\right|_{\text{repop}} & = &
\Gamma'\left[K_{\nu'_{\tau}} - {1 \over 2}(S_0 - S_z) \right].
\end{eqnarray}
The completes the specification of the evolution 
equations for ${\bf S}$ and $S_0$.

Finally, we have to specify the time rate of change of ${\bf Q}$ 
and $Q_0$ under the influence of $\nu_{\tau} \leftrightarrow 
\nu'_{\mu}$ oscillations, $\nu_{\mu}
\leftrightarrow \nu'_{\tau}$ oscillations and repopulation. 
We note first that $Q_{x,y}$ are unaffected by these processes:
\begin{equation}
\left. {\partial Q_{x,y} \over \partial t} \right|_{\nu_{\tau} 
\to \nu'_{\mu}} = 
\left. {\partial Q_{x,y} \over \partial t} \right|_{\nu_{\mu} 
\to \nu'_{\tau}} =
\left. {\partial Q_{x,y} \over \partial t} \right|_{\text{repop}} = 0.
\end{equation}
Then, from Eq.(\ref{constraints}) we see that
\begin{eqnarray}
\left. {\partial Q_0 \over \partial t} \right|_{\nu_{\tau} 
\to \nu'_{\mu}} & = & {1 \over 2}
\left. {\partial \over \partial t} (P_0 - P_z) \right|_{\nu_{\tau} \to
\nu'_{\mu}} = - {1 \over 2} \left. {\partial P_z \over \partial t} 
\right|_{\nu_{\tau} \to \nu'_{\mu}},
\nonumber\\
\left. {\partial Q_z \over \partial t} \right|_{\nu_{\tau} \to \nu'_{\mu}} & = & 
- {1 \over 2} \left. {\partial \over \partial t} (P_0 - P_z) 
\right|_{\nu_{\tau} \to
\nu'_{\mu}} = + {1 \over 2} \left. {\partial P_z \over \partial t} 
\right|_{\nu_{\tau} \to \nu'_{\mu}},
\nonumber\\
\left. {\partial Q_0 \over \partial t} \right|_{\nu_{\mu} \to 
\nu'_{\tau}} & = & {1 \over 2}
\left. {\partial \over \partial t} (S_0 + S_z) \right|_{\nu_{\mu} \to
\nu'_{\tau}} = + {1 \over 2} \left. {\partial S_z \over \partial t} 
\right|_{\nu_{\mu} \to \nu'_{\tau}},
\nonumber\\
\left. {\partial \over \partial t} Q_z \right|_{\nu_{\mu} \to 
\nu'_{\tau}} & = & 
{1 \over 2} \left. {\partial \over \partial t} (S_0 + S_z) 
\right|_{\nu_{\mu} \to
\nu'_{\tau}} = + {1 \over 2} \left. {\partial S_z \over \partial t} 
\right|_{\nu_{\mu} \to \nu'_{\tau}}.
\end{eqnarray}
Finally,
\begin{eqnarray}
\left. {\partial \over \partial t} {1 \over 2}(Q_0 + Q_z) 
\right|_{\text{repop}} & = & 
\Gamma \left[ K_{\nu_{\mu}} - {1 \over 2} (Q_0 + Q_z) \right],\nonumber\\
\left. {\partial \over \partial t} {1 \over 2} (Q_0 - Q_z) 
\right|_{\text{repop}} & = & 
\Gamma' \left[ K_{\nu'_{\mu}} - {1 \over 2} (Q_0 - Q_z) \right],
\end{eqnarray}
specifies the remaining repopulation equations. 

This completes the list of quantum kinetic 
equations for our system. Of course, because of
the constraints in Eq.(\ref{constraints}), some of 
these equations are redundant.

To use these equations one needs, (i) equations 
connecting neutrino asymmetries with
chemical potentials, and (ii) an evolution equation for 
the temperature $T'$ of the mirror
plasma. The chemical potentials are calculated in exactly 
the same way as discussed in
Sec.III. The $T'$ evolution equation is obtained by extending the 
two-flavour derivation of
Sec.III in the obvious way. As before, it is 
best to first imagine that the mirror
electroweak interactions are artificially switched off. 
The energy density in mirror states
is then entirely due to the $\nu'_{\mu}$ and $\nu'_{\tau}$ species:
\begin{eqnarray}
\rho_{\nu'_{\mu}} + \rho_{\nu'_{\tau}} & = &
\int_0^{\infty} \left(N_{\nu'_{\mu}} + N_{\overline{\nu}'_{\mu}} +
N_{\nu'_{\tau}} + N_{\overline{\nu}'_{\tau}} \right) p dp\nonumber\\
& = & {1 \over 2} \int_0^{\infty} \left( P_0 - P_z 
+ \overline{P}_0 - \overline{P}_z + S_0 -
S_z + \overline{S}_0 - \overline{S}_z \right) p N^{\text{eq}}(p,T,0) dp.
\end{eqnarray}
With mirror electroweak interactions now switched 
on, this energy density is distributed
amongst all of the relevant mirror species. The rate 
at which energy density is being
transferred from the ordinary to the mirror sector is thus
\begin{equation}
\left. {d\rho' \over dt}\right|_{\text{osc}} 
= {1 \over 2} \int_0^{\infty} \left. 
{\partial \over \partial t}\right|_{\text{osc}}
\left( P_0 - P_z + \overline{P}_0 -
\overline{P}_z + S_0 -
S_z + \overline{S}_0 - \overline{S}_z \right) p N^{\text{eq}}(p,T,0) dp,
\end{equation}
where $\rho'$ is the total energy density in mirror species, and
\begin{equation}
\left. {d \over dt}\right|_{\text{osc}} \equiv
\left. {d \over dt} \right|_{\nu_{\tau} \to \nu'_{\mu}} +
\left. {d \over dt} \right|_{\nu_{\mu} \to \nu'_{\tau}} +
\left. {d \over dt} \right|_{\nu_{\mu} \to \nu'_{\mu}}.
\end{equation}
Introducing $\gamma_{\rho}$ as
per Eq.(\ref{gammarho}) and using the QKEs we obtain,
\begin{equation}
{d \gamma_{\rho} \over dt} = - {1 \over 2\rho} \int_0^{\infty} 
\left[ \beta_{\tau\mu'}
\left( P_y + \overline{P}_y \right) 
+ \beta_{\mu\mu'} \left( Q_y + \overline{Q}_y \right)
+ \beta_{\mu\tau'} \left( S_y + \overline{S}_y \right)
\right] p N^{\text{eq}}(p,T,0) dp
\end{equation}
as the $T'$ evolution equation, where
\begin{equation}
\beta_{\tau\mu'} \equiv {\delta m^2_{\tau\mu'} \over 2p} 
\sin 2\theta_{\tau\mu'},\quad
\beta_{\mu\mu'} \equiv {\delta m^2_{\mu\mu'} \over 2p},\quad
\beta_{\mu\tau'} \equiv {\delta m^2_{\mu\tau'} \over 2p} 
\sin 2\theta_{\tau\mu'}.
\end{equation}
Note that $\beta_{\tau\mu'} \simeq -\beta_{\mu\tau'}$ because 
$\delta m^2_{\tau\mu'} \simeq
-\delta m^2_{\mu\tau'}$ for the parameter space of interest.

We first present the main result of numerically solving the above
equations. After doing so, we will provide a physical description of
what lies behind the mathematics. The main result is displayed 
in Fig.2, which shows the region of
$(\delta m^2_{\tau\mu'},\ \sin^2 2\theta_{\tau\mu'})$ parameter 
space which is consistent with Big Bang Nucleosynthesis, 
for various values of $\delta m^2_{\mu\mu'}$ motivated by the
atmospheric neutrino anomaly. The allowed region lies above 
the relevant solid line (which corresponds to a particular 
$\delta m^2_{\mu\mu'}$), and to the left of the
dash-dotted line. The solid lines arise from solving 
the QKEs, while the dash-dotted line is the upper bound 
quoted in Eq.(\ref{thetarange}) applied to the lepton
number creating mode $\nu_{\tau} \leftrightarrow \nu'_{\mu}$. 
For the sake of definiteness, we have adopted 
$\delta N_{\nu,\text{eff}} \stackrel{<}{\sim} 0.6$ as the BBN bound 
on the expansion rate of the universe (expressed as an equivalence to 
additional relativistic neutrino flavours, as
is customary). Of course, at the present time there is 
some confusion regarding the value of this bound, due 
to conflicting primordial element abundance measurements. 
The value of $0.6$ was chosen for illustrative purposes only. 
The position of the dash-dotted line depends on
the $\delta N_{\nu,\text{eff}}$ chosen. The solid lines, on 
the other hand, define sharp transition regions.
Below the lines, the mirror sector comes into 
thermal equilibrium because of the eventual copious
production of $\nu'_{\mu}$ from $\nu_{\mu} \to \nu'_{\mu}$ 
oscillations. 
Above the lines, essentially no mirror matter 
is produced by this oscillation 
mode. Of course, the closer one
gets to the dash-dotted line, the more mirror matter is 
produced by the $\nu_{\tau} \to
\nu'_{\mu}$ mode. For $\sin^2 2\theta_{\tau\mu'} 
\stackrel{<}{\sim} 10^{-5}$ the results
obtained here using the QKEs are almost identical to 
those obtained earlier using the static
approximation. The results differ at large values 
of the mixing angle mainly because lepton number
is created rapidly enough to spoil the validity of 
the static approximation.\footnote{A
complete discussion of the static approximation can be found 
in Refs.\cite{fv1,bvw}. In particular, it
was shown in Ref.\cite{bvw} that the static approximation is an 
adiabatic-like approximation for
partially incoherent oscillations in the small vacuum 
mixing angle parameter space regime.}

{\it The importance of Fig.2 lies in its demonstration that 
the $\nu_{\mu} \to \nu'_{\mu}$
solution in the EPM to the atmospheric neutrino problem is 
cosmologically consistent for a large region of oscillation 
parameter space.} 
Furthermore, most of this region sees the $\nu_{\tau}$
having a cosmologically interesting mass. In particular, note that 
the hot dark matter region marked as a shaded band in Fig.2 
has a significant overlap with the allowed region
from BBN (we will discuss more about the dark matter
region in section VI). It is also interesting to note 
that the BBN allowed 
region implied by the $\nu_{\mu} \to \nu'_{\mu}$ solution to the 
atmospheric neutrino problem is larger than the
corresponding region\cite{fv1,f} obtained when $\nu'_{\mu}$ is 
replaced by a strictly sterile neutrino.
The production of sterile neutrinos tends
to delay the onset of the rapid exponential growth\cite{fv1}
which means that by the time it occurs the $\nu_\mu \leftrightarrow
\nu'_\mu$ oscillations can destroy $L^{(\mu \mu')}$ more 
efficiently since they are not damped so much by the collisions.
In the case of the mirror neutrino 
$\nu'_{\mu}$, the mirror electroweak
interactions have the effect of reducing their number density, 
so the production of lepton number is not delayed.
Of course a $\nu_\tau$ in the eV mass region is
currently being searched for in the short baseline Nomad/Chorus
experiments. Such experiments are extremely important
to test for the eV tau neutrino which is suggested
by Figure 2. Unfortunately, we cannot predict 
$\sin^2 2\theta_{\tau \mu}$, so these experiments will
either discover $\nu_\tau \leftrightarrow \nu_\mu$ oscillations
or constrain $\sin^2 2\theta_{\tau \mu}$.
 
Before closing this section, we will discuss some of 
the numerical details of performing the
above computation, which will entail also a discussion of 
the physics of the result.

MSW resonances play a key role in the evolution of the system. 
It is instructive to examine the connection between the 
neutrino asymmetries and the resonance momenta of the three
important two-flavour modes within our system. Note, first 
of all, that the effective
potentials of the three modes 
depend on different linear combinations of neutrino
asymmetries:
\begin{eqnarray}
L^{(\tau\mu')} & = & 2 L_{\nu_{\mu}} + 3 L_{\nu_{\tau}} - 
L_{\nu'_{\mu}},\nonumber\\
L^{(\mu\mu')} & = & 3 L_{\nu_{\mu}} + 2 L_{\nu_{\tau}} - 
L_{\nu'_{\mu}},\nonumber\\
L^{(\mu\tau')} & = & 4 L_{\nu_{\mu}} + 3 L_{\nu_{\tau}} + L_{\nu'_{\mu}},
\end{eqnarray}
where we have set $L_{\nu_e} = L_{\nu'_e} = 0$, and we have 
used conservation of lepton number\footnote{
Of course the sum of lepton numbers need not be exactly
zero. However, we can set the sum to zero without loss
of generality provided that the sum is not large.}
\begin{equation}
L_{\nu_e} + L_{\nu_{\mu}} + L_{\nu_{\tau}} + L_{\nu'_e} 
+ L_{\nu'_{\mu}} + L_{\nu'_{\tau}} = 0,
\end{equation}
to eliminate $L_{\nu'_{\tau}}$. The lepton number creating mode 
$\nu_{\tau} \to \nu'_{\mu}$
generates a nonzero $L_{\nu_{\tau}}$,
which means that $L^{(\mu\mu')}$ is also nonzero. The 
latter quantity then suppresses
$\nu_{\mu} \to \nu'_{\mu}$ oscillations, provided that 
it grows sufficiently quickly for a
sufficiently long period of time. This is not inevitable, 
because the effect of the lepton
number destroying mode $\nu_{\mu} \to \nu'_{\mu}$ is to try 
to destroy $L^{(\mu\mu')}$
through the creation of nonzero values for $L_{\nu_{\mu}}$ and 
$L_{\nu'_{\mu}}$ to compensate the
nonzero $L_{\nu_{\tau}}$ and $L_{\nu'_{\mu}}$ produced
by the $\nu_\tau \to \nu'_\mu$ oscillations. 
The essence of the calculation presented above is the
determination of when $L^{(\mu\mu')}$ is driven to zero, and 
when it is not. This depends on
the oscillation parameters, as summarised in Fig.2.

The resonance momentum for the $\nu_{\tau} \to \nu'_{\mu}$ mode 
and its antimatter analogue are given by
\begin{eqnarray}
{p_{\tau\mu'} \over T} & = & {1 \over 2}\left[ {a_0 L^{(\tau\mu')} 
\over b_0 T^2} + 
\sqrt{ \left( {a_0 L^{(\tau\mu')} \over b_0 T^2} \right)^2 
+ {4 |\delta m^2_{\tau\mu'}| 
\cos 2\theta_{\tau\mu'} \over b_0 T^6} }\right],\nonumber\\
{\overline{p}_{\tau\mu'} \over T} & = & {1 \over 2}\left[ 
- {a_0 L^{(\tau\mu')} \over b_0 T^2}
+  \sqrt{ \left( {a_0 L^{(\tau\mu')} \over b_0 T^2} \right)^2 + 
{4 |\delta m^2_{\tau\mu'}|
\cos 2\theta_{\tau\mu'} \over b_0 T^6} } \right].
\end{eqnarray}
The $p$- and $T$-independent quantities $a_0$ and $b_0$ are defined through,
\begin{equation}
a(p) \equiv - a_0 {T^3 L^{(\alpha \beta')} p \over 
\delta m^2_{\alpha\beta'} },\quad
b(p) \equiv - b_0 { T^4 p^2 \over \delta m^2_{\alpha\beta'} },
\label{70}
\end{equation}
leading to
\begin{equation}
a_0 = {4\sqrt{2}\zeta(3)G_F \over \pi^2},\qquad b_0 =
{4\sqrt{2}\zeta(3)G_FA_{\alpha} \over \pi^2 M_{W}^2}.
\end{equation}
In the following we will consider the $L^{(\tau\mu')} > 0$ case
for definiteness.\footnote{A discussion of the overall sign of
the asymmetries created can be found in Ref.\cite{fv1}.} 
Before the rapid exponential creation of lepton number (that is
for $T > T_c$), the neutrino and 
antineutrino resonance momenta for
the lepton number creating modes are equal.
As $L^{(\tau\mu')}$ gets exponentially created, the 
neutrino resonance momentum
$p_{\tau\mu'}$
moves rapidly to infinity, while the antineutrino resonance momentum
$\overline{p}_{\tau\mu'}$ remains at a value of order $T$. 
Numerical calculations show that
$\overline{p}_{\tau\mu'}/T$ typically takes a value 
in the range $0.2 - 0.6$ 
at $T \simeq T_c/2$ where $T_c$ is the critical temperature
at which lepton number creation begins. These observations 
are important, because they mean that $\nu_{\tau}
\to \nu'_{\mu}$ oscillations are unimportant after the 
creation of lepton number, while
$\overline{\nu}_{\tau} \to \overline{\nu}'_{\mu}$ 
oscillations remain very important. This
is simply because the neutrino resonance momentum has moved 
to the tail of the Fermi-Dirac
distribution, while the antineutrino resonance 
momentum is within the body of the
distribution. (Note that these observations will play 
a central role in the next section.)

The resonance momenta for the $\nu_{\mu} \to \nu'_{\mu}$ 
and $\overline{\nu}_{\mu} \to
\overline{\nu}'_{\mu}$ modes are given by
\begin{eqnarray}
{p_{\mu\mu'} \over T} & = & {a_0 L^{(\mu\mu')} \over b_0 T^2},\quad 
{\overline{p}_{\mu\mu'} \over T} = 0\quad \text{if}\ 
L^{(\mu\mu')} > 0;\nonumber\\
{p_{\mu\mu'} \over T} & = & 0,\quad 
{\overline{p}_{\mu\mu'} \over T} = - {a_0 L^{(\mu\mu')} \over 
b_0 T^2}\quad \text{if}\
L^{(\mu\mu')} < 0.
\end{eqnarray}
In the region of parameter space where $L^{(\mu\mu')}$ 
is not driven to zero, we see that
$p_{\mu\mu'}/T$ gets driven to infinity (staying 
with the $L_{\nu_{\tau}} > 0$ case), while
$\overline{p}_{\mu\mu'}/T$ stays at zero. In the 
region of parameter space where
$L^{(\mu\mu')}$ gets destroyed, we see that $p_{\mu\mu'}/T$ 
moves from zero to a finite value
as $L_{\nu_{\tau}}$ gets created, then moves back 
towards zero as the compensating
$L_{\nu_{\mu}}$ is induced. There is a sharp
transition between these two possibilities for the evolution 
of $p_{\mu\mu'}/T$, with the
boundary given by the solid lines in Fig.2.
Above the solid line $L^{(\mu \mu')}$ is created early enough
and is large enough so that the $\nu'_\mu$ is never
significantly populated by $\nu_\mu \leftrightarrow \nu'_\mu$
oscillations, these oscillations being 
heavily suppressed by the matter effects resulting from the
large $L^{(\mu \mu')}$. Below the solid line the
the $\nu'_\mu$ states would eventually become populated
by $\nu_\mu \leftrightarrow \nu'_\mu$ oscillations
in the temperature range $6 \stackrel{<}{\sim} T/\text{MeV} \stackrel{<}{\sim}
10$. Furthermore, the other mirror particles would also become populated
due to the mirror weak interactions, which would effectively double
the energy density of the universe prior to the BBN epoch.

The $\nu_{\mu} \to \nu'_{\tau}$ and $\overline{\nu}_{\mu} \to
\overline{\nu}'_{\tau}$ resonance momenta are given by
\begin{eqnarray}
{p_{\mu\tau'} \over T} & = & {1 \over 2}\left[ 
{a_0 L^{(\mu\tau')} \over b_0 T^2} \pm 
\sqrt{ \left( {a_0 L^{(\mu\tau')} \over b_0 T^2} \right)^2 - 
{4 |\delta m^2_{\mu\tau'}| 
\cos 2\theta_{\tau\mu'} \over b_0 T^6} }\right],\nonumber\\
{\overline{p}_{\mu\tau'} \over T} & = & {1 \over 2}\left[ - 
{a_0 L^{(\mu\tau')} \over b_0 T^2}
\pm  \sqrt{ \left( {a_0 L^{(\mu\tau')} \over b_0 T^2} \right)^2 
- {4 |\delta m^2_{\mu\tau'}|
\cos 2\theta_{\tau\mu'} \over b_0 T^6} } \right].
\end{eqnarray}
Because the sign of $\delta m^2_{\mu\tau'}$ is positive,
we see a qualitatively different behaviour for 
these resonance momenta compared to their
mirror reflections in the $\nu_{\tau} + \nu'_{\mu}$ 
subsystem. Before the creation of lepton
number, there are no solutions to the resonance 
conditions. If lepton number evolves to the
point where
\begin{equation}
\left( { a_0 L^{(\mu\tau')} \over b_0 T^2 } \right)^2 
= { 4 |\delta m^2_{\mu\tau'}| \over b_0 T^6 }
\end{equation}
then (taking the $L^{(\mu\tau')} > 0$ case)
$\nu_{\mu} \leftrightarrow \nu'_{\tau}$
comes on resonance at a finite value of the momentum, while 
$\overline{\nu}_{\mu} \leftrightarrow \overline{\nu}'_{\tau}$ never 
comes on resonance. In the region of
parameter space where $L_{\nu_{\tau}}$ dominates, it is 
easy to show that this point occurs when
\begin{equation}
{p_{\mu\tau'} \over T} 
\simeq {1 \over 3}{p_{\mu\mu'} \over T} \approx 2.
\end{equation}
Thus by the time sufficient $L_{\nu_\tau}$ and $L_{\nu'_\mu}$ asymmetries
have been generated by $\nu_\tau \leftrightarrow \nu'_\mu$ oscillations
for the $\nu'_\tau \leftrightarrow \nu_\mu$ 
oscillations to have a resonance momentum, the
$\nu'_\mu \leftrightarrow \nu_\mu$ oscillation resonance momentum
is already into the tail of the distribution (for the parameter space where
negligible $L_{\nu_\mu}$ has been created by $\nu_\mu
\leftrightarrow \nu'_\mu$ oscillations).\footnote{Actually, in the 
psuedo-Dirac case\cite{pd} where
there are no mirror interactions
the $\nu_\mu + \nu'_\tau$ oscillation system is more important.
The reason is that, in the psuedo-Dirac case, these oscillations
destroy {\it exactly} the same combination of lepton numbers as does the $\nu_{\mu}
\leftrightarrow \nu'_{\mu}$ mode, that is,
in this case $L^{(\mu \tau')} = L^{(\mu \mu')}$.
Thus, in the pseudo-Dirac alternative to the mirror scenario,
the $\nu_\mu \leftrightarrow \nu'_\tau$ oscillations can help
the $\nu_\mu \leftrightarrow \nu'_\mu$ oscillations destroy
$L^{(\mu \mu')}$, which means that the `allowed region' can
be significantly reduced.}
For low temperatures, $T \stackrel{<}{\sim} T_c/3$, the
$b-$term in the matter potential can be 
approximately neglected and the resonance momentum
can be derived from
$a(p) \simeq \cos2\theta_{\mu\tau'}$, leading to
\begin{equation}
{p_{\mu \tau'} \over T} \simeq {\delta m^2_{\mu \tau'}\cos2\theta_{\mu
\tau'} 
\over a_0 T^4 L^{(\mu \tau')}}.
\end{equation}
Observe that the effect of the $\nu_\mu \leftrightarrow \nu'_\tau$
oscillations is to 
decrease $|L^{(\mu \tau')}|$ and hence
to {\it increase} $p_{\mu \tau'}/T$.
By the time $T \sim T_c/2$, the resonance momentum has reached $p_{\mu \tau'}/T \sim 15$.
We mention this here because it will be important in
the following section.

The observations about the evolution of resonance momenta 
made above are relevant to the
numerical integration of the quantum kinetic equations.
Because this integration is CPU time consuming, we employ the
useful time saving approximation 
of integrating the oscillation and collision driven 
aspects of the evolution  in the region around the MSW 
resonances. Since the precise details and justification of this
have been covered in Ref.\cite{f}, we will not repeat the 
discussion here. 

\section{Low temperature and BBN epochs: effect of oscillations on light element abundances}

\subsection{Introduction}

The primordial deuterium to hydrogen ($D/H$) 
ratio can be used to give a sensitive
determination of the baryon to photon ratio $\eta$ 
which, given the estimated primordial
$^4$He mass fraction, can be used to infer the 
effective number of light neutrino flavours
$N_{\nu, \text{eff}}$ during the BBN epoch. This value 
can then be compared with the
predictions for $N_{\nu,\text{eff}}$ from various models 
of particle physics to find out
which ones are compatible with standard BBN. 
For example, the minimal standard model
predicts $N_{\nu,\text{eff}} = 3$. At the present time, most 
estimates favour $N_{\nu,\text{eff}} < 3.6$ and some
estimates favour $N_{\nu, \text{eff}} < 3.0$\cite{bbn1}.  
Of course, even if a model of particle physics is shown to be
incompatible with BBN, this does not necessarily mean that 
the model is incorrect, since it is also possible that one 
of the standard assumptions of BBN may not be correct\cite{ka}.

For gauge models with mirror or sterile neutrinos, one in general expects
$N_{\nu,\text{eff}} \neq 3$.
In fact, $N_{\nu,\text{eff}}$ may be less than three or greater than
three.  The prediction for $N_{\nu,\text{eff}}$ depends on the 
oscillation parameters in a given model.
One possible consequence of ordinary-mirror (or ordinary-sterile)
neutrino oscillations is the
excitation of mirror neutrino states, which 
typically leads to an increase in the expansion rate of
the universe and thereby also increases $N_{\nu,\text{eff}}$.
Another possible consequence of ordinary-mirror
neutrino oscillations is the
dynamical generation of an electron-neutrino asymmetry. 
This also has important implications for BBN,
as it directly
affects the reaction rates which determine the neutrino to proton ($n/p$) 
ratio just before nucleosynthesis.
If the electron neutrino asymmetry is positive then it will
decrease $N_{\nu,\text{eff}}$, while if it
is negative then it will increase $N_{\nu,\text{eff}}$.
 
The neutron to nucleon ratio, $X_n (t)$, is related to 
the primordial Helium mass fraction,
$Y_P$, by\footnote{For a review of helium synthesis, see for example
Ref.\cite{weinberg}.}
\begin{equation}
Y_P = 2X_n
\end{equation}
just before nucleosynthesis.
The evolution of $X_n (t)$ is governed by the equation,
\begin{equation}
\frac{dX_n}{dt}=-\lambda(n\rightarrow p)X_n +
\lambda(p\rightarrow n)(1-X_n),
\end{equation}
where the reaction rates are approximately
\begin{eqnarray}
\label{eq:rates}
\lambda(n\rightarrow p)\simeq \lambda(n + \nu_e \rightarrow p + e^-) 
+ \lambda(n + e^+ \rightarrow p + \overline{\nu}_e), \nonumber \\
\lambda(p\rightarrow n) \simeq \lambda(p + e^- \rightarrow n + \nu_e) 
+ \lambda(p + \overline{\nu}_e \rightarrow n + e^+),
\end{eqnarray}
depend on the momentum distributions of the species involved.
The processes in Eq.(\ref{eq:rates}) for 
determining $n\leftrightarrow p$ are only important
for temperatures above about $0.4$ MeV. Below this temperature the 
weak interaction rates
freeze out and neutron decay becomes the dominant 
factor affecting the $n/p$ ratio.  An
excess of $\nu_e$ over $\overline{\nu}_e$, due
to the creation of a positive $L_{\nu_e}$ would 
change the rates for the processes in Eq.(\ref{eq:rates}). 
The effect of this would be to reduce the n/p ratio, and hence
reduce $Y_P$.
Neutron decay is not significantly altered by 
lepton asymmetries.   It is quite well known that
a small change in $Y_P$ due to the modification of 
$\nu_e$ and $\overline \nu_e$ distributions
does not impact significantly on the 
other light element abundances (see for example Ref.\cite{olive}).
A small modification to the expansion rate, 
using the convenient unit $N_{\nu,\text{eff}}$, 
primarily affects only $Y_P$, with\cite{walker}
\begin{equation}
\delta Y_P \simeq 0.012\times \delta N_{\nu,\text{eff}}.
\label{walk}
\end{equation}
In Appendix A we describe in detail how we compute the effect
on $Y_P$ due to the modified $\nu_e$ and $\overline \nu_e$ distributions.

In two previous papers \cite{fv2,bfv}, we studied
the implications for BBN of oscillations within two distinct 
four-neutrino-flavour models
which featured the three ordinary neutrinos and one sterile
neutrino. In Ref.\cite{fv2}, a model with
the mass hierarchy $m_{\nu_{\tau}}\gg
m_{\nu_{\mu}}, m_{\nu_e}, m_{\nu_s}$ was 
considered. In this case 
$\overline{\nu}_\tau \leftrightarrow \overline{\nu}_s$ oscillations
resulted in an excess of $\nu_{\tau}$ over
$\overline{\nu}_{\tau}$ (in the case
where $L_{\nu_\tau} > 0$) thereby generating a large tau-neutrino 
asymmetry.  It was shown that if
$\overline{\nu}_{\tau}-\overline{\nu}_e$ oscillations also occurred, 
then some of the tau-neutrino asymmetry was reprocessed into an 
electron-neutrino asymmetry. The effective number of
neutrino flavours found in Ref.\cite{fv2} was either $2.5$ or $3.4$,  
depending on the ambiguity for the sign of 
the asymmetry and hence the prediction
for $N_{\nu,\text{eff}}$ (Ref.\cite{fv1} discusses the sign ambiguity issue).  For a
positive asymmetry, $\delta N_{\nu,\text{eff}} \simeq -0.5$ 
was obtained over a range of
mass differences $|\delta m_{\tau s}^2|\sim 10-1000\ \text{eV}^2$, 
while for a negative
asymmetry the result was $\delta N_{\nu,\text{eff}} \simeq +0.4$.  
Later, in a separate paper with Bell \cite{bfv}, 
we considered another four-neutrino model where
$\nu_\tau$ and $\nu_\mu$ were taken to be
approximately maximal combinations of
two nearly degenerate mass eigenstates, $\nu_1$ and $\nu_2$,
with $m_{\nu_1}, m_{\nu_2} \gg  m_{\nu_e}, m_{\nu_s}$.   
In that case,
we found $N_{\nu,\text{eff}} \simeq 2.7$ or $3.1$
depending on the sign of the asymmetry.

As the above paragraph illustrates, the prediction
for $N_{\nu,\text{eff}}$ is a model dependent quantity.
In the next section we will estimate $N_{\nu,\text{eff}}$ in the EPM
for various illustrative parameter ranges.

\subsection{Low temperature neutrino asymmetry evolution in the EPM: Case 1}

We now study the ``low temperature'' evolution of the
number distributions and lepton numbers in the
EPM. As discussed in Sec.II, by ``low temperature'' we mean the
regime succeeding the exponential growth epoch.
In this regime, the evolution of the neutrino ensemble is dominated by
coherent effects, because the $T^5$ decrease in the damping function $D$ renders
negligible the decohering effect of collisions. Repopulation, however, is still important.

Consider, for the moment, two-flavour small angle ordinary-mirror
oscillations $\nu_\alpha \leftrightarrow
\nu'_\beta$. In the case of undamped evolution, we know from numerical integration of the
exact quantum kinetic equations that the
adiabatic approximation is
valid provided that $\sin^2 2\theta_{\alpha\beta'} \stackrel{>}{\sim} 10^{-10}$.
Now, coherent 
adiabatic MSW transitions completely convert $\nu_\alpha \leftrightarrow    
\nu'_\beta$ at the resonance   
momentum of these states.  For 
adiabatic two-flavour neutrino oscillations in the early universe it 
is then quite easy to see that the rate of change of lepton number is 
governed by the simple equation\cite{fv2},
\begin{equation}
{dL_{\nu_\alpha} \over dT} = -{dL_{\nu'_\beta} \over dT} = 
-X\left| {d(p_{\text{res}}/T) \over dT}\right|,
\label{he}
\end{equation}
where
\begin{equation}
X = {T \over n_\gamma}\left( N_{\bar \nu_\alpha} - N_{\bar \nu'_\beta}
\right),
\end{equation}
and the case $L_{\nu_\alpha} > 0$ has been considered
(so that the resonance occurs for antineutrinos).
Equation (\ref{he}) relates the rate of change 
of lepton number to the speed of the
resonance momentum through the neutrino distribution. 
Reference \cite{bvw} provides a
detailed discussion of how this equation can be derived from 
Eq.(\ref{wed}) for the case of
adiabatic evolution with a narrow resonance 
width.\footnote{Note that when collisional
decoherence in neglected, the QKEs produce standard 
Schr\"odinger-like MSW evolution with
repopulation effects added via a Boltzmann 
approach.} Equation (\ref{he}) can be simplified using 
\begin{equation}
{d(p_{\text{res}}/T) \over dT} = {\partial (p_{\text{res}}/T)
\over \partial T} + {\partial (p_{\text{res}}/T) \over 
\partial L_{\nu_\alpha}} {dL_{\nu_\alpha} \over dT},
\label{presderiv}
\end{equation}
 from which it follows that
\begin{equation}
{dL_{\nu_\alpha} \over dT} = 
- {dL_{\nu'_\beta} \over dT} = 
{fX{\partial (p_{\text{res}}/T) \over \partial T}
\over 1 - fX{\partial (p_{\text{res}}/T) \over \partial L_{\nu_\alpha}}}  
\ = \ {-4fXp_{\text{res}}/T^2 \over
1 - {2 f X p_{\text{res}} \over T[L^{(\alpha)} - L'^{(\beta)}]}},
\label{presevol}
\end{equation}
where $f = 1$ for 
$d(p_{\text{res}}/T)/dt > 0$ 
(that is for $d(p_{\text{res}}/T)/dT < 0$) 
and $f = -1$ for
$d(p_{\text{res}}/T)/dt < 0$. For the multi-flavour case under 
analysis, coupled equations
based on Eq.(\ref{presevol}) will be used. 

Of course the evolution of the lepton number can
also be described using the QKEs. As mentioned above,
they give the same answer provided that the
evolution is adiabatic. In the case of non-adiabatic
evolution, the QKEs should be used instead of the simple Eq.(\ref{he}).
For our study of the low temperature evolution of the 
number distributions and lepton numbers in the EPM,
we will make use of the adiabatic approximation encoded in Eq.(\ref{presevol}).
In fact, it turns out that the evolution of the system
in the EPM model is quite complicated. For
example, three-flavour effects cannot be ignored,
so solving the problem using the Quantum Kinetic
Equations would be extremely complicated and
(CPU) time consuming.

We first consider the parameter region
\begin{equation}
m_{\nu_{\tau +}} \simeq m_{\nu_{\tau-}} \gg
m_{\nu_{\mu +}},  m_{\nu_{\mu -}},
m_{\nu_{e +}},  m_{\nu_{e -}},
\label{1}
\end{equation}
with
\begin{equation}
m_{\nu_{\mu +}},  m_{\nu_{\mu -}},
m_{\nu_{e +}},  m_{\nu_{e -}} \ll 1 \ \text{eV}.
\label{2}
\end{equation}
We will call this ``Case 1''.
(Later on we will consider another case, Case 2, where $m_{\nu_{\mu \pm}} \sim \text{eV}$
as suggested by the LSND results.)
In Case 1, the following oscillation modes all have 
approximately the same $|\delta m^2|$, which we denote as 
$\delta m^2_{\text{large}}$:
\begin{eqnarray}
\nu_\tau &\leftrightarrow & \nu'_e, \quad \nu_\tau \leftrightarrow 
\nu'_\mu, \quad \nu'_\tau \leftrightarrow \nu_e, \quad 
\nu'_\tau \leftrightarrow \nu_\mu, \nonumber \\
\nu_\tau &\leftrightarrow & \nu_e, \quad \nu_\tau \leftrightarrow 
\nu_\mu, \quad \nu'_\tau \leftrightarrow \nu'_e, \quad \nu'_\tau 
\leftrightarrow \nu'_\mu.
\label{3}
\end{eqnarray}
Note that $\delta m^2_{\text{large}} \simeq m^2_{\nu_{\tau\pm}}$. 
All the other oscillation modes have much
smaller $\delta m^2$ values.
In fact, for Case 1, we will consider the parameter 
space region where the $\delta
m^2$ values of all the other oscillation modes are small enough so that
they can be approximately neglected for temperatures
$T \stackrel{>}{\sim} 0.4 \ \text{MeV}$.
This last condition means that these modes will
not affect the neutron/proton ratio and hence 
cannot significantly affect BBN.

In the following discussion we consider the
case $L_{\nu_\tau} > 0$ for definiteness. This means that
the $\overline{\nu}_\tau \leftrightarrow \overline{\nu}'_e$ and
$\overline{\nu}_\tau \leftrightarrow \overline{\nu}'_\mu$ oscillations
generate $L_{\nu_\tau}$ while the
other oscillations reprocess some of this asymmetry
into other flavours. Of course a crucial issue for
BBN is to find out how much of this asymmetry is 
reprocessed to the electron neutrinos, and at
what temperature this occurs.

In order to use Eq.(\ref{presevol}),
we have to employ the resonance conditions to determine the resonance momenta as
functions of temperature and the neutrino asymmetries. We begin by noting
that the sum of the ordinary and mirror lepton
numbers are conserved by the oscillations, and
we will suppose that they sum to zero\footnote{
Of course our results do not depend significantly on this
assumption. For example, if we put the 
sum in Eq.(\ref{5}) equal to a number of the order of
the baryon asymmetry then the resulting analysis will
change very little.}
\begin{equation}
L_{\nu_e} + L_{\nu_\mu} + L_{\nu_\tau}
+ L_{\nu'_e} + L_{\nu'_\mu} + L_{\nu'_\tau} = 0.
\label{5}
\end{equation}
Furthermore, we take as initial conditions that 
\begin{equation}
L_{\nu_\mu} \simeq L_{\nu_e},\quad
L_{\nu'_\mu} \simeq L_{\nu'_e}.
\label{zzz}
\end{equation}
We will show that this assumption is robust shortly.
With the above initial conditions,
it follows that the eight oscillation modes of Eq.(\ref{3})
can be classified together into four groups of two, each group having
approximately the same resonance momentum:
\begin{eqnarray}
&\text{Group 1:}&\quad \overline 
\nu_\tau \leftrightarrow \overline \nu'_e,\quad 
\overline \nu_\tau \leftrightarrow \overline \nu'_\mu,
\quad p_{\text{res}} \equiv P_1,\nonumber \\
&\text{Group 2:}&\quad \nu'_\tau \leftrightarrow \nu_e, 
\quad \nu'_\tau \leftrightarrow 
\nu_\mu, \quad p_{\text{res}} \equiv P_2,\nonumber \\
&\text{Group 3:}&\quad \overline \nu_\tau \leftrightarrow 
\overline \nu_e, \quad 
\overline \nu_\tau \leftrightarrow \overline \nu_\mu,
\quad p_{\text{res}} \equiv P_3,\nonumber \\
&\text{Group 4:}&\quad \overline \nu'_\tau \leftrightarrow 
\overline \nu'_e,\quad  
\overline \nu'_\tau \leftrightarrow \overline \nu'_\mu,
\quad p_{\text{res}} \equiv P_4.
\label{4}
\end{eqnarray}
In the Low Temperature Epoch, the $b-$term in the effective
potential can be approximately neglected because it decreases as $T^6$. This
means that the resonance condition is approximately
$a(p) = \pm \cos 2\theta \simeq \pm 1$ for small angle oscillations.
For $\nu_\alpha \leftrightarrow \nu'_\beta$ small angle oscillations,
the resonance momentum is therefore to a good approximation given by
\begin{equation}
{p_{res} \over T} = {\delta m^2_{\alpha\beta'} \over 
a_0 T^4 L^{(\alpha\beta')}},
\label{7}
\end{equation}
where we have used the notation defined earlier in Eq.(\ref{70}).
For the four groups of oscillation modes in Eq.(\ref{4}),
\begin{equation}
{P_i \over T} = {\delta m^2_{\text{large}} \over a_0 T^4 L_i},
\label{8}
\end{equation}
where $i=1,\ldots,4$ and
\begin{eqnarray}
L_1 &\equiv& L^{(\tau e')} = {7 \over 2}L_{\nu_\tau} +
5L_{\nu_e} + {1 \over 2}L_{\nu'_\tau}, \nonumber \\
L_2 &\equiv& - L^{(e \tau')} = 2L_{\nu_\tau} + 5L_{\nu_e}
- L_{\nu'_{\tau}},
\nonumber \\
L_3 &\equiv& L^{(\tau)} - L^{(e)} = L_{\nu_\tau} - L_{\nu_e},
\nonumber \\
L_4 &\equiv& L'^{(\tau)} - L'^{(e)} = L_{\nu'_\tau} - L_{\nu'_e} = 
{1 \over 2} L_{\nu_{\tau}} + L_{\nu_e} + {3 \over 2} L_{\nu'_{\tau}}.
\label{9}
\end{eqnarray}
Note that Eqs.(\ref{5}) and (\ref{zzz}) have been used in the above
equation to express $L_{\nu_\mu}$, $L_{\nu'_\mu}$ and
$L_{\nu'_e}$ in terms of the $L_{\nu_e}$, $L_{\nu_\tau}$ and
$L_{\nu'_\tau}$.

In the following discussion we will focus on the parameter 
space region where all of the oscillations
are approximately adiabatic. This is extraordinarily 
helpful, because adiabatic transitions are independent of 
the vacuum mixing angles (so long as the mixing
angles are much smaller than one). This means that 
generic outcomes can be calculated for a reasonably large 
range of parameters, rather than having to consider small points 
in oscillation parameter space on a case by case basis. As 
noted earlier, two flavour subsystems in this epoch of 
the early universe evolve adiabatically
provided that the relevant $\sin^2 2\theta \stackrel{>}{\sim} 10^{-10}$.
This is not a very stringent requirement.
In particular, there will be a large
range of parameters where the evolution is both
adiabatic and satisfies the experimental and
cosmological constraints.
Given the parameter region of Eqs.(\ref{1}) and (\ref{2}), 
consistency with BBN constrains
$\sin^2 2\theta_{\tau \mu'}$ to be\footnote{Note 
that $\sin^2 2\theta_{\tau \mu'} = 
\sin^2 2\theta_{\tau'\mu}$ from the parity symmetry.},
\begin{equation}
\sin^2 2\theta_{\tau \mu'} \stackrel{<}{\sim} \ \text{few} \
\times 10^{-4} \left( {\text{eV}^2 \over \delta m^2_{\text{large}}}\right)^{1 \over 2},
\end{equation}
while Nomad/Chorus constrain $\sin^2 2\theta_{\tau \mu}$ 
to be\cite{nomch}
\begin{equation}
\sin^2 2\theta^{\tau \mu} \stackrel{<}{\sim} 10^{-3}
\ \text{for} \ \delta m^2_{\text{large}} \stackrel{>}{\sim} 40 \ \text{eV}^2.
\end{equation}

We now discuss the effects of 
each of the four groups of modes in Eq.(\ref{4}).

\begin{enumerate}
\item {\it The $\overline \nu_\tau \leftrightarrow \overline \nu'_e$ and $\overline
\nu_\tau \leftrightarrow \overline \nu'_\mu$ Group 1 modes:}
These modes have negative $\delta m^2$ values and thus
create the relevant lepton numbers $L_{\nu_\tau}$, $L_{\nu'_e}$ and $L_{\nu'_{\mu}}$. 
It is important to understand that if these two
modes have slightly different resonance momenta, say $P_1^a$ and $P_1^b$, then
they generate lepton numbers so that $P_1^a \to P_1^b$. This is tantamount to ensuring
that the initial conditions of Eq.(\ref{zzz}) hold, provided that the difference
between the initial values of $L_{\nu_e}$ and $L_{\nu_{\mu}}$ is 
not too great. To see that the
resonance momenta are dynamically driven to coincide, assume that $P_1^a > P_1^b$.  This
means that the $\overline \nu_\tau \leftrightarrow \overline \nu'_e$
resonance momentum preceeds the $\overline \nu_\tau \leftrightarrow 
\overline \nu'_\mu$ resonance momentum. Now,
the $\overline \nu_\tau \leftrightarrow \overline \nu'_e$ oscillations
convert all of the resonant $\overline \nu_\tau$'s into $\overline \nu'_e$'s.
The closely following $\overline \nu_\tau \to \overline \nu'_\mu$ 
resonance has a much weaker effect, since there are no $\overline 
\nu_\tau$'s left to convert into $\overline \nu'_\mu$ states. (For 
this to be true the resonance momenta must be close enough 
so that the converted $\overline \nu_\tau$ states 
do not get completely refilled by the weak interactions before 
the trailing resonance momentum $P_1^b$ passes their
momentum value.) Because of the disparity in raw material for 
processing, $L_{\nu'_e}$ is
created much more rapidly then $L_{\nu'_\mu}$. According to Eq.(\ref{8}),
this in turn means that $P_1^a$ increases more slowly
relative to $P_1^b$ and thus $P_1^b \to P_1^a$.
Obviously, if we had started with $P_1^b > P_1^a$, then
we also would have found that the
evolution of lepton numbers is such that $P_1^a \to P_1^b$.
Because the dynamics drives
the two resonances in this group to approximately coincide, 
the system cannot be described in terms
of two-flavour oscillations.
Instead, three-flavour effects amongst $\overline \nu_{\tau}$, $\overline
\nu'_e$ and $\overline \nu'_{\mu}$
effect the adiabatic conversion
\begin{equation}
|\overline \nu_\tau \rangle \leftrightarrow 
{1 \over \sqrt{2}}(|\overline \nu'_e\rangle + |\overline \nu'_\mu\rangle).
\end{equation}
This means that as $P_1$ sweeps through the $\overline \nu_\tau$
momentum distribution, 
\begin{eqnarray}
N_{\overline \nu_e'}(P_1) &\to& {1 \over 2}\left[
{N_{\overline \nu_e'}(P_1) \over 2}
+ {N_{\overline \nu_\mu'}(P_1) \over 2}
+ N_{\overline \nu_\tau}(P_1)
\right], \nonumber \\
N_{\overline \nu_\mu'}(P_1) &\to& {1 \over 2}\left[
{N_{\overline \nu_e'}(P_1) \over 2}
+ {N_{\overline \nu_\mu'}(P_1) \over 2}
+ N_{\overline \nu_\tau}(P_1)
\right], \nonumber \\
N_{\overline \nu_\tau}(P_1) &\to& {1 \over 2}\left[
N_{\overline \nu'_e}(P_1) + N_{\overline \nu'_\mu}(P_1) \right].
\label{p1}
\end{eqnarray}
In our numerical work the continuous momentum distribution for each flavour
is replaced by a finite number of `cells' on a logarithmically
spaced mesh. As the momentum $P_1$ passes
a cell, the number density in the cell
is modified according to Eq.(\ref{p1}).\footnote{Note that it is legitimate to consider
probabilities, as encoded in the number density distributions, rather than probability
amplitudes in effecting the conversion. This is because fully adiabatic transitions are
sufficiently `classical'.} Of course weak interactions will repopulate
these cells as they thermalise the
neutrino momentum distributions.
We will discuss this later.

\item {\it The $\nu'_\tau \leftrightarrow \nu_e$ and 
$\nu'_\tau \leftrightarrow \nu_\mu$ Group 2
modes.} These modes have positive $\delta m^2$ values. 
At quite high temperatures, where the Group 1 oscillation modes
are exponentially creating $L_{\nu_\tau}$, the Group 2 oscillation modes
generate $L_{\nu'_\tau}$, $L_{\nu_e}$ and
$L_{\nu_\mu}$ such that $L^{'(\tau)} - L^{(e)} \to 0$ and
$L^{'(\tau)} - L^{(\mu)} \to 0$.
As already discussed in section IV,
this makes $P_2/T \gg 1$ at the onset of the low temperature epoch. 
Our numerical work shows
that the initial value of $P_2/T$ is typically about $15$ and
decreasing by the time $T \sim T_c/2$.
The subsequent evolution of $P_2/T$ is a little complicated, but can be roughly understood 
from Eqs.(\ref{presderiv}) and (\ref{8}), which combine
to produce
\begin{equation}
{d(P_2/T) \over dT} = - {P_2 \over T} \left[ {4 \over T} + 
{1 \over L_2}{d L_2\over dT}  \right],
\end{equation}
or equivalently,
\begin{equation}
{d(P_2/T) \over dt} \simeq {P_2 \over T}\left[
{5.5T^2 \over M_P} - {1 \over L_2}{dL_2 \over dt}\right].
\label{P2deriv}
\end{equation}
By the time $T \sim T_c/2$, the group 2 modes have driven $L_2$ to be quite small.
The second term in the right-hand side of
the above equation therefore dominates, making $P_2/T$
a decreasing function of time.
So, at the start of the low temperature epoch,
$P_2/T$ slowly decreases, converting
$\nu_e$'s and $\nu_\mu$'s to $\nu'_\tau$'s as it does so.
As for the Group 1 modes above, it is easy to see that if the
two Group 2 modes had slightly different resonance 
momenta, $P_2^a$ and $P_2^b$, then
the dynamics forces $P_2^a \to P_2^b$.
The effect of the three-flavour $\nu_e-\nu_\mu-\nu'_\tau$ 
subsystem is to convert $|\nu'_\tau \rangle$ to 
${1 \over \sqrt{2}}(|\nu_e\rangle + |\nu_\mu\rangle)$.
So, as $P_2$ moves (backward) through the neutrino momentum distribution, 
\begin{eqnarray}
N_{\nu_e}(P_2) & \to & {1 \over 2}\left[
{N_{\nu_e}(P_2) \over 2}
+ {N_{\nu_\mu}(P_2) \over 2}
+ N_{\nu'_\tau}(P_2)\right],\nonumber \\
N_{\nu_\mu}(P_2) & \to & {1 \over 2}\left[
{N_{\nu_e}(P_2) \over 2} +
{N_{\nu_\mu}(P_2) \over 2} +
N_{\nu'_\tau}(P_2)\right],\nonumber \\
N_{\nu'_\tau}(P_2) & \to & {1 \over 2}\left[
N_{\nu_e}(P_2) + N_{\nu_\mu}(P_2)\right].
\label{p2}
\end{eqnarray}
The effect of this conversion is to generate significant $L_{\nu_e}$ and   
$L_{\nu_\mu}$ asymmetries which are negative in sign
(given that we have taken $L_{\nu_\tau} > 0$ for definiteness).
As the evolution unfolds, at some temperature
$P_2/T$ changes direction and begins to increase again. 
This is due to the gradual increase in $L_2$ which eventually makes
the second term on the right-hand side of Eq.(\ref{P2deriv}) smaller
in magnitude than the first term. The full analysis, incorporating
all of the modes simultaneously, requires a numerical treatment. 
We compute the minimum $P_2/T$ to be $\sim 6$, which is still 
in the tail of the distribution. The upshot of this somewhat 
complicated evolution is that
the $P_2$ resonance momentum does not sweep through
the entire momentum distribution, but rather,
it sweeps through a significant part of the high momentum tail.
In the temperature regime where
$P_2/T$ makes the return journey from its minimum value
back to high values, the adiabatic MSW transitions
have little effect because they simply swap the almost equal
number densities of $\nu'_\tau$ and $\nu_{e,\mu}$ that were created by adiabatic
transitions before
the turnaround.\footnote{Actually, in this region, the two resonance momenta
$P_2^{a,b}$ are no longer dynamically driven to 
coincide, making the oscillations
somewhat more complicated. However, if the $\nu'_\tau$
tail is fully populated from the previous evolution
of the system then this complication matters very little.}
This is of course only true provided that the momentum
distribution of $\nu'_\tau$ states does not get
significantly modified by the mirror weak interactions, an issue
we will discuss in more detail later.

\item {\it The $\overline \nu_\tau \leftrightarrow \overline \nu_e$ 
and $\overline \nu_\tau \leftrightarrow \overline \nu_\mu$ 
Group 3 modes.}
These modes, being ordinary-ordinary,
are slightly different in character to the ordinary-mirror
modes.  Their effect is to reprocess 
some of the $L_{\nu_\tau}$ into $L_{\nu_e}$ and $L_{\nu_\mu}$.
In the early stages of lepton number creation, all of the 
ordinary-ordinary modes are unimportant, 
because they simply swap flavours with almost identical number
density distributions. However, eventually the 
$L_{\nu_\tau}$ asymmetry created by the Group 1 modes is 
large enough to distort the $\overline \nu_\tau$ 
momentum distribution so that the attendant reduction in
the $\overline \nu_\tau$ number density relative to that 
for $\overline \nu_{e,\mu}$ allows
$\overline \nu_\tau \leftrightarrow \overline \nu_{e,\mu}$ 
oscillations to induce nontrivial dynamics: the depletion 
of $\overline \nu_e$ and $\overline \nu_\mu$
states at the resonance momentum $P_3$. 
This effect becomes significant when $L_{\nu_\tau}$
becomes quite large, which occurs roughly when 
$P_1/T \stackrel{>}{\sim} 2$. Now, from Eq.(\ref{9}) it 
is evident that $P_3 \sim {7 \over 2}P_1$ using the fact that 
$L_{\nu_{\tau}}$ is the largest lepton number in the system. 
This has the important consequence that
the overall effect of the Group 3 oscillations is not very large
because by the time $L_{\nu_\tau}$ is large, $P_3$ is
already well into the tail of the momentum distribution. 
This is fortunate, because these oscillations are more 
complicated to describe.  Unlike the Group 1 and 2 modes,
it is easy to see that if the resonance momenta, $P_3^a$ 
and $P_3^b$, of the two modes are slightly different, 
then they do {\it not} subsequently evolve to coincide. 
(The reason is that if, say, the $\overline \nu_\tau \leftrightarrow 
\overline \nu_e$ resonance momentum preceeds the $\overline \nu_\tau 
\leftrightarrow \overline \nu_\mu$ resonance momentum 
then the $\overline \nu_\tau \leftrightarrow 
\overline \nu_e$ oscillations act to reduce 
$L^{(\tau e)}$ thereby {\it increasing} the rate at which this
resonance momentum moves relative to the $\overline \nu_\tau
\leftrightarrow \overline \nu_\mu$ resonance momentum.)
Note, however, that because of the more influential Group
2 oscillations, it follows that $P_3^a$ and $P_3^b$ are at 
least {\it approximately} equal.  What precisely happens will 
then depend on the width of these resonances and whether they overlap
or not. This means that the effects will be dependent
on the values of the relevant oscillation parameters.
In our numerical work we will assume that
the resonances overlap, so that
$|\overline \nu_\tau\rangle \leftrightarrow {1 \over \sqrt{2}}(|\overline \nu_e \rangle 
+ |\overline \nu_\mu \rangle)$.  In this case,
\begin{eqnarray}
N_{\overline \nu_e}(P_3) &\to& {1 \over 2}\left[
{N_{\overline \nu_e}(P_3)\over 2} 
+ {N_{\overline \nu_\mu}(P_3)\over 2} 
+  N_{\overline \nu_\tau}(P_3)\right],
\nonumber \\
N_{\overline \nu_\mu}(P_3) &\to& {1 \over 2}\left[
{N_{\overline \nu_e}(P_3) \over 2}
+ {N_{\overline \nu_\mu}(P_3) \over 2}
+  N_{\overline \nu_\tau}(P_3)\right],
\nonumber \\
N_{\overline \nu_\tau}(P_3) &\to& {1 \over 2}\left[
N_{\overline \nu_e}(P_3) +  N_{\overline \nu_\mu}(P_3)\right].
\label{p3}
\end{eqnarray}
We stress that, were the above assumption prove to
be invalid, our numerical results would not be
greatly affected because the Group 3 modes have a relatively 
weak effect for the reasons discussed above.
Finally we note that the $L_{\nu_e}$ and $L_{\nu_\mu}$ 
asymmetries created by the Group 3 modes have
the opposite sign to the $L_{\nu_e}$ and $L_{\nu_\mu}$
asymmetries generated by the Group 2 modes.

\item {\it The $\overline \nu'_\tau \leftrightarrow \overline \nu'_e$ 
and $\overline\nu'_\tau \leftrightarrow \overline\nu'_\mu$ 
Group 4 modes.}  
These mirror-mirror modes can
be neglected because $P_4$
is always greater than $P_1$, and
thus the $\overline \nu'_{e,\mu}$ 
states are approximately empty (as is
$\overline \nu'_\tau$) when the $P_4$ resonance momentum
moves through.
\end{enumerate} 

Having understood to some extent the effect of each group 
of oscillations, it is now time to solve the complete system 
of coupled equations for the various lepton numbers. These
are obtained by a straightforward generalisation of the 
two-flavour case given in Eq.(\ref{presevol}). They are:
\begin{eqnarray}
{dL_{\nu_\tau}  \over dT} &=&
-X_1 \left|{d(P_1/T)\over dT}\right| - 
X_3\left|{d(P_3/T) \over dT}\right|, \nonumber \\
{dL_{\nu_e} \over dT} &=& 
{1 \over 2}X_3\left|{d(P_3/T) \over dT}\right| + 
{1 \over 2}X_2\left|{d(P_2/T)\over dT}\right|, \nonumber \\
{dL_{\nu_\mu} \over dT} &=& {dL_{\nu_e} \over dT},\quad 
{dL_{\nu'_\tau} \over dT} = -X_2\left|{d(P_2/T) \over dT}\right|,
\nonumber \\
{dL_{\nu'_e} \over dT} &=& {dL_{\nu'_\mu} \over dT} = -{1 \over 2}
\left({dL_{\nu_\tau} \over dT} + {dL_{\nu'_\tau} \over dT} + 
{dL_{\nu_e} \over dT} + {dL_{\nu_\mu}\over dT}\right), 
\label{1a}
\end{eqnarray}
where
\begin{eqnarray}
X_1 \equiv {T \over n_\gamma}\left(
N_{\overline \nu_\tau}(P_1) -
{1 \over 2}[N_{\overline \nu'_e}(P_1) 
+ N_{\overline \nu'_\mu}(P_1)]\right), \nonumber \\
X_2 \equiv {T \over n_\gamma}\left(
{1 \over 2}[N_{\nu_e}(P_2) + N_{\nu_\mu}(P_2)] -
N_{\nu'_\tau}(P_2) \right), \nonumber \\ 
X_3 \equiv {T \over n_\gamma}\left(
N_{\overline \nu_\tau} (P_3) -
{1 \over 2}[N_{\overline \nu_e}(P_3) + N_{\overline \nu_\mu}
(P_3)] \right). \nonumber \\
\label{2a}
\end{eqnarray}
Expanding out Eq.(\ref{1a}) we find
\begin{eqnarray}
y_1 {dL_{\nu_\tau} \over dT} &=& \alpha + \beta {dL_{\nu_e} \over dT} + 
\gamma {dL_{\nu'_\tau} \over dT},
\nonumber \\
y_2 {dL_{\nu_e} \over dT} &=& \delta + \rho {dL_{\nu_\tau} 
\over dT} + \zeta {dL_{\nu'_\tau} \over dT},
\nonumber \\
y_3 {dL_{\nu'_\tau} \over dT} &=& \eta + \theta {dL_{\nu_e} \over dT} + 
\phi {dL_{\nu_\tau} \over dT},
\label{3a}
\end{eqnarray}
where
\begin{eqnarray}
y_1 &\equiv& 1 - f_1X_1{\partial(P_1/T) \over \partial L_{\nu_\tau}}
- f_3X_3 {\partial(P_3/T) \over \partial L_{\nu_\tau}}
= 1 + {7f_1X_1 P_1 \over 2TL_1} + 
{f_3X_3P_3 \over TL_3}, \nonumber \\
y_2 &\equiv& 1 + {1 \over 2}f_3X_3{\partial(P_3/T) 
\over \partial L_{\nu_e}}
+ {1 \over 2}f_2X_2 {\partial(P_2/T) \over \partial L_{\nu_e}}
= 1 + {f_3X_3 P_3 \over 2TL_3}
- {5f_2X_2 P_2 \over 2TL_2}, \nonumber \\
y_3 &\equiv& 1 - f_2X_2{\partial (P_2/T) \over \partial L_{\nu'_\tau}}
= 1 - {f_2X_2 P_2 \over TL_2},
\nonumber \\
\alpha &\equiv& f_1X_1{\partial (P_1/T) \over \partial T} 
+ f_3X_3{\partial (P_3/T) \over \partial T} = -4f_1X_1P_1/T^2
- 4f_3X_3P_3/T^2,
\nonumber \\ 
\beta &\equiv& f_1X_1 {\partial (P_1/T) \over \partial L_{\nu_e}}
+ f_3X_3 {\partial (P_3/T) \over \partial L_{\nu_e}} = 
{-5f_1X_1 P_1 \over TL_1} + {f_3X_3 P_3 \over TL_3},
\nonumber \\
\gamma &\equiv& f_1X_1{\partial (P_1/T) 
\over \partial L_{\nu'_\tau}} = 
{-f_1X_1 P_1 \over 2TL_1},
\nonumber \\
\delta &\equiv& -{1 \over 2}f_3X_3 {\partial (P_3/T) \over \partial T} 
- {1 \over 2}f_2X_2{\partial (P_2/T) \over \partial T} 
= 2f_3X_3P_3/T^2+2f_2X_2P_2/T^2, 
\nonumber \\
\rho &\equiv& -{1 \over 2}f_3X_3{\partial (P_3/T) \over \partial 
L_{\nu_\tau}} - {1 \over 2}f_2X_2 {\partial (P_2/T) 
\over \partial L_{\nu_\tau}} = 
{f_3X_3P_3 \over 2TL_3} +
{f_2X_2P_2 \over TL_2}, 
\nonumber 
\\
\zeta &\equiv& -{1 \over 2}f_2X_2{\partial (P_2/T) 
\over \partial L_{\nu'_\tau}} = -{f_2X_2P_2 \over 2TL_2},
\nonumber \\
\eta &\equiv& f_2X_2{\partial (P_2/T) \over \partial T}
= -4f_2X_2 P_2/T^2,
\nonumber \\
\theta &\equiv& f_2X_2 {\partial (P_2/T) \over \partial L_{\nu_e}}= 
-{5f_2X_2P_2 \over TL_2},
\nonumber \\
\phi &\equiv& f_2X_2 {\partial (P_2/T) \over \partial L_{\nu_\tau}} 
= -{2f_2X_2P_2 \over TL_2},
\label{4a}
\end{eqnarray}
and $f_i = 1$ for $d(P_i/T)/dt > 0$ and $f_i = -1$
for $d(P_i/T)/dt < 0$ ($i=1,2,3$).
Solving Eq.(\ref{3a}) we find,
\begin{eqnarray}
{dL_{\nu_e} \over dT} &=& {(\delta y_3 + \zeta \eta)(y_1 y_3 - \gamma \phi)
+ (\rho y_3 + \phi\zeta)(\alpha y_3 + \gamma \eta) \over
(y_2 y_3 - \zeta \theta)(y_1 y_3 - \gamma \phi) - 
(\rho y_3 + \phi\zeta)(\beta y_3 + \gamma \theta)},
\nonumber \\
{dL_{\nu_\tau} \over dT} &=& {\alpha y_3 + \gamma \eta +
(\beta y_3 + \gamma \theta){dL_{\nu_e} \over dT}\over
y_1 y_3 - \gamma \phi},
\nonumber \\
{dL_{\nu'_\tau} \over dT} &=& {1 \over y_3}\left[
\eta + \theta {dL_{\nu_e} \over dT} + \phi {dL_{\nu_\tau}
\over dT}\right].
\label{5a}
\end{eqnarray}
We compute the number densities 
incorporating Eqs.(\ref{p1}-\ref{p3}).
The repopulation and thermalisation of
the neutrino momentum distributions is
taken into account using the same expression 
as in the high temperature epoch:
\begin{equation}
{\partial \over \partial t} {N_{\nu_\alpha}(p) \over N^{\text{eq}}(p,T,0)}
\simeq \Gamma_{\alpha}(p)
\left[{N^{\text{eq}}(p,T,\mu_{\nu_{\alpha}}) \over N^{\text{eq}}(p,T,0)}   
- {N_{\nu_\alpha}(p) \over N^{\text{eq}}(p,T,0)} \right],
\label{dfd}
\end{equation}
where $\Gamma_{\alpha}(p)$ is the total collision
rate. The chemical potentials are computed from 
the lepton numbers as per Sec.III.

Observe that the oscillations will necessarily generate a 
significant number of $\overline
\nu'_e, \overline \nu'_\mu$ and $\nu'_\tau$ mirror neutrino 
species. We will make the
simplifying assumption that there is negligible 
thermalisation of these mirror neutrinos. By
this we mean that the mirror weak interactions of 
these mirror states are weak enough to not
appreciably modify the mirror neutrino momentum 
distributions. We will discuss later the
circumstances required for this is to be a valid 
approximation, and the expected effects when it is not valid.

In solving Eq.(\ref{5a}) initial conditions for 
$L_{\nu_\alpha}$ and $L_{\nu'_{\alpha}}$ for each $\alpha$ must be
specified at a temperature, $T_{\text{low}}$, that serves as 
the initial point for the low temperature epoch. 
For definiteness we take $T_{\text{low}} = T_c/2$, where $T_c$ 
is the critical point during the high temperature epoch at which 
the explosive growth of $L_{\nu_{\tau}}$
begins. Our results are quite insensitive to the precise
value taken for $T_{\text{low}}$ so long as it is
high enough for the lepton numbers to be still much less than 1. (This 
issue is discussed more fully in Ref.\cite{fv2}.)
The lepton numbers are related to the resonance momenta 
by using Eqs.(\ref{8}) and (\ref{9}),
\begin{eqnarray}
L_{\nu_e} &=& {\omega \over 24}\left({2 \over P_1} +  {1\over P_2}
- {9 \over P_3}\right), \nonumber \\
L_{\nu_\tau} &=& {\omega \over 24}\left({2 \over P_1} +  {1\over P_2}
+ {15 \over P_3}\right), \nonumber \\
L_{\nu'_\tau} &=& {\omega \over 24}\left({14 \over P_1} -  {17\over P_2}
- {15 \over P_3}\right), 
\end{eqnarray}
where $\omega \equiv {\delta m^2_{\text{large}}\over a_0 T^3}$. 
(The Group 4 oscillation modes have been neglected for
reasons discussed earlier.)
Thus, specifying the values $P_i/T$ at $T=T_{\text{low}}$ completely 
fixes the values of the lepton numbers at that temperature. From our numerical 
work, we find that the $T \sim T_{\text{low}}$ values of the 
resonance momenta $P_1$ and $P_2$ are given approximately by
\begin{equation}
P_1/T \sim 0.3,\qquad  P_2/T \sim 15.
\end{equation}
These values are approximately independent of the 
vacuum oscillation parameters
so long as the various mixing angles obey 
$\sin^2 2\theta \stackrel{>}{\sim} 10^{-10}$
and provided the $\delta m^2$'s lie in the range of interest.
Also note that our subsequent numerical work is not
very sensitive to the precise initial values of $P_1/T$ and $P_2/T$
provided that $P_1/T$ is small (less than about
$0.6$) and $P_2/T$ is large (greater than about $10$).
We also need to specify the initial values of 
the signs $f_i$. We take $f_1 = f_3 = 1$ and 
$f_2 = -1$ at $T=T_{\text{low}}$.
Subsequently $f_i$ are evaluated from the previous time step.

In the region during and just
after the exponential growth, the 
initial production of $L_{\nu_e}$ and $L_{\nu_\mu}$ due
to the oscillations $\overline \nu_\tau \leftrightarrow  
\overline \nu_e$ and $\overline \nu_\tau \leftrightarrow \overline \nu_\mu$ 
is suppressed because the number densities of all the ordinary 
neutrino flavours are almost equal. At $T = T_{\text{low}}$ 
we find that the creation of lepton number
due to these oscillations is approximately negligible.
This means that the main contribution to $L_{\nu_e}$,
$L_{\nu_\mu}$ and $L_{\nu'_\tau}$ at
$T \sim T_{\text{low}}$
is from $\nu'_\tau \leftrightarrow \nu_{e,\mu}$ oscillations,
and thus $L_{\nu'_\tau} \simeq -2L_{\nu_e}$.  It follows 
that the initial value for $P_3/T$ can be approximately 
related to the initial values of $P_1/T$ and $P_2/T$ by
\begin{equation}
{1 \over P_3} \simeq {18 \over 33P_1} - {15 \over 33P_2}.
\end{equation}

We have solved this system of equations for the 
illustrative example of $\delta m^2_{\text{large}} = 50 \ 
\text{eV}^2$. In Fig.3 we show the evolution of the four
resonance momenta $P_i/T$. The evolution of $L_{\nu_\alpha}$ and
$L_{\nu'_\alpha}$ for the same $\delta m^2_{\text{large}}$ 
parameter choice is plotted in Fig.4 for the 
$L_{\nu_\tau} > 0$ case.

Let us now turn to the implications of the oscillations for
BBN.  The change in $Y_P$ due to the neutrino oscillations
can be separated into two contributions,
\begin{equation}
\delta Y_P = \delta_1 Y_P + \delta_2 Y_P,
\end{equation}
where $\delta_1 Y_P$ is the change due
to the effect of the modified electron neutrino 
momentum distributions on the reaction rates, and
$\delta_2 Y_P$ is
due to the change in the energy density (or equivalently
the change in the expansion rate of the universe).
The former effect can be determined by  
numerically integrating the rate equations 
for the processes given in Eq.(\ref{eq:rates}) 
using the modified electron neutrino momentum 
distributions $N_{\nu_e}$ and $N_{\overline \nu_e}$
as discussed in Appendix A. 
The latter contribution can be computed
from the momentum distributions of the ordinary and
mirror neutrinos through
\begin{equation}
\delta_2 Y_P \simeq 0.012 \left(
{1 \over 2\rho_0} \sum_{\alpha = 1}^3 
\int_0^{\infty} [N_{\nu_\alpha}(p) +
N_{\overline \nu_\alpha}(p) + N_{\nu'_\alpha}(p) +
N_{\overline \nu'_\alpha}(p)]pdp - 3\right),
\end{equation}
where 
\begin{equation}
\rho_0 \equiv \int^{\infty}_0 N^{\text{eq}}(p,T,0) pdp 
= {7 \pi^2 \over 240} T^4,
\label{Weylrho}
\end{equation}
is the energy density of a Weyl fermion at equilibrium with 
zero chemical potential.  [Recall that Eq.(\ref{walk}) 
can be used to express $\delta Y_P$, $\delta_1 Y_P$ and 
$\delta_2 Y_P$ in terms of effective neutrino number,
$\delta N_{\nu,\text{eff}}$, $\delta_1 N_{\nu,\text{eff}}$ and $\delta_2    
N_{\nu,\text{eff}}$, respectively.]
To calculate $\delta_2 Y_P$, we numerically determine the 
momentum distributions at $T = 0.5\ \text{MeV}$. 
Because of the approximate kinetic decoupling of neutrinos
for temperatures below about $3-4$ MeV,
large contributions\footnote{By `large
contributions' we mean $\delta_2 N_{\nu,\text{eff}}
\stackrel{>}{\sim} 0.10$.} to $\delta_2 Y_P$, should they exist,
must have been generated earlier. A temperature of 
$0.5\ \text{MeV}$ is therefore a safe
place to evaluate the final $\delta_2 Y_P$.

Recall that there is an ambiguity concerning the sign 
of the $L_{\nu_\tau}$ lepton
asymmetry. We have considered the $L_{\nu_{\tau}} > 0$ case above
for definiteness, 
but $L_{\nu_\tau} < 0$ is equally likely a priori. (See Refs.\cite{fv1} 
for further discussion of this.) For the negative
$L_{\nu_{\tau}}$ case, the roles of particles and anti-particles 
are reversed for the modes quoted in Eq.(\ref{4}) and 
subsequent equations.  One consequence of this is
that the signs of all the other asymmetries are also reversed. 
The quantity $\delta_1 Y_P$ will obviously be significantly affected by 
this ambiguity in sign, while $\delta_2 Y_P$ will not be 
affected at all. This means that we have two possible values 
for the overall change in the effective number of neutrino 
flavours during BBN. The results of the numerical
work is presented in Figs.5 and 6. Figure 5 treats the 
$L_{\nu_\tau} < 0$ case (which it turns out means that 
$L_{\nu_e} > 0$), while Fig.6 displays the $L_{\nu_\tau} > 0$ case
(which implies that $L_{\nu_e} < 0$).

Observe that $\delta_1 Y_P$ is not very large. The main 
contribution to it is from the modification of the high 
momentum tail of the $\nu_e$ distribution due to the 
Group 2 $\nu'_\tau
\leftrightarrow \nu_e$ oscillations. This is partially offset 
by the modification of the $\overline \nu_e$ distribution 
due to $\overline \nu_\tau \leftrightarrow \overline \nu_e$
oscillations.
It is also evident that $\delta_2 Y_P$ is close to zero for 
$\delta m^2_{\text{large}} \stackrel{<}{\sim} 300\ \text{eV}^2$.  
This is simply because the generation of mirror states,
which is dominated by the $\nu_\tau \leftrightarrow \nu'_e$ and 
$\nu_\tau \leftrightarrow \nu'_\mu$ modes, occurs below the 
kinetic decoupling temperature for $\nu_\tau$'s. This
means that the $\nu_\tau$ states which have converted into 
mirror states are not repopulated.  For larger values of 
$\delta m^2_{\text{large}}$, the $\nu_\tau$ states begin to get repopulated,
and the energy density increases accordingly.
We should also emphasise that our calculations contain
approximations. The most important are that the
repopulation is handled approximately via Eq.(\ref{dfd})
and we have neglected mirror thermalisation.
Thus, our results have a theoretical uncertainty, which
we estimate to be of order $\delta N_{\nu,\text{eff}} \sim 0.3$
(see following discussion).

We now discuss in more detail the effects of mirror
neutrino thermalisation. Recall that in the 
foregoing computations, we have included mirror neutrino
thermalisation via the quantity $\gamma_{\rho} \equiv (T'/T)^4$ 
during the High Temperature Epoch, but neglected it during 
the Low Temperature Epoch. (Remember that mirror
weak interaction rates increase with temperature.) We now 
discuss when this approximation is valid and the 
expected effects when it is not.

Given our division of the evolution of the system into 
High and Low Temperature Epochs, it is convenient to also classify 
mirror weak interactions into two categories. The first
category consists of the interactions of the mirror 
neutrinos generated during the Low Temperature Epoch with 
the background mirror neutrinos, electrons and positrons left over
from the preceding High Temperature Epoch (henceforth 
designated MHTB for ``mirror high-T background''). The second 
category is the elastic collisions of the $\overline \nu'_e$,
$\overline \nu'_\mu$ and $\nu'_\tau$ neutrinos generated 
during the Low Temperature Epoch with themselves. We now 
estimate each of these thermalisation rates.

The interaction rate of a mirror
neutrino $\nu'_{\alpha}$ of momentum $p$ with 
the MHTB is approximately given by,
\begin{equation}
\Gamma_m (p) = \gamma_\rho \Gamma (p) \simeq \gamma_\rho 
y_{\alpha}G_F^2T^5 \left({p \over 3.15 T}\right). 
\end{equation}
Now, from the discussion above [see Eq.(\ref{4})], the 
additional mirror neutrino states created during the Low
Temperature Epoch consist of the flavours 
$\overline \nu'_e$, $\overline \nu'_\mu$ and $\nu'_\tau$,
Their interactions with the MHTB can be approximately
neglected if
\begin{equation}
{\Gamma_m (p) \over H} \stackrel{<}{\sim} 1 \quad  \Rightarrow \quad 
\gamma_\rho y_{\alpha}G_F^2 {M_P
\over 5.5}\left({p \over 3.15T}\right)T^3 \stackrel{<}{\sim} 1.
\label{MHTBbound}
\end{equation}
As summarised in Fig.3, the resonance momentum $P_2/T$ 
for the modes producing $\nu'_\tau$'s is always higher than 
the resonance momentum $P_1/T$ for the modes producing $\overline
\nu'_e$ and $\overline \nu'_\mu$.
The mirror thermalisation effects will therefore be
most important for the $\nu'_\tau$ states. It is also 
clear that the relatively high momentum $\nu'_\tau$ 
states are produced at a higher temperature than $\overline 
\nu'_e$ and $\overline \nu'_\mu$ states of a corresponding 
momentum. The temperature range of interest for the 
$\nu'_\tau$ lies between $T_{\text{low}} \simeq T_c/2$ and 
the temperature $T_{\text{min}}$ at which $P_2/T$ reaches its
minimum value of $\simeq 6$. 

We can easily numerically compute $T_{\text{min}}$ to obtain,
\begin{equation}
{T_{\text{min}} \over \text{MeV}} \simeq 0.70 
\left({\delta m^2_{\text{large}} \over
\text{eV}^2}\right)^{1 \over 4}.
\label{digedo}
\end{equation}
We now estimate the effects of thermalisation 
by considering the interaction rate
for a $\nu'_\tau$ with a typical momentum
of $p/T \sim 8$ at a temperature around or slightly 
higher than $T_{\text{min}}$. Although
the constraint Eq.(\ref{MHTBbound}) on $\gamma_\rho$ is 
stronger for higher values of $T$,
the number of $\nu'_\tau$'s produced is lower, so 
the effects of their thermalisation will
be correspondingly weaker. The choices made for $p/T$ and $T$ 
as input for Eq.(\ref{MHTBbound}) represent a 
reasonable ``compromise'' driven by these considerations. 
So, we estimate that the interactions of
$\nu'_\tau$ with the MHTB can be neglected
provided that
\begin{equation}
\gamma_\rho \stackrel{<}{\sim} {1 \over 3 T_{\text{min}}^3} 
\sim  \left({\text{eV}^2 \over \delta m^2_{\text{large}}}\right)^{3 \over 4}.
\end{equation}

We now estimate 
the thermalisation rate, due to elastic collisions with themselves
and with the $\overline \nu'_{e,\mu}$ states produced during 
the Low Temperature Epoch, of the mirror $\nu'_\tau$
states produced during the Low Temperature Epoch. We will 
collectively call the mirror neutrino/antineutrino states 
produced during the Low Temperature Epoch as the ``mirror low-T
background'' (MLTB). Let us denote the collision rate 
for a $\nu'_\tau$ of momentum $p$ 
with the $\overline \nu'_{e,\mu}$ ($\nu'_\tau$) 
component of the MLTB by $\Gamma_1 (p)$ [$\Gamma_2 (p)$].
The relevant collision rates can be
obtained from Ref.\cite{ekt} to yield
\begin{eqnarray}
\Gamma_1 (p) & \simeq & 0.13 G_F^2 T^5 \left(
{\rho_{\overline \nu'_e} + \rho_{\overline \nu'_\mu} \over \rho_0}
\right)\left({p \over 3.15T}\right),
\nonumber \\
\Gamma_2 (p) & \simeq & 0.77 G_F^2T^5\left(
{\rho_{\nu'_\tau} \over \rho_0}
\right)\left({p \over 3.15T}\right),
\end{eqnarray}
where $\rho_0$ is defined in Eq.(\ref{Weylrho}).
So, the interactions of the $\nu'_\tau$ with the
$\overline \nu'_{e,\mu}$ ($\nu'_\tau$) 
MLTB can be approximately neglected provided that
\begin{eqnarray}
{\Gamma_1 \over H} &\stackrel{<}{\sim}& 1 \quad \Rightarrow 
\quad 0.13 G_F^2 T^3 {M_P \over 5.5}
\left( {\rho_{\overline \nu'_e} + \rho_{\overline \nu'_\mu} \over 
\rho_0} \right)\left({p \over 3.15T}\right)
\stackrel{<}{\sim} 1, \nonumber \\
{\Gamma_2 \over H} &\stackrel{<}{\sim}& 1 \quad \Rightarrow 
\quad 0.77 G_F^2 T^3{M_P \over 5.5}
\left( {\rho_{\nu'_\tau} \over \rho_0}
\right)\left({p \over 3.15T}\right) \stackrel{<}{\sim} 1.
\label{diu}
\end{eqnarray}
Our numerical work shows that
$(\rho_{\overline \nu'_e} + \rho_{\overline \nu'_\mu})/\rho_0 \ll 1$
until quite low temperatures $T/\text{MeV} \stackrel{<}{\sim} 
(\delta m^2_{\text{large}}/\text{eV}^2)^{1/4}$.
In fact the second condition in Eq.(\ref{diu}), from $\nu'_\tau \nu'_\tau$ elastic
collisions, is the
more stringent requirement. For this case we can estimate 
the collision rate by considering
a $\nu'_\tau$ of typical momentum $p/T \sim 8$, which is produced
at a temperature
$T/\text{MeV} \sim 1.3 (\delta m^2_{\text{large}}/\text{eV}^2)^{1/4}$. 
The ratio of energy
densities required is estimated from adiabatic conversion as
$P_2/T$ evolves from its initial value to about $8$: 
\begin{equation}
{\rho_{\nu'_\tau} \over \rho_0} 
\simeq {T^4 \over 2\pi^2 \rho_0}\int^{\infty}_{8}
{y^3 dy \over 1 + e^y} \simeq 0.04.
\end{equation}
Using these numbers
we estimate from Eq.(\ref{diu}) that $\nu'_\tau \nu'_\tau$
elastic collisions can be approximately neglected provided that
\begin{equation}
0.05 \left( {\delta m^2_{\text{large}} \over \text{eV}^2}
\right)^{3/4} \stackrel{<}{\sim} 1
\quad \Rightarrow \quad \delta m^2_{\text{large}}
\stackrel{<}{\sim} 50 \ \text{eV}^2.
\end{equation}
It turns out that this numerical bound is not very 
sensitive to what we choose for a
``typical'' $P_2/T$, so it is fairly robust.
Thus, in summary, we conclude that the thermalisation of the mirror
neutrinos can be approximately neglected provided that
\begin{equation}
\gamma_\rho \stackrel{<}{\sim} 
\left({\text{eV}^2 \over \delta m^2_{\text{large}}}\right)^{3 \over 4} \quad
\text{and} \quad
\delta m^2_{\text{large}} \stackrel{<}{\sim} 50 \ \text{eV}^2.
\label{thermbounds}
\end{equation}

Let us now discuss what happens when there is
significant thermalisation of the mirror
neutrinos.  Let us first consider the
case of the $\nu'_\tau$ states. If they are thermalised, 
the $\nu'_\tau$ distribution will
be close to an equilibrium distribution given by
\begin{equation}
N^{\text{eq}}_{\nu'_\tau} = {1 \over 2\pi^2}{p^2 \over 1 
+ \exp\left({p-\mu_{\nu'_\tau} \over T_{\nu'_\tau} }\right)}.
\end{equation}
The elastic $\nu'_\tau$ collisions with the
background $\nu'_\tau$ conserve both
the number density $n_{\nu'_\tau}$ and energy density 
$\rho_{\nu'_\tau}$, so these quantities can be used to determine
the two parameters $\mu_{\nu'_\tau}$ and $T_{\nu'_\tau}$. Now,
\begin{eqnarray}
n_{\nu'_\tau} &=& \int^{\infty}_0 N_{\nu'_\tau} dp
\simeq {T^3_{\nu'_\tau} \over \pi^2}e^{{\mu_{\nu'_\tau} \over
T_{\nu'_\tau}}}, \nonumber \\
\rho_{\nu'_\tau} &=& \int^{\infty}_0 N_{\nu'_\tau} pdp
\simeq {3T^4_{\nu'_\tau} \over \pi^2}e^{{\mu_{\nu'_\tau} \over
T_{\nu'_\tau}}}.
\end{eqnarray}
Thus,
\begin{equation}
T_{\nu'_\tau} = {\rho_{\nu'_\tau} \over 3n_{\nu'_\tau}},
\quad \mu_{\nu'_\tau} = T\ln\left( {27\pi^2 n^4_{\nu'_\tau} \over
\rho^3_{\nu'_\tau}}\right).
\end{equation}
Hence for any $T$, we can compute $T_{\nu'_\tau}$ and $\mu_{\nu'_\tau}$
by computing $\rho_{\nu'_\tau}$ and $n_{\nu'_\tau}$ from $N_{\nu'_\tau}$.
To estimate the effects of the repopulation we use the
equation,
\begin{equation}
\left.
{d \over dt} {N_{\nu'_{\tau}}(p) \over N^{\text{eq}}(p,T,0)} \right|_{\text{repop}} 
\simeq \Gamma_{2}(p)
\left[{N^{\text{eq}}_{\nu'_\tau}(p) \over
N^{\text{eq}}(p,T,0)}
- {N_{\nu'_\tau}(p) 
\over N^{\text{eq}}(p,T,0)}\right],
\end{equation}
where $\Gamma_{2}(p)$ is the elastic $\nu'_\tau \nu'_\tau$ collision
rate quoted earlier. 

The most important effect of the thermalisation of the $\nu'_\tau$ occurs
when $P_2/T$ evolves from $P_2/T|_{\text{min}}$ to infinity. 
For the example of Figs.3 and 4, namely $\delta m^2_{\text{large}} 
= 50 \ \text{eV}^2$, we have computed $N^{\text{eq}}_{\nu'_\tau}$ 
at the temperature $T_{\text{min}}$. This is shown in Fig.7
as a function of $p/T$ in order to compare this distribution 
with the distribution of $\nu'_\tau$ states which would exist 
in the absence of $\nu'_\tau$ thermalisation. In this
latter case, the MSW transitions that occurred during the 
previous evolution of $P_2/T$ from about $15$ down to about $6$, 
populated the $\nu'_\tau$ states from the tail of the
$\nu_{e,\mu}$ distributions (which have approximately 
negligible chemical potentials in this
region). Furthermore, ordinary weak interactions 
repopulated the depleted $\nu_{e,\mu}$
tails. This means, in the absence of mirror thermalisation, 
the journey back from $P_2/T|_{\text{min}}$ to infinity is 
dynamically inert as the oscillating species always
have approximately equal number densities in the resonance region. 
However, if the $\nu'_\tau$'s are thermalised this is not the case.

Computing the evolution of the system when $P_2/T$ evolves 
from $P_2/T|_{min}$ back to high values appears to be problematic. 
The problem is that in this region the resonance momenta
$P_2^a$ and $P_2^b$ for the modes $\nu'_\tau \leftrightarrow \nu_e$ 
and $\nu'_\tau \leftrightarrow \nu_\mu$, respectively, are 
not dynamically driven to coincide. Thus, in this case, we might 
expect different results depending on which resonance momentum goes
first.  Since the previous evolution of the system was such that 
the two resonance momenta coincided, it is not clear which 
resonance momentum will in fact go first. For instance, the
result may well depend on a statistical fluctuation, and therefore 
may be different in different regions of the universe. The physical 
implication of this would be a spatially dependent $Y_p$ distribution.

We have made some numerical estimates using the 
prescription given in Eq.(\ref{p2}). Our numerical results 
indicate that the overall affect of $\nu'_\tau$ thermalisation is not
unacceptably large, typically 
about $\delta N_{\nu,\text{eff}} \sim 0.3$ [for
the entire range of interest in $\delta m^2_{large}$]. 
For $L_{\nu_\tau} > 0$ the effect is positive, that is
$\delta N_{\nu,\text{eff}} \sim +0.3$, while for 
$L_{\nu_\tau} < 0$ the effect is negative, that is
$\delta N_{\nu,\text{eff}} \sim -0.3$.
This essentially results in a theoretical 
error of this magnitude for the
parameter space region which violates the 
bounds in Eq.(\ref{thermbounds}).

\subsection{Low temperature neutrino asymmetry evolution in the EPM: Case 2}

We now consider the case where the $\nu_\mu$ and $\nu'_\mu$ masses are
not negligible. 
This is of considerable interest since $m_{\nu_{\mu\pm}} \sim 1 \ \text{eV}$
is expected if the LSND anomaly \cite{lsnd} is due to neutrino oscillations.     
We consider the mass hierarchy,
\begin{equation}
m_{\nu_{\tau +}} \simeq m_{\nu_{\tau -}} \gg
m_{\nu_{\mu +}} \simeq  m_{\nu_{\mu -}} \gg
m_{\nu_{e +}},  m_{\nu_{e -}}.
\label{1b}
\end{equation}
In Case 2, there are two $\delta m^2$ scales. The modes
listed in Eq.(\ref{3})
have the large $\delta m^2 \equiv \delta m^2_{\text{large}}$,
while
\begin{equation}
\nu_\mu \leftrightarrow \nu'_e, \quad 
\nu_\mu \leftrightarrow \nu_e,  \quad
\nu'_\mu \leftrightarrow \nu_e, \quad 
\nu'_\mu \leftrightarrow \nu'_e,
\label{2b}
\end{equation}
plus the associated antiparticle modes have the
smaller $\delta m^2 \equiv \delta m^2_{\text{small}}$. The specification of the parameter
space of interest is completed by taking the remaining modes, $\nu_\alpha \leftrightarrow
\nu'_\alpha$ for $\alpha = e,\mu,\tau$, to have a negligible $\delta m^2$ (much
smaller than $\delta m^2_{\text{small}}$).

The four modes in Eq.(\ref{2b}) typically have distinct
resonance momenta which we denote as follows:
\begin{eqnarray}
\overline \nu_\mu &\leftrightarrow & 
\overline \nu'_e, \quad p_{\text{res}} = p_1,\nonumber \\
\nu'_\mu &\leftrightarrow & \nu_e, \quad p_{\text{res}} = p_2,\nonumber \\
\overline \nu_\mu &\leftrightarrow & \overline \nu_e, \quad p_{\text{res}} = p_3,\nonumber
\\
\overline \nu'_\mu &\leftrightarrow & \overline \nu'_e, \quad p_{\text{res}} = p_4.
\label{3b}
\end{eqnarray}
The resonance momenta for each of these
modes can be obtained from Eq.(\ref{7}),
\begin{equation}
{p_i \over T} = {\delta m^2_{\text{small}} \over a_0 T^4 L_i},
\label{4b}
\end{equation}
where $i=1,\ldots,4$ and
\begin{eqnarray}
L_1 &\equiv& L^{(\mu)} - L'^{(e)} = 2L_{\nu_\mu} +
L_{\nu_\tau} + L_{\nu_e}  - 2L_{\nu'_e} - L_{\nu'_\mu} -
L_{\nu'_\tau}, \nonumber \\
L_2 &\equiv& -(L'^{(\mu)} - L^{(e)}) = -2L_{\nu'_\mu} -
L_{\nu'_e} - L_{\nu'_\tau} + 2L_{\nu_e}
+ L_{\nu_\mu} + L_{\nu_\tau}, \nonumber \\
L_3 &\equiv& L^{(\mu)} - L^{(e)} = L_{\nu_\mu} - L_{\nu_e},
\nonumber \\
L_4 &\equiv& L'^{(\mu)} - L'^{(e)} = L_{\nu'_\mu} - L_{\nu'_e}.
\label{5b}
\end{eqnarray}
Observe that we use the lower case $p_i$ notation for
the $\delta m^2_{\text{small}}$ modes of Eq.(\ref{3b})
and the uppercase $P_i$ notation for the $\delta m^2_{\text{large}}$
modes of Eq.(\ref{4}).

Since $\delta m^2_{\text{small}} \ll \delta m^2_{\text{large}}$,
it follows that $p_{1,2} \ll P_i$. (Note however that $p_3$ and $p_4$ start out
being infinitely large because of the $L_{\nu_e} = L_{\nu_{\mu}}$ and 
$L_{\nu'_e} = L_{\nu'_{\mu}}$ conditions. As we will explain shortly,
these two modes have little effect). For our numerical
work, we will consider the parameter space region 
where the hierarchy between $\delta
m^2_{\text{small}}$ and $\delta m^2_{\text{large}}$
is great enough so that $p_{1,2}/T \stackrel{<}{\sim} 0.5$ when
$P_i/T \stackrel{>}{\sim} 10$. Numerically,
this corresponds to $\delta m^2_{\text{large}} \stackrel{>}{\sim} 50
\delta m^2_{\text{small}}$.
Let us denote by $T = T_x$ the temperature
at which the $P_i/T$ are all greater than 10. From Fig.3 and Eq.(\ref{7}) it is easy to see
that $T_x$ is given by 
\begin{equation}
{T_x \over \text{MeV}} \simeq 
0.3 \left({\delta m^2_{\text{large}} \over \text{eV}^2}\right)^{1 \over 4}. 
\label{iop}
\end{equation}
The evolution of the neutrino ensemble during the Low Temperature 
Epoch for Case 2 therefore
breaks up into two temperature regions:
$T \stackrel{>}{\sim} T_x$ and $T \stackrel{<}{\sim} T_x$.
When $T \stackrel{>}{\sim} T_x$, the 
evolution of the system is dominated by
the $\delta m^2_{\text{large}}$ modes of Eq.(\ref{4}),
because the $\delta m^2_{\text{small}}$ modes are
negligible due to their very small resonance momenta. 
In this temperature region
the evolution of the lepton numbers and number densities
can be evaluated using the equations of the previous 
section. When $T \stackrel{<}{\sim} T_x$,
the $\delta m^2_{\text{large}}$ modes are no
longer effective, because they all have
$P_i/T \stackrel{>}{\sim} 10$.
The $\delta m^2_{\text{small}}$ modes begin
to become important.
For $\delta m^2_{\text{small}} \sim 1 \ \text{eV}^2$,
it follows from Eq.(\ref{4b}) that $p_{1,2}/T \sim 1$ for $T \sim 1 \
\text{MeV}$. This means that for $\delta m^2_{\text{small}}$
in the LSND range, the oscillation modes of Eq.(\ref{2b}) become
important while the BBN reactions
$n \leftrightarrow p$ are still rapid. 
So, these oscillations
can potentially influence BBN and therefore
should not be ignored.

It turns out that it is not possible to use the adiabatic 
approximation [as encoded in Eq.(\ref{presevol})] to work out 
the effects of the oscillation subsystem given in Eq.(\ref{2b}). 
This is because of the structure 
of Eq.(\ref{4b}). For example, $\nu'_\mu
\leftrightarrow \nu_e$ oscillations create significant 
$L_{\nu_e}$ asymmetry, as $p_2/T$
sweeps through the $\nu_e$ momentum distribution. 
The $L_{\nu_e}$ asymmetry becomes so large
that $L_2 \to 0$. This makes the rate of change of $p_2/T$ 
very large and the system is no
longer adiabatic. Because of this complication, we will 
analyse the effects of the modes in
Eq.(\ref{2b}) using the Quantum Kinetic Equations.

We start integrating the QKEs for the subsystem of Eq.(\ref{2b}) 
at $T = T_x$ with the values of the number distributions 
$N_{\nu_\alpha}$ and $N_{\nu'_\alpha}$, and the lepton
numbers $L_{\nu_\alpha}$ and $L_{\nu'_\alpha}$ ($\alpha = 
e,\mu,\tau$) obtained from the
evolution equations of the previous Subsection. We will not 
explicitly write down the QKEs for this subsystem here, 
because their form is obvious once the contents of Sec.IV above have
been understood. Nevertheless for completeness we will
include them in Appendix B. 
 The reader, however, should not the following points:
\begin{enumerate}
\item We utilise the
approximation that the $x$ and $y$ components of 
the polarisation vectors for the modes of
Eq.(\ref{2b}) vanish at $T = T_x$. While this is not 
expected to actually be the case, the
subsequent evolution is not sensitive to the 
particular choices made for the initial values.
This is because the resonance momenta for the subsystems are 
either very small or very large, so the very first
stage of the evolution after $T = T_x$ is fairly unimportant. 
Furthermore, correct values for the $x$ and $y$ components are 
quickly generated by the QKEs soon after $T = T_x$. 
\item Repopulation and thermalisation of the mirror 
neutrino ensembles have been neglected,
consistent with our treatment of the evolution between the 
end of the High Temperature Epoch and $T = T_x$. 
Actually, this is an excellent approximation in this regime, since for
typical interesting parameter choices the mirror sector 
temperature is quite low.
\item The ordinary-ordinary $\overline \nu_\mu 
\leftrightarrow \overline \nu_e$ and mirror-mirror 
$\nu'_\mu \leftrightarrow \nu'_e$ modes can, to a good
approximation, actually be omitted. Recall that the 
resonance momenta $p_3$ and $p_4$ are initially very large, 
much larger than $p_1$ and $p_2$ respectively. Subsequent
evolution maintains this hierarchy in the resonance momenta. 
The $\overline \nu_\mu \leftrightarrow \overline \nu'_e$ mode, 
with resonance momentum $p_1$, is strongly
reprocessing lepton number as the resonance moves through 
the body of the $\overline \nu_\mu$ distribution. 
The coupled $\overline \nu_\mu \leftrightarrow \overline \nu_e$ mode,
on the other hand, sees its resonance momentum $p_3$ remain in 
the tail of the distribution, where it is ineffective because 
of the essentially identical number densities of $\overline
\nu_\mu$ and $\overline \nu_e$ in the tail. Remember that 
because weak interaction rates after typical values of $T_x$ 
are getting quite weak, there is little thermalisation of the
reprocessed $e$ and $\mu$ asymmetries created at low 
momenta $p_1$. In other words, the
$p_3$ resonance barely ``knows'' the asymmetry is there.
\end{enumerate}

In Figs.8,9 we plot the evolution of the resonance
momenta, $P_i/T$ and $p_i/T$ for an 
illustrative example. We choose $\delta m^2_{\text{large}} = 
50 \ \text{eV}^2$, $\delta m^2_{\text{small}} = 1\
\text{eV}^2$ and all of the vacuum mixing angles 
to be $10^{-8}$. We emphasise that our results should be 
approximately independent of the vacuum mixing angles so long as $10^{-10}
\stackrel{<}{\sim} \sin^2 2\theta \ll 1$. In Fig.10
we plot the evolution of all of the asymmetries for the same example. 

The effect of the oscillations on BBN is given in Figs.11-16. 
In Fig.11 we have plotted $N_{\nu,\text{eff}}$ versus 
$\delta m^2_{\text{small}}$, with $\delta m^2_{\text{large}} =
50\ \text{eV}^2$ and $L_{\nu_\tau} < 0$. Figures 12 and 13 
are similar except $\delta m^2_{\text{large}} = 200$ and 
$800 \ \text{eV}^2$, respectively. Figures 14-16 are the same
as Figs.11-13 except that the opposite sign asymmetries have 
been considered. {\it It is very important to note that the 
effect of a nonzero $\delta m^2_{\text{small}}$ is
considerable. In particular, as Figs.11-13 illustrate, 
$N_{\nu, \text{eff}}$ depends sensitively on 
$\delta m^2_{small}$, with
negative corrections to $N_{\nu,\text{eff}}$ equal to about 
one effective neutrino flavour are achieved for $\delta
m^2_{\text{small}}$ values in the few $\text{eV}^2$ range.} 
In Figs.14-16 observe that $L_{\nu_e}$ has a large 
effect even at very low values compared to the corresponding effect
in Figs.11-13. This asymmetry is due to the neutron/proton 
mass difference.

Recall that Figs.11-16 have not included the
effects of mirror neutrino
thermalisation. As already discussed,
these effects should be significant 
for the modes with $\delta m^2 = \delta m^2_{\text{large}}$
if $\delta m^2_{\text{large}} \stackrel{>}{\sim} 50 \ \text{eV}^2$. 
The thermalisation of mirror neutrinos should
have negligible effect for the modes with $\delta m^2 = \delta m^2_{\text{small}}$
because they are only important
when the temperature is typically less than about 1 MeV.
As discussed earlier, our rough estimate of the effect of
the mirror thermalisation is about $\delta N_{\nu,\text{eff}} \sim 0.3$.
For $L_{\nu_\tau} < 0$ ($L_{\nu_\tau} > 0$), the effect of
mirror thermalisation should be to {\it decrease}
({\it increase}) $\delta N_{\nu,\text{eff}}$ by of order $-0.3$
($+0.3$).

As mentioned above, the numerical results
were obtained using the parameter choice $\sin^2 2\theta_{\mu
e'} = 10^{-8}$. Actually we expect the results 
to be quite insensitive to $\sin^2 2\theta_{\mu e'}$ so long as
$\sin^2 2\theta_{\mu e'} \ll 1$. The reason is that the
amount of $L_{\nu_e}$ that gets created 
is already close to the maximal amount possible.
That is, after its rapid creation (which is at $T \sim 0.6\ \text{MeV}$
in the example in Figure 8), the quantity $L^{(e \mu')} \sim 0$. Increasing $\sin^2
2\theta_{\mu e'}$
cannot increase the amount of $L_{\nu_e}$ much since
it is already close to the maximum possible. Also, 
it cannot be created much earlier. Thus, the results
shown in Figs.11-16 should be approximately independent
of $\sin^2 2\theta_{\mu e'}$.\footnote{
We have numerically checked this by looking at 
the case $\sin^2 2\theta_{\mu e'} = 10^{-7}$ and found
almost identical results. We have also checked smaller
$\sin^2 2\theta_{\mu e'}$. For $\sin^2 2\theta_{\mu e'} 
\stackrel{<}{\sim} 10^{-9}$  the oscillations begin
to become so non-adiabatic that the oscillations
start to become less effective.} 

Finally, we should remark that the results of this
section indicate that the
bounds obtained in Sec.IV can
be evaded somewhat. The reason is that even if
$\delta N_{\nu,\text{eff}} \simeq 1.5$ from
the high temperature population of mirror
states from the $\nu_\tau \to \nu'_\mu$ oscillations,
this can be compensated by a $\delta N_{\nu,\text{eff}}
\sim -1.0$ from the low temperature generation of 
a large $L_{\nu_e}$ asymmetry.

\section{Implications for hot dark matter}

In the scenario considered in this paper,
where $\nu_\mu \leftrightarrow \nu'_\mu$ oscillations solve the
atmospheric neutrino anomaly,
a BBN bound of $N_{\nu,\text{eff}} \stackrel{<}{\sim} 3.6$ implies
$m_{\nu_\tau} \stackrel{>}{\sim} 1\ \text{eV}$ for
$|\delta m^2_{\text{atmos}}| \simeq 10^{-2.5} \ \text{eV}^2$ 
(see Fig.2).  Neutrino masses in the
eV range have long been considered cosmologically interesting, because
they would make a significant
contribution to the energy density of the universe.
In the standard Big Bang model, the contribution of
massive standard neutrinos to the energy density is given by
the well known formula,
\begin{equation}
\Omega_{\nu} = {\sum_\alpha m_{\nu_\alpha}
\over h^2 92\ \text{eV}},
\label{dff}
\end{equation}
where $h$ is the usual cosmological parameter 
parameterising the uncertainty in the Hubble constant.
Thus, neutrinos in the eV mass range are a well known 
and well motivated candidate for hot dark matter.

Before the advent of information at high red shift values\cite{high}, 
large scale structure formation
studies strongly favoured a hot plus cold dark matter mixture 
with $\Omega_{\nu} \simeq 0.20 - 0.25$ \cite{sch}. While recent 
work incorporating the new high redshift large scale
structure data has reduced the need for a hot dark matter 
component, it remains an interesting possibility. Given that 
$\nu_\tau$ masses greater than a few eV or so are well 
motivated from the combined requirements of the atmospheric neutrino
anomaly and BBN, we see that the existence of neutrino 
hot dark matter is a generic prediction of the EPM.

There is an interesting complication in the hot dark matter 
story for the EPM (and models with sterile neutrinos) which 
we now discuss. We will take by way of concrete
example that only the $\nu_\tau$ (and $\nu'_\tau$) has an eV 
scale mass. Again for the sake of the example, we will 
consider the neutrino mass range required by what was the favoured 
hot plus cold dark matter scenario \cite{sch}, 
\begin{equation}
3 \ \text{eV} \stackrel{<}{\sim} m_{\nu_\tau} \stackrel{<}{\sim}
7 \ \text{eV},
\label{hdmrange}
\end{equation}
even though the present situation is less clear. The point we want to make is that
whatever a ``favoured neutrino hot dark matter mass range'' might be at any given time,
the situation is modified somewhat in the case of the 
EPM model. The reason is that the 
$\nu_\tau \leftrightarrow \nu'_\mu$ and $\nu_\tau \leftrightarrow 
\nu'_e$ oscillations
generate such a large $L_{\nu_{\tau}}$
that the total number of tau neutrinos is actually significantly 
reduced. For the parameter region
\begin{equation}
10 \stackrel{<}{\sim} \delta m^2_{\text{large}}/\text{eV}^2
\stackrel{<}{\sim} 300,
\label{ken}
\end{equation}
the final value of $L_{\nu_\tau}$ is about $0.27$.
The large final lepton number occurs because
about $70\%$ of the anti-neutrinos have
been depleted (for the $L_{\nu_\tau} > 0$ case) 
which means that the total number of tau neutrinos plus tau antineutrinos 
is roughly $0.65$ of the standard expectation.
(Note that the total number of neutrinos has not changed much:
the missing heavy tau antineutrinos have just been
converted into light mirror states.)
Also note that a small number of $\nu'_\tau$ are also
generated by the oscillations, and it turns out that
the total number of $\nu_\tau$ and $\nu'_\tau$ (plus antiparticle) states
is about $0.70$ of the standard expectation.
The effect of this is to change the ``favoured hot dark 
matter mass range'' from what the expectation would be 
in the absence of mirror (or sterile) neutrinos.
We can guess that in the context of the EPM
the ``favoured'' tau neutrino mass is actually about $50\%$ 
larger than the naive expectation. (A full large scale 
structure computation would need to be performed to fully 
explore the consequences of a depleted $\nu_\tau$ distribution.)  
Thus, in the EPM model, the hypothetical favoured mass 
range of Eq.(\ref{hdmrange}) becomes instead 
\begin{equation}
5 \ \text{eV} \stackrel{<}{\sim} m_{\nu_{\tau \pm}} \stackrel{<}{\sim}
10 \ \text{eV}.
\end{equation}
This means that  
$\delta m^2_{\text{large}}$ is expected to be in the range
\begin{equation}
25\ \text{eV}^2 \stackrel{<}{\sim} \delta m^2_{\text{large}} \stackrel{<}{\sim} 
100 \ \text{eV}^2,
\label{reg}
\end{equation}
provided of course that the scenario of Eq.(\ref{hdmrange}) is correct.
The hypothetical hot dark matter region of Eq.(\ref{reg}) 
is the shaded band on Fig.2. From this Figure, we see that there is 
considerable overlap
between the BBN allowed region and the hot dark matter region.

Finally, note that structure formation outcomes in 
hot plus cold dark matter models are generically sensitive 
to the number of eV neutrino flavours, not just to
$\Omega_{\nu}$. These studies typically assume that the 
number of eV neutrino flavours (usually taken to be degenerate 
in mass) at the epoch of matter-radiation equality is an 
integer. It is important to understand that this is only true 
provided that mirror or sterile neutrinos do not exist. Indeed, 
as we have just explained above, we expect $N_{\nu}^{\text{heavy}} 
\simeq 0.70$ in the EPM in the parameter space 
region of Eqs.(\ref{1},\ref{2}) and (\ref{ken}).

\section{Implications for the cosmic microwave background}

During the next decade or so, high precision measurements of the 
anisotropy of the cosmic microwave background (CMB)
will be performed by several experiments (such as the 
PLANCK and MAP missions). These satellites should be able to measure 
detailed spectral properties of 
the electromagnetic radiation in the universe at the
epoch of photon-matter decoupling \cite{turn}.
In this context it is important to note that mirror and sterile
neutrinos can leave their `imprint' on the cosmic
microwave background \cite{hr}. This information will 
complement knowledge obtained from BBN because, (i) BBN and 
photon decoupling take place at different epochs, and (ii) BBN is
sensitive to both the expansion rate and the direct 
effect of $L_{\nu_e}$ on nuclear reaction rates whereas the 
CMB is insensitive to the direct effects of the asymmetry.
Because of point (ii) we have to distinguish 
between {\it expansion rate} and {\it
effective} neutrino flavour counting. So, it is 
useful to introduce the quantities $N_{\nu}^{\text{light}}$ and 
$N_{\nu}^{\text{heavy}}$ which effectively count the 
number of light neutrino and heavy neutrino flavours, 
respectively, at the epoch of photon decoupling. These quantities, 
which quantify expansion rates, are to be used
in conjunction with $N_{\nu,\text{eff}}$ which contains both 
expansion rate and $L_{\nu_e}$ information. It is important to 
appreciate that the number of {\it relativistic} neutrino
flavours may be different at the time of photon decoupling 
compared to BBN.  So, in this context, `light'  means 
much less than about an eV, making these neutrinos 
relativistic at the epoch of photon decoupling, 
and `heavy' means more than about an eV, making those neutrinos 
approximately non-relativistic. 
Of course in the minimal standard model of particle physics
with its three massless neutrinos, $N_{\nu,\text{eff}}
= N^{\text{light}}_{\nu} = 3$ and $N^{\text{heavy}}_{\nu} = 0$.
However, in models with sterile or mirror neutrinos,
$N_{\nu,\text{eff}} \neq N_{\nu}^{\text{light}}$ and
$N_{\nu}^{\text{heavy}} \neq 0$ in general. 
It is also important to appreciate that in the
EPM (or in models with sterile neutrinos), none of these 
quantities is in general an integer.

The CMB implications of the EPM depend on the neutrino parameter region.    
If we take by way of example the mass hierarchy of Eqs.(\ref{1}) and (\ref{2}), with 
$m_{\nu_{\tau \pm}} \stackrel{>}{\sim} 1\ \text{eV}$ (as
suggested by Fig.2) then
\begin{equation}
N_{\nu}^{\text{heavy}} \simeq {n_{\nu_\tau} + n_{\nu'_\tau} 
+ \text{antiparticles}\over 2n_0},\quad
N_{\nu}^{\text{light}} = {\rho_{\nu_e} + \rho_{\nu_\mu} + 
\rho_{\nu'_e} + \rho_{\nu'_\mu}  + \text{antiparticles} \over 2\rho_0},
\end{equation} 
where $n_i$ ($\rho_i$) is the mass (energy) density of species $i$ with 
$n_0$ ($\rho_0$) being the mass (energy) density of a Weyl fermion distribution with zero 
chemical potential.
Taking $\delta m^2_{\text{large}}$ in the
range Eq.(\ref{ken}) we find that
\begin{equation}
N_{\nu}^{\text{heavy}} \approx 0.70,\quad
N_{\nu}^{\text{light}} \approx 2.3.
\end{equation}
This should be distinguishable from the minimal standard model expectation.

We conclude by emphasising that in general
the precise measurements of the CMB may well prove to 
be quite useful in distinguishing between various competing
explanations of the neutrino anomalies, since each
model should leave quite a distinctive imprint on the CMB. 

\section{Conclusion}

The Exact Parity Model is theoretically well motivated by the neurotic desire of some to
have the full Lorentz Group as an exact symmetry of nature. It is very interesting that this
model can, essentially as a byproduct, provide an elegant explanation of the atmospheric and
solar neutrino problems in a way that is fully compatible with the LSND results. In this
paper, we explored the novel cosmological phenomena implied by the existence of mirror
neutrinos.

We focussed on the parameter space region
\begin{equation}
m_{\nu_{e+}} \simeq m_{\nu_{e-}} \stackrel{<}{\sim} m_{\nu_{\mu +}} \simeq m_{\nu_{\mu -}}
\stackrel{<}{\sim} m_{\nu_{\tau +}} \simeq m_{\nu_{\tau -}}
\end{equation}
with all intergenerational vacuum mixing angles obeying
\begin{equation}
10^{-10} \stackrel{<}{\sim} \sin^2 2\theta \ll 1.
\end{equation}
The mass splittings amongst the $e$-like and $\mu$-like states were chosen to solve the
solar and atmospheric neutrino problems, respectively. 
The evolution of the neutrino and
mirror neutrino ensembles was then calculated for the 
cosmological epoch between $T =
m_{\mu}$ and Big Bang Nucleosynthesis. Generic 
outcomes were obtained for significant
regions of parameter space because (i) {\it some} of the 
final neutrino asymmetries turned out to be
independent of the oscillation parameters for a range of 
those parameters, and (ii) {\it many} of the modes were 
adiabatic and hence independent of vacuum mixing angles.

The most important specific conclusions were:
\begin{enumerate}
\item The $\nu_\mu \to \nu'_\mu$ solution to the atmospheric 
neutrino problem is consistent
with Big Bang Nucleosynthesis for the parameter space region 
illustrated in Fig.2. The
$\nu_\tau$ mass implied by this region makes the $\nu_\tau$ a 
hot dark matter particle. This calculation improves on that 
discussed in Ref.\cite{fv3} 
through the use of Quantum Kinetic Equations. 
\item The effect of EPM neutrino oscillations on the 
primordial Helium abundance has been
computed. We find that a large change to the effective 
number of neutrino flavours during
Big Bang Nucleosynthesis is produced for a range of 
parameters. In particular, a change
equivalent to adding or removing about one neutrino 
flavour is obtained when the $\nu_e -
\nu_\mu$ mass splitting is in the LSND range.
\end{enumerate}

\acknowledgments{
R.F. would like to thank S. Blinnikov for interesting
correspondence and for sending him a copy of one of
his papers.
R.F. is an Australian Research Fellow. R.R.V. 
is supported by the Australian Research Council.}

\newpage
\appendix

\section{Details of the Helium abundance computation}

The modification of the $\nu_e$ and $\overline \nu_e$ distributions
due to the creation of $L_{\nu_e}$ affects
Big Bang Nucleosynthesis.
This is primarily due to the modification of the
$n \leftrightarrow p$ reaction rates.
The result of this is a modification of the neutron/proton
ratio. The most important observable
effect of a small change to the neutron/proton
ratio is a modification to the prediction for the Helium
mass fraction $Y_p$. This effect can be expressed
as a change in the predicted $N_{\nu,\text{eff}}$
through the well known relation $\delta Y_p \simeq
0.012\delta N_{\nu,\text{eff}}$.\footnote{Of course we are not saying 
that this equivalence is exact.
It is not. The change in $Y_p$ due to the modification of the
$\nu_e$ and $\overline \nu_e$ distributions cannot be
exactly represented as a change in $N_{\nu,\text{eff}}$.
This is because these two effects will have different
impacts on the other primordial element abundances.
However, because a small modification in
the $\nu_e$ and $\overline \nu_e$ distributions,
or a small change in $N_{\nu,\text{eff}}$, {\it primarily} affects
$Y_p$, our use of the relation $\delta Y_p \simeq 0.012
\delta N_{\nu,\text{eff}}$ is reasonable.
We prefer to express our results in terms of 
$\delta N_{\nu,\text{eff}}$ rather than $\delta Y_p$ just because 
$\delta N_{\nu,\text{eff}}$ is a more familiar unit.}
In computing the modification of $Y_p$ due
to the modified neutrino distributions,
$N_{\nu_e}$ and $N_{\overline \nu_e}$, we do not
need to use a full nucleosynthesis code.
The reason is that the effects of the modified
neutrino distributions are only important for
temperatures $T \stackrel{>}{\sim} 0.4 \ \text{MeV}$,
well before nucleosynthesis actually occurs.
A review of standard Helium synthesis which
we found useful was Ref.\cite{weinberg}.
Our approach and notation follows this treatment quite closely.

Recall that the primordial Helium mass fraction, $Y_P$,
is related to the ratio of neutrons to nucleons, $X_n$, by
$Y_P = 2X_n$ just before nucleosynthesis. 
$X_n$ is governed by the differential equation
\begin{equation}
- {dX_n \over dt} = \lambda (n \to p)X_n
- \lambda (p \to n)(1 - X_n),
\label{lank}
\end{equation}
where
\begin{eqnarray}
\lambda (n \to p) &\equiv& 
\lambda (n + \nu_e \to p + e^-) + 
\lambda (n + e^+ \to p + \overline \nu_e) 
+ \lambda (n \to p + e^- + \overline \nu_e),
\nonumber \\
\lambda (p \to n) &\equiv& 
\lambda (p + e^- \to n + \nu_e) + 
\lambda (p + \overline \nu_e \to n + e^+) 
+ \lambda (p \to n + e^+ + \nu_e).
\end{eqnarray}
The rates for these processes are given by
\begin{eqnarray}
\lambda (n + \nu_e \to p + e^-) &=& A\int 
{v_e E_e^2 \stackrel{\sim}{N}_{\nu_e}  \over
1 + \exp(-E_e/T)} dp_{\nu},
\nonumber \\
\lambda (n + e^+ \to p + \overline \nu_e) &=&
A\int {p^2_e (p_{\nu}^2 - \stackrel{\sim}{N}_{\overline \nu_e})
\over 1 + \exp(E_e/T)}dp_e,
\nonumber \\
\lambda (n \to p + e^- + \overline \nu_e) &=&
A\int {v_e E_e^2(p_{\nu}^2 - \stackrel{\sim}{N}_{\overline \nu_e}) 
\over 1 + \exp(-E_e/T)} dp_{\nu},
\nonumber \\
\lambda (p + e^- \to n + \nu_e) &=& A\int
 {p^2_e (p_{\nu}^2 - \stackrel{\sim}{N}_{\nu_e}) \over
1 + \exp(E_e/T)} dp_e, \nonumber \\ 
\lambda (p + \overline \nu_e \to n + e^+) &=&
A\int {v_e E_e^2 \stackrel{\sim}{N}_{\overline \nu_e} 
\over 1 + \exp(-E_e/T)} dp_{\nu}, \nonumber \\
\lambda (p + e^- + \overline \nu_e \to n) &=&
A\int {v_e E^2_e \stackrel{\sim}{N}_{\overline \nu_e} \over 
1 + \exp(E_e/T)}dp_{\nu},
\end{eqnarray}
where $v_e = p_e/E_e$ is the velocity of the electron 
(we use $\hbar = c =1$ throughout) and $\stackrel{\sim}{N}_{\nu}$ is
related to the neutrino distribution functions by
$\stackrel{\sim}{N}_{\nu} \equiv 2\pi^2 N_{\nu}$. 
The constant $A$ can be expressed in
terms of the vector and axial vector coupling
constants of the nucleon\cite{weinberg}, 
\begin{equation}
A = {g_V^2 + 3g_A^2 \over 2\pi^3}.
\end{equation}
Also, $E_e$ and $E_{\nu}$ are related by
\begin{eqnarray}
E_e - E_{\nu} = Q \quad \text{for} \quad n + \nu_e \leftrightarrow
p + e^-, \nonumber \\
E_{\nu} - E_e = Q \quad \text{for} \quad n + e^+ \leftrightarrow
p + \overline \nu_e, \nonumber \\
E_{\nu} + E_e = Q \quad \text{for} \quad n \leftrightarrow p + e^- + \overline \nu_e,
\label{sri}
\end{eqnarray}
where $Q \equiv m_n - m_p \simeq 1.293 \ \text{MeV}$.
The integrals of Eq.(\ref{sri}) are taken
over all positive values of $p_{\nu}$ and $p_{e}$
allowed by these relations.

In order to compute $Y_P$ we need to know the time
when nucleosynthesis occurs and neutron decay
ceases. This is handled approximately
by simply stopping the evolution of $X_n$ at
a point where agreement with the expected value
of $Y_P \sim 0.24$ occurs, which we find to be
roughly when $t \approx 300$ seconds.
This approximation does not affect the accuracy of
our results at all since we are only interested
in the difference between $Y_P$ using the modified
$\nu_e$ and $\overline \nu_e$ distributions
and $Y_P$ using the standard distributions (i.e.
Fermi-Dirac distributions with zero chemical potentials).
Thus, to a excellent approximation, the modification of
$Y_P$ due to the non-standard neutrino distributions
has the form
\begin{equation}
\delta Y_p \simeq 2\delta X_n(t=300s),
\end{equation}
where $\delta X_n(t=300s)$ is the difference between 
$X_n(t=300s)$ computed using
the neutrino momentum distributions $N_{\nu_e}$ and $N_{\overline \nu_e}$
and $X_n(t=300s)$ using the standard momentum distributions.
Of course the distributions $N_{\nu_e}$ and $N_{\overline \nu_e}$ typically
depend on the time, so that the evolution of $X_n^{(0)}(t)$
must be computed concurrently with the 
evolution of $N_{\nu_e}$ and $N_{\overline \nu_e}$.
In solving the differential equation 
Eq.(\ref{lank}), we employ the usual the initial condition 
$X_n = 0.5$.

We have checked our code against some previous calculations.
For example, in Ref.\cite{olive} they consider
the case of a time independent neutrino chemical
potential (taken to arise from some unknown
physics at high temperature). From
Figure 2 of Ref.\cite{olive}, they find that
$\delta Y_p \simeq -0.020$ for $\mu_{\nu}/T = -\mu_{\overline \nu}/T
\simeq 0.09$ (for constant $\eta$).
Our code also gives exactly the same results under the same conditions.

Finally note that at low temperatures $T \stackrel{<}{\sim} m_e$,
the $e^+ e^-$ annihilation process increases the temperature of
the photons relative to the neutrinos. It also affects the 
time-temperature relation. In our numerical work, we take these effects
into account using the equations given in
Ref.\cite{weinberg} (suitably modified to incorporate three light
neutrino flavours instead of two).
Of course this detail actually does not affect our results much,
since most of the effects of neutrino asymmetries are only
important for temperatures $T \stackrel{>}{\sim} m_e$.
Nevertheless, following E. Hillary (private
communication) we include it because it is there.

\section{The quantum kinetic equations for the modes with $\delta m^2 = \delta
m^2_{\text{small}}$ in Case 2 of Section V}

This Appendix deals with the case defined by Eq.(\ref{1b}), where
there are a class of modes having $\delta m^2_{\text{large}}$ [see Eq.(\ref{3})] and
another class of modes having
$\delta m^2_{\text{small}}$ [see Eq.(\ref{2b})]. As 
discussed in Sec.V, the $\delta m^2_{\text{large}}$ and $\delta
m^2_{\text{small}}$
modes approximately decouple from each other provided that
$\delta m^2_{\text{large}} \stackrel{>}{\sim} 50 \delta m^2_{\text{small}}$.
The evolution of the $\delta m^2_{\text{large}}$ oscillations can
be evaluated using the adiabatic formalism of
Sec.VB. The $\delta m^2_{\text{small}}$ modes
can be neglected initially because their resonance momenta
satisfy $p_{1,2}/T \ll 1$. By the time
$T = T_x$ [see Eq.(\ref{iop})], the $\delta m^2_{\text{small}}$ modes
begin to be important, and provided that $\delta m^2_{\text{large}}
\stackrel{>}{\sim} 50 \delta m^2_{\text{small}}$ is satisfied, the
$\delta m^2_{\text{large}}$ modes can be neglected because
$P_i/T \stackrel{>}{\sim} 10$. To compute the effects of the 
$\delta m^2_{\text{small}}$ modes we must numerically integrate
the quantum kinetic equations. Thus, we start the quantum kinetic
equations at $T=T_x$ with the initial values of 
$N_{\nu_\alpha}$, $N_{\nu'_{\alpha}}$, $L_{\nu_\alpha}$ and $L_{\nu'_\alpha}$
obtained from the previous evolution involving the $\delta m^2_{\text{large}}$
modes. 

In this Appendix we do not follow exactly the notation of Sec.IV.
We adopt an equivalent but simplifing change of variables which
is very useful for complicated coupled-mode systems
such as the one we are currently dealing with.

For each of the four oscillation modes, we assign a
density matrix $P^i_{x,y,z,0}$ ($i=1,\ldots,4$).
In solving this system, it is convenient to use
the variables $P^i_x, P^i_y, N_{\nu_{\alpha}},
N_{\overline \nu_{\alpha}}, N_{\nu'_{\alpha}}$ and $N_{\overline \nu'_{\alpha}}$
rather than the variables $P^i_0, P^i_x, P^i_y$ and $P^i_z$.
The $P^i_{0,z}$  are related to the $N$'s as follows:
\begin{eqnarray}
P^1_0(p) = {N_{\overline \nu_\mu}(p) + N_{\overline \nu'_e}(p) \over 
N^{\text{eq}}(p,T,0)},\quad P^1_z(p) = 
{N_{\overline \nu_\mu}(p) - N_{\overline \nu'_e}(p) \over
N^{\text{eq}}(p,T,0)}, \nonumber \\
P^2_0 (p) = {N_{\nu_e}(p) + N_{\nu'_\mu}(p) \over
N^{\text{eq}}(p,T,0)},\quad 
P^2_z (p) = {N_{\nu_e}(p) - N_{\nu'_\mu}(p)\over
N^{\text{eq}}(p,T,0)}, \nonumber \\
P^3_0(p)  = {N_{\overline \nu_\mu}(p) + N_{\overline \nu_e}(p) \over
N^{\text{eq}}(p,T,0)},\quad 
P^3_z (p) = {N_{\overline \nu_\mu} (p) - N_{\overline \nu_e} (p) \over
N^{\text{eq}}(p,T,0)}, \nonumber \\
P^4_0 (p) = {N_{\overline \nu'_\mu} (p) + N_{\overline \nu'_e}(p)
\over N^{\text{eq}}(p,T,0)},\quad 
P^4_z (p) = {N_{\overline \nu'_\mu} (p) - N_{\overline \nu'_e} (p) \over
N^{\text{eq}}(p,T,0)}. 
\label{6b}
\end{eqnarray}
The evolution of the number densities has a contribution
from coherent effects and a contribution from the subsequent
repopulation. Thus,
\begin{equation}
{dN_{\nu_\alpha} \over dt} =
\left.  {dN_{\nu_\alpha} \over dt}\right|_{\text{osc}} + 
\left.  {dN_{\nu_\alpha} \over dt}\right|_{\text{repop}},\quad 
{dN_{\overline \nu_\alpha} \over dt} =
\left.  {dN_{\overline \nu_\alpha} \over dt}\right|_{\text{osc}} + 
\left.  {dN_{\overline \nu_\alpha} \over dt}\right|_{\text{repop}}. 
\end{equation}
The contribution from coherent effects can be broken up among
the four modes as follows:
\begin{eqnarray}
\left.  {dN_{\nu_e} \over dt}\right|_{\text{osc}} &=&
\left.  {dN_{\nu_e} \over dt}\right|_{\nu'_\mu \leftrightarrow
\nu_e}, \quad
\left.  {dN_{\overline \nu_e} \over dt}\right|_{\text{osc}} =
\left.  {dN_{\overline \nu_e} \over dt}\right|_{\overline \nu_\mu \leftrightarrow
\overline \nu_e}, \nonumber \\
\left.  {dN_{\nu_\mu} \over dt}\right|_{\text{osc}} &=& 0,\quad
\left.  {dN_{\overline \nu_\mu} \over dt}\right|_{\text{osc}} =
\left. {dN_{\overline \nu_\mu} \over dt}\right|_{\overline \nu_\mu \leftrightarrow
\overline \nu_e} + 
\left. {dN_{\overline \nu_\mu} \over dt}\right|_{\overline \nu_\mu \leftrightarrow
\overline \nu'_e}, \nonumber \\
\left.  {dN_{\nu'_e} \over dt}\right|_{\text{osc}} &=& 0,\quad
\left.  {dN_{\overline \nu'_e} \over dt}\right|_{\text{osc}} =
\left.  {dN_{\overline \nu'_e} \over dt}\right|_{\overline \nu_\mu \leftrightarrow
\overline \nu'_e} + 
\left. {dN_{\overline \nu'_e} \over dt}\right|_{\overline \nu'_\mu \leftrightarrow
\overline \nu'_e}, \nonumber \\
\left.  {dN_{\nu'_\mu} \over dt}\right|_{\text{osc}} &=& 
\left. {dN_{\nu'_\mu} \over dt}\right|_{\nu'_\mu \leftrightarrow \nu_e}, \quad 
\left.  {dN_{\overline \nu'_\mu} \over dt}\right|_{\text{osc}} = 
\left.  {dN_{\overline \nu'_\mu} \over dt}\right|_{\overline \nu'_\mu \leftrightarrow 
\overline \nu'_e},
\end{eqnarray}
with 
\begin{eqnarray}
\left. {dN_{\overline \nu_\mu} \over dt}\right|_{\overline \nu_\mu 
\leftrightarrow \overline \nu'_e} &=&
- \left. {dN_{\overline \nu'_e} \over dt}\right|_{\overline 
\nu_\mu \leftrightarrow \overline \nu'_e} = {1 \over 2}
\int  \beta_1 \overline P^1_y 
N^{\text{eq}}(p,T,0)dp,
\nonumber \\
\left.  {dN_{\nu'_\mu} \over dt}\right|_{\nu'_\mu \leftrightarrow \nu_e}   
&=& - {dN_{\nu_e} \over dt}|_{\nu'_\mu \leftrightarrow \nu_e} =
-{1 \over 2} \int \beta_2 P_y^2  
N^{\text{eq}}(p,T,0)dp,
\nonumber \\
\left.  {dN_{\overline \nu_\mu} \over dt}\right|_{\overline 
\nu_\mu \leftrightarrow \overline \nu_e} &=&
- \left. {dN_{\overline \nu_e} \over dt}\right|_{\overline 
\nu_\mu \leftrightarrow \overline \nu_e} = {1 \over 2}
\int \beta_3\overline P^3_y N^{\text{eq}}(p,T,0)dp,
\nonumber \\
\left.  {dN_{\overline \nu'_\mu} \over dt}\right|_{\overline 
\nu'_\mu \leftrightarrow \overline \nu'_e} &=&
- \left. {dN_{\overline \nu'_e} \over dt}\right|_{\overline 
\nu'_\mu \leftrightarrow \overline \nu'_e} = {1 \over 2}
\int \beta_4\overline P^4_y 
N^{\text{eq}}(p,T,0)dp,
\label{uuu}
\end{eqnarray}
where $\beta_i = -\delta m^2_{\text{small}}\sin 2\theta_i/2p$ for
$i=1,3,4$ and $\beta_i = \delta m^2_{\text{small}}\sin 2\theta_i/2p$ for
$i=2$.  The rate of change of the number densities due to
repopulation is handled approximately via the equation,
\begin{equation}
\left. {d \over dt} {N_{\nu_\alpha}(p) \over N^{\text{eq}}(p,T,0)}
\right|_{\text{repop}} 
\simeq \Gamma_{\alpha}(p)
\left[{N^{\text{eq}}(p,T,\mu) \over 
N^{\text{eq}}(p,T,0)} - {N_{\nu_\alpha}(p) \over 
N^{\text{eq}}(p,T,0)} \right]
\end{equation}
where $\Gamma_{\alpha}(p)$ is the total collision
rate, and $N^{\text{eq}}(p,T,\mu)$ is the equilibrium
distribution which is a function of
the chemical potentials which can
be computed from the lepton numbers (see Sec.III for further
discussion).

The rates of change of the lepton numbers are given by
\begin{eqnarray}
{dL_{\nu_e} \over dt} &=&
\left.  {dL_{\nu_e} \over dt}\right|_{\overline \nu_\mu \leftrightarrow 
\overline \nu_e} +
\left.  {dL_{\nu_e} \over dt}\right|_{\nu'_\mu 
\leftrightarrow \nu_e}, \nonumber \\
{dL_{\nu_\mu} \over dt} &=&
\left.  {dL_{\nu_\mu} \over dt}\right|_{\overline \nu_\mu 
\leftrightarrow \overline \nu'_e} +
\left. {dL_{\nu_\mu} \over dt}\right|_{\overline \nu_\mu 
\leftrightarrow \overline \nu_e}, \nonumber \\
{dL_{\nu'_e} \over dt} &=&
\left. {dL_{\nu'_e} \over dt}\right|_{\overline \nu_\mu 
\leftrightarrow \overline \nu'_e} +
\left. {dL_{\nu'_e} \over dt}\right|_{\overline \nu'_\mu 
\leftrightarrow \overline \nu'_e}, \nonumber \\
{dL_{\nu'_\mu} \over dt} &=&
\left. {dL_{\nu'_\mu} \over dt}\right|_{\nu'_\mu \leftrightarrow \nu_e} +
\left. {dL_{\nu'_\mu} \over dt}\right|_{\overline \nu'_\mu 
\leftrightarrow \overline \nu'_e}, 
\label{7b}
\end{eqnarray}
where 
\begin{eqnarray}
\left. {dL_{\nu_\mu} \over dt}\right|_{\overline \nu_\mu 
\leftrightarrow \overline \nu'_e} &=&
- \left. {dL_{\nu'_e} \over dt}\right|_{\overline \nu_\mu 
\leftrightarrow \overline \nu'_e} =
-{1 \over 2n_\gamma}\int 
\beta_1\overline P^1_y 
N^{\text{eq}}(p,T,0)dp,
\nonumber \\
\left. {dL_{\nu'_\mu} \over dt}\right|_{\nu'_\mu \leftrightarrow \nu_e} &=&
- \left. {dL_{\nu_e} \over dt}\right|_{\nu'_\mu \leftrightarrow \nu_e} =
-{1 \over 2n_\gamma}\int \beta_2 P_y^2 N^{\text{eq}}(p,T,0)dp,
\nonumber \\
\left. {dL_{\nu_\mu} \over dt}\right|_{\overline \nu_\mu 
\leftrightarrow \overline \nu_e} &=&
- \left. {dL_{\nu_e} \over dt}\right|_{\overline \nu_\mu 
\leftrightarrow \overline \nu_e} =
-{1 \over 2n_\gamma}\int 
\beta_3\overline P^3_y 
N^{\text{eq}}(p,T,0)dp,
\nonumber \\
\left. {dL_{\nu'_\mu} \over dt}\right|_{\overline \nu'_\mu 
\leftrightarrow \overline \nu'_e} &=&
- \left. {dL_{\nu'_e} \over dt}\right|_{\overline \nu'_\mu 
\leftrightarrow \overline \nu'_e} =
-{1 \over 2n_\gamma}\int \beta_4\overline P^4_y 
N^{\text{eq}}(p,T,0)dp.
\label{8b}
\end{eqnarray}
Actually it turns out that the effect of the $\overline \nu'_\mu \leftrightarrow
\overline \nu'_e$ mode can be neglected because 
$N_{\overline \nu'_\mu}(p_4) \simeq N_{\overline \nu'_{e}}(p_4)$.
This is because the modes with $\delta m^2 = \delta m^2_{\text{large}}$ create
approximately equal numbers of $\overline \nu'_\mu$ 
and $\overline \nu'_e$ states.
Also, the $\overline \nu_\mu \leftrightarrow \overline \nu'_e$
modes always have a lower resonance 
momentum than the $\overline \nu'_\mu \leftrightarrow 
\overline \nu'_e$ modes.
This means that the change in $N_{\overline \nu'_e}(p)$ due to
$\overline \nu_\mu \leftrightarrow \overline \nu'_e$
modes does not occur until the
$\overline \nu'_\mu \leftrightarrow \overline \nu'_e$ resonance 
momentum has already passed by.
By similar reasoning, the $\overline \nu_\mu \leftrightarrow \overline 
\nu_e$ modes can also neglected to a good approximation.

\newpage
\centerline{\large {\bf Figure Captions}}
\vskip 0.5cm
\noindent
Figure 1: $|L_{\nu_\tau}|/h$ (where $h \equiv T^3_{\nu}/T_{\gamma}^3$)
versus temperature for $\nu_\tau
\leftrightarrow \nu'_\mu$ oscillations with 
$\delta m^2 = -50\ \text{eV}^2$ and $\sin^2 2\theta = 10^{-8}$.
\vskip 0.5cm
\noindent
Figure 2: Region of parameter space in the $\sin^2 2\theta_{\tau \mu'},
-\delta m^2_{\tau \mu'}$ plane where $L_{\nu_\tau}$ is
generated rapidly enough so that the $\nu_\mu \leftrightarrow
\nu'_\mu$ oscillations cannot significantly populate the
$\nu'_\mu$ states (for $T \stackrel{>}{\sim} 0.4 \ \text{MeV}$).  
This region, which in the figure is denoted
by the ``Allowed Region'', includes all of 
the parameter space above the solid line(s). 
The top, middle and bottom solid lines correspond to 
the atmospheric $\delta m^2$ values of 
$\delta m^2_{\mu \mu'}/\text{eV}^2 = 10^{-2},\ 10^{-2.5}$ and
$10^{-3}$ respectively.
The dashed-dotted line is the nucleosynthesis bound 
Eq.(\ref{thetarange}), which takes $\delta N_{\nu, \text{eff}}
\stackrel{<}{\sim} 0.6$ for definiteness, and the shaded region is the hot dark
matter region indicted from
some studies of structure formation (see Sec.VI).

\vskip 0.5cm
\noindent
Figure 3: Evolution of the resonance momenta $P_i/T$, for
the example with $\delta m^2_{\text{large}} = 50 \ \text{eV}^2$.
The solid line, long-dashed line, short dashed line, and dashed-dotted
line correspond to $P_1/T, P_2/T, P_3/T$ and $P_4/T$ respectively.

\vskip 0.5cm
\noindent
Figure 4: Evolution of the lepton numbers for the same
example as Fig.3.

\vskip 0.5cm
\noindent
Figure 5: $\delta N_{\nu, \text{eff}}$ versus $\delta m^2_{\text{large}}$
for Case 1 [see Eqs.(\ref{1},\ref{2})].
The dashed line is the contribution $\delta_1 N_{\nu,\text{eff}}$
due to the effects of the $L_{\nu_e}$ asymmetry while
the dashed-dotted line is the contribution $\delta_2 N_{\nu,\text{eff}}$
due to the change in the expansion rate. The solid line
is the total contribution $\delta_1 N_{\nu,\text{eff}} +
\delta_2 N_{\nu,\text{eff}}$.
This figure considers the case $L_{\nu_\tau} < 0$.

\vskip 0.5cm
\noindent
Figure 6: Same as Figure 5 except $L_{\nu_\tau} > 0$ is considered.

\vskip 0.5cm
\noindent
Figure 7: $N^{\text{eq}}_{\nu'_\tau}$ 
(bottom solid line) at the 
temperature $T = T_{\text{min}}$ (see text) for the example 
of Figs.3 and 4. The top solid line is the expected distribution
of $\nu'_\tau$'s if the thermalisation due to the mirror
weak interactions is neglected.
The unit along the vertical axis is $\text{MeV}^2$.

\vskip 0.5cm
\noindent
Figure 8: Evolution of the resonance momenta $P_i/T$ and 
$p_i/T$ for the example with $\delta m^2_{\text{large}} = 50 \ \text{eV}^2$ and
$\delta m^2_{\text{small}} = 1\ \text{eV}^2$.
The bold lines on the right of the figure correspond
to the $\delta m^2_{\text{large}}$ modes and
the thin lines on the left of the figure correspond to
the $\delta m^2_{\text{small}}$ modes.
For the bold lines,
the solid line, long-dashed line, short-dashed line, and dashed-dotted
line correspond to $P_1/T, P_2/T, P_3/T$ and $P_4/T$ respectively.
For the thin lines,
the solid line, long-dashed line and dashed-dotted
line correspond to $p_1/T, p_2/T$ and $p_3/T \simeq p_4/T$
respectively.

\vskip 0.5cm
\noindent
Figure 9: Magnified version of the left-hand side of Figure 8.

\vskip 0.5cm
\noindent
Figure 10: Evolution of the lepton numbers for the same
example as Figs.8 and 9.

\vskip 0.5cm
\noindent
Figure 11: $N_{\nu,\text{eff}}$ versus $\delta m^2_{\text{small}}$
with $\delta m^2_{\text{large}} = 50 \ \text{eV}^2$. The case $L_{\nu_\tau} < 0$
(which it turns out implies $L_{\nu_e} > 0$) has been considered.
The dashed line is the contribution $\delta_1 N_{\nu,\text{eff}}$
due to the effects of the $L_{\nu_e}$ asymmetry while
the dashed-dotted line is the contribution $\delta_2 N_{\nu,\text{eff}}$
due to the change in the expansion rate. The solid line
is the total contribution $\delta_1 N_{\nu,\text{eff}} +
\delta_2 N_{\nu,\text{eff}}$.

\vskip 0.5cm
\noindent
Figure 12: Same as Figure 11 except $\delta m^2_{\text{large}} = 200\ \text{eV}^2$.

\vskip 0.5cm
\noindent
Figure 13: Same as Figure 11 except $\delta m^2_{\text{large}} = 800\ \text{eV}^2$.

\vskip 0.5cm
\noindent
Figure 14: Same as Figure 11 except 
$L_{\nu_\tau} > 0$ (and hence $L_{\nu_e} < 0$) is considered.

\vskip 0.5cm
\noindent
Figure 15: Same as Figure 12 except 
$L_{\nu_\tau} > 0$ (and hence $L_{\nu_e} < 0$) is considered.

\vskip 0.5cm
\noindent
Figure 16: Same as Figure 13 except 
$L_{\nu_\tau} > 0$ (and hence $L_{\nu_e} < 0$) is considered.

\newpage
\epsfig{file=f1.eps,width=15cm}
\newpage
\epsfig{file=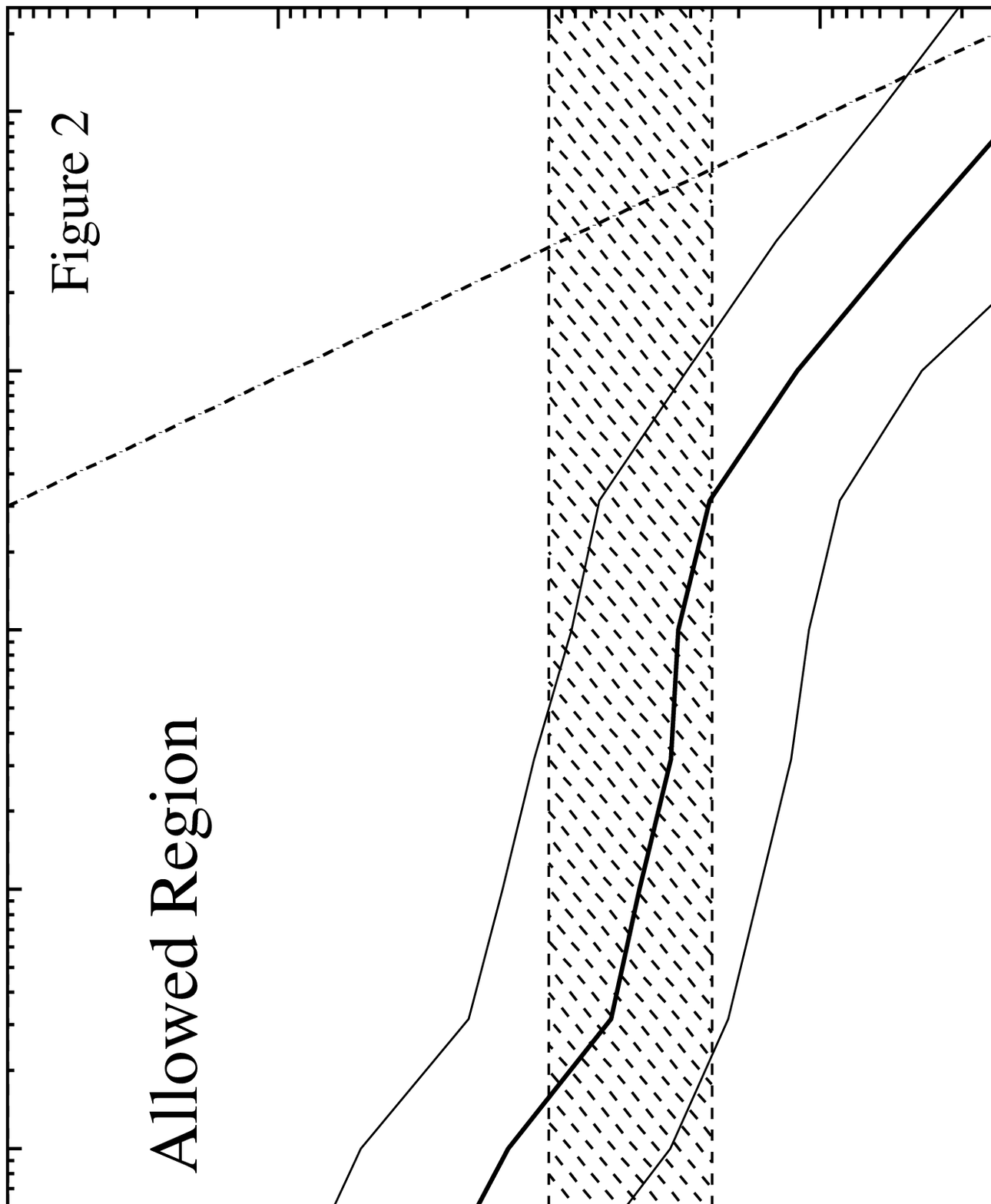,width=15cm}
\newpage
\epsfig{file=f3.eps,width=15cm}
\newpage
\epsfig{file=f4.eps,width=15cm}
\newpage
\epsfig{file=f5.eps,width=15cm}
\newpage
\epsfig{file=f6.eps,width=15cm}
\newpage
\epsfig{file=f7.eps,width=15cm}
\newpage
\epsfig{file=f8.eps,width=15cm}
\newpage
\epsfig{file=f9.eps,width=15cm}
\newpage
\epsfig{file=f10.eps,width=15cm}
\newpage
\epsfig{file=f11.eps,width=15cm}
\newpage
\epsfig{file=f12.eps,width=15cm}
\newpage
\epsfig{file=f13.eps,width=15cm}
\newpage
\epsfig{file=f14.eps,width=15cm}
\newpage
\epsfig{file=f15.eps,width=15cm}
\newpage
\epsfig{file=f16.eps,width=15cm}
\end{document}